\documentclass[prb,twocolumn,superscriptaddress,floatfix,longbibliography]{revtex4-2}

\usepackage{amsmath,amssymb,bm}
\usepackage{mathtools}
\usepackage{graphicx}
\usepackage{xcolor}
\usepackage{tikz}
\usetikzlibrary{arrows.meta}
\usepackage[colorlinks=true,linkcolor=blue,citecolor=blue,urlcolor=blue]{hyperref}

\newcommand{\gpi}{g_{\rm d}^{(\pi)}}
\newcommand{\mupred}{\mu_{\rm pred}}
\newcommand{\Vd}{V_{\mathrm d}}

\newcommand{\gd}{g_{\mathrm d}}
\newcommand{\cv}{c}

\newcommand{\Tr}{T_1}
\newcommand{\SG}{\mathrm{SG}}
\begin{document}

\title{Controlling truncation errors in loop tensor network renormalization}

\author{Rikuto Fushio}
\email{fushio-rikuto835@g.ecc.u-tokyo.ac.jp}
\affiliation{Department of Applied Physics, University of Tokyo, Tokyo 113-8656, Japan}
\author{Kokin Cho}
\email{cho-kokin446@g.ecc.u-tokyo.ac.jp}
\affiliation{Department of Physics, University of Tokyo, Tokyo 113-0033, Japan}
\author{Yuta Yamamoto}
\email{4345823097@g.ecc.u-tokyo.ac.jp}
\affiliation{Department of Physics, University of Tokyo, Tokyo 113-0033, Japan}
\author{Akira Otsuki}
\email{akira-ots@g.ecc.u-tokyo.ac.jp}
\affiliation{Department of Applied Physics, University of Tokyo, Tokyo 113-8656, Japan}
\author{Masahito Yamazaki}
\email{yamazaki@hep-th.phys.s.u-tokyo.ac.jp}
\thanks{Corresponding author.}
\affiliation{Department of Physics, University of Tokyo, Tokyo 113-0033, Japan}
\affiliation{Kavli Institute for Physics and Mathematics of the Universe (Kavli IPMU), UTIAS, University of Tokyo, Chiba 277-8583, Japan}
\affiliation{Center for Data-Driven Discovery (CD3), Kavli IPMU, UTIAS, University of Tokyo, Kashiwa, Chiba 277-8583, Japan}
\affiliation{Trans-Scale Quantum Science Institute, University of Tokyo, Tokyo 113-0033, Japan}
\affiliation{Center for Interdisciplinary Theoretical and Mathematical Sciences (iTHEMS), RIKEN, Saitama 351-0198, Japan}

\begin{abstract}
  At finite bond dimension $\chi$, tensor network renormalization (TNR) of a critical model eventually becomes gapped.
  For LoopTNR with $O(2)$ symmetry on the $\mathsf{c}=1$ critical line of the six-vertex model, parametrized by the Luttinger parameter $K$, we show that the correlation length $\xi$ of the gapped phase is set by the seed of the dimerization injected by the truncation, and depends on $\chi$ only through it.
  Read from a weak-coupling window early in the same run, the seed predicts $\xi=10^3$--$10^{16}$ via the sine-Gordon mass--coupling relation with no free parameter, to within a few percent for $K\le1.5$ and about $10\%$ at $K=1.75$; since the seed fluctuates strongly with $\chi$, $\xi$ is not monotonic in $\chi$.
  An alternation of the vertex weights shifts the seed; with its sign read from a $\pi$ twist, one short calibration predicts a counterterm that delays the breakdown in 16 of the 17 compensated groups of a campaign fixed in advance.
  A correction of the coarse-grained tensors, repeated during the flow, keeps the runs at $K=1$ and $1.5$ critical up to step 160, the last step computed, and exposes the marginal truncation error that remains.
  Our findings illustrate a systematic procedure for controlling the relevant truncation error, and thereby extending the critical window, in tensor network renormalization at finite bond dimension.
\end{abstract}

\maketitle

\section{Introduction}
\label{sec:intro}

A critical point is a fixed point of the renormalization-group (RG) flow, and the conformal field theory (CFT) at the fixed point characterizes universal properties that do not depend on the microscopic details of the system.
Tensor network renormalization implements the RG numerically in real space~\cite{Levin2007,Xie2012,Evenbly2015,Yang2017}.
At criticality the coarse-grained tensors approach a fixed-point tensor, from which the scaling dimensions~\cite{Gu2009,Cardy1984} and operator product expansion (OPE) coefficients~\cite{UedaYamazaki2023} of the CFT can be extracted.
At any finite bond dimension $\chi$, however, each truncation introduces a small error, and the RG flow amplifies its components along the relevant directions that the imposed symmetries allow, so that the run leaves the fixed point after a finite RG time.
When such a perturbation opens a gap, the correlation length of the approximate model being simulated sets the scale at which the spectrum departs from its critical form.
For matrix product states this length follows the finite-entanglement scaling $\xi_\chi\propto\chi^{\kappa_{\rm FE}}$~\cite{Tagliacozzo2008,Pollmann2009,Pirvu2012}.
The scaling has been interpreted as the effect of a relevant perturbation selected by the symmetries of the ansatz~\cite{Schneider2024,Huang2024}, and the same viewpoint has been applied to real-space tensor network renormalization: in the Ising and three-state Potts models the finite-$\chi$ spectra scale with a correlation length that is a power of $\chi$~\cite{Ueda2023}, and for the Ising model level-dependent crossover lengths also follow a power law in the bond dimension~\cite{HuangCY2023}.
It is natural to ask whether the correlation length of a real-space RG is likewise a smooth function of $\chi$ and, if it is not, what determines it and whether it can be controlled.

In a companion paper~\cite{Paper1a} we identified the truncation errors of loop tensor network renormalization (LoopTNR)~\cite{Yang2017} on the $\mathsf{c}=1$ critical line of the six-vertex model, a compact boson with tunable Luttinger parameter $K$.
With $O(2)$ symmetry imposed on the tensors, the dominant error is the dimerization operator $\Vd$ of dimension $K$.
It is allowed because LoopTNR preserves neither the one-site translation $\Tr$ nor the reflection $R_{\rm v}$ of the lattice, and its contribution to the levels grows by $2^{2-K}$ per RG step at second order.
The operator and its growth rate do not depend on $\chi$.
Its initial amplitude, which we call the seed, depends on $\chi$ and fluctuates by up to two orders of magnitude between neighboring $\chi$.
As a result the correlation length is not monotonic in $\chi$: at $K=1$, increasing $\chi$ from $24$ to $27$ shortens it by a factor of $17$, from $7.8\times10^5$ to $4.6\times10^4$ lattice spacings (Appendix~\ref{app:seed}).
For $K<2$ the dimerization is relevant, so every run with a nonzero seed eventually leaves the critical line.
The long-distance theory is then the sine-Gordon model, in which the mass of the lightest charged particle, a soliton, is related to the coupling by an exact formula~\cite{Zamolodchikov1995} that has been used for the dimerization gap of XXZ chains~\cite{Takayoshi2010,Hikihara2017}.

One of our main results is that this formula, applied to the seed read in the weak-coupling window, predicts the correlation length of the gapped phase with no free parameter (Sec.~\ref{sec:xi}).
Since the coupling is read early in the same run and the power of the coupling in the formula follows from its dimension, this tests the amplitude of the mass--coupling relation, one number for each $K$.
At a fixed window cut the median of the predicted over the measured length deviates from unity by less than $1\%$ for $K\le1.5$ and by $3\%$ at $K=1.75$, over $\xi=10^3$--$10^{16}$ lattice spacings, with a systematic uncertainty of $1$--$3\%$ for $K\le1.5$ and about $10\%$ at $K=1.75$; individual runs with $\chi\ge24$ deviate by up to $5\%$ for $K\le1.5$ and $11\%$ at $K=1.75$.
It accounts for the irregular dependence of $\xi$ on $\chi$, and hence of the number of usable RG steps, which is fixed by $\xi$.

A second main result is that the seed can be shifted and cancelled at fixed $\chi$, so that the critical window is lengthened without increasing $\chi$ (Secs.~\ref{sec:control} and \ref{sec:signed}).
An alternation of the vertex weights $a$ and $b$, which carries the lattice quantum numbers of the dimerization, shifts the seed linearly and moves $\xi$ as predicted.
A staggered vertex weight, which differs from the dimerization by a reflection parity, does not measurably couple to it.
Tuned to cancel the seed, the alternation delays the breakdown of the conformal spectrum by 6 to 38 steps at fixed $\chi$.
A $\pi$ twist of the $U(1)$ symmetry in the readout makes the seed visible at first order, together with its sign, so that one short calibration predicts this counterterm (Sec.~\ref{sec:signed}).
After refinement with a few further short runs, the compensated run at $K=1.5$ and $\chi=40$ outlasts an uncompensated run at $\chi=96$ at a few percent of its cost.
The compensation lengthens the critical window but leaves the levels within it as accurate as those of the uncompensated run at the same $\chi$.
It is therefore useful when many RG steps, i.e., large systems, are needed, but it does not make the scaling dimensions more precise.

Third, when the correction is repeated during the flow, a run at $K=1$ or $1.5$ stays on the critical line up to step 160, the last step computed, with a slowly drifting $K$; the repeated correction thereby exposes the marginal error that remains once the dimerization is removed (Sec.~\ref{sec:iterative}).
Finally, for $K\le1.5$ the seed accumulates mostly over the first steps after the first truncation, followed by a weaker injection that acts like a constant source and supplies about half of it at $K=1.75$; it depends strongly on the loop optimization and is not a sum of independent contributions of individual steps (Sec.~\ref{sec:branches}).

This paper builds on Ref.~\cite{Paper1a}.
We use the same code, notation, and readout conventions, which are summarized in Sec.~\ref{sec:setup} together with the definition of the seed.
Section~\ref{sec:xi} tests the prediction of $\xi$, Sec.~\ref{sec:control} the dependence on $\chi$ and the control by the alternation, Sec.~\ref{sec:signed} the signed readout and the one-shot compensation, Sec.~\ref{sec:iterative} the correction during the RG, and Sec.~\ref{sec:branches} where the seed is injected.
Section~\ref{sec:discussion} summarizes and discusses the results.

\section{Setup}
\label{sec:setup}
\subsection{Model and dimerization}

We study the six-vertex model~\cite{Lieb1967,Baxter1982} with weights $(a,b,\cv)=(1,1,\cv)$, which define the F model~\cite{Rys1963}, in the notation of Ref.~\cite[Sec.~II]{Paper1a}.
The model is critical for $0<\cv<2$, and we parametrize it by
\begin{equation}
  \Delta_{6\mathrm v}\coloneqq1-\frac{\cv^2}{2},\qquad
  K\coloneqq\frac{\pi}{2\arccos\Delta_{6\mathrm v}},
  \label{eq:K}
\end{equation}
so that $K>1/2$, with the free-fermion point at $K=1$ and the BKT transition of the F model at $K=1/2$~\cite{vanBeijeren1977}.
In the continuum it becomes a compact boson~\cite{Kadanoff1979,Nienhuis1984,Alcaraz1988} whose vertex operators $V_{e,m}$, with dual charge $e$ and $U(1)$ charge $m$, have conformal spin $em$ and scaling dimension
\begin{equation}
  x_{e,m}=e^2K+\frac{m^2}{4K} ,
  \label{eq:spectrum}
\end{equation}
with the normalization and conventions of Ref.~\cite[Sec.~II]{Paper1a}.
Charge conjugation maps $(e,m)\to(-e,-m)$, one-site translation $\Tr$ acts as $V_{e,m}\to(-1)^eV_{e,m}$, and the reflection $R_{\rm v}$ about a line through a column of vertices maps $V_{e,m}\to(-1)^eV_{-e,m}$~\cite[App.~A]{Paper1a}.
The dimerization field
\begin{equation}
  \Vd\coloneqq\frac{V_{1,0}+V_{-1,0}}{\sqrt2}
  \label{eq:Vd}
\end{equation}
of dimension $K$ is therefore $U(1)$ neutral and $C$ even, but odd under both $\Tr$ and $R_{\rm v}$.
In the XXZ chain it corresponds to the bond alternation~\cite{Takayoshi2010,Giamarchi2004}, as checked in Ref.~\cite[App.~B]{Paper1a}.
If $O(2)$ and $\Tr$ are both preserved, no relevant perturbation is allowed for $K>1/2$.
LoopTNR, however, coarse-grains a network with a two-tensor unit cell, on which the original $\Tr$ has no local representation, and it does not impose $R_{\rm v}$ either.
The dimerization $\Vd$ is therefore allowed even when $O(2)$ is imposed, and it is relevant for $K<2$~\cite{Paper1a}.

\subsection{LoopTNR and readout}
We use LoopTNR~\cite{Yang2017} as implemented in TNRKit.jl~\cite{TNRKit} with $O(2)$ imposed and the code and parameters of the main runs of Ref.~\cite{Paper1a} (Appendix~\ref{app:numerics}).
Each RG step rescales lengths by $\sqrt2$ and rotates the lattice by $135^\circ$~\cite{TNRKitCode}.
At step $n$ the coarse-grained tensors are contracted into a cylinder of circumference $2\sqrt2\,a_n$ and transfer length $\sqrt2\,a_n$ ($a_n=2^{n/2}a_0$), and the leading eigenvalues $\lambda_i$ of its transfer matrix give the scaling dimensions~\cite{Cardy1984,Gu2009}
\begin{equation}
  x_i(n)\coloneqq-\frac{1}{2\pi\operatorname{Im}\tau_n}\ln\left|\frac{\lambda_i}{\lambda_0}\right| ,\qquad \tau_n=i/2 .
  \label{eq:readout}
\end{equation}
We follow the relative deviation of the lowest charged level, $x_{0,1}=1/(4K)$,
\begin{equation}
  X(n)\coloneqq1-4Kx_{0,1}(n).
  \label{eq:X}
\end{equation}
We measure lengths in units of $a_0$ and write $L\coloneqq2\sqrt2\,(\sqrt2)^n$ for the circumference of the readout cylinder.
We write $g_j(L)\coloneqq\tilde g_j(L/2\pi)^{2-x_j}$ for the dimensionless couplings of operators $\Phi_j$ of dimension $x_j$ in the Hamiltonian $H=H_{\rm CFT}+\sum_j\tilde g_j\int_0^Ldy\,\Phi_j(y)$ on a circle of circumference $L$, which generates the transfer matrix along the tube~\cite[Sec.~III B]{Paper1a}.
To second order, the dimerization shifts the charged level by $\delta x_{0,1}=\alpha_{0,1}(K)\,\gd^2$, with $\alpha_{0,1}(K)$ tabulated in Ref.~\cite[App.~B]{Paper1a} ($\alpha_{0,1}=4\pi^2\ln2$ at $K=1$).
Since $\gd(L)\propto L^{2-K}$, this contribution grows by
\begin{equation}
  r_{\rm d}=2^{2-K}
  \label{eq:rate}
\end{equation}
per step.
After a short-distance transient, $X$ remains at a plateau of order $10^{-4}$ until the dimerization grows out of it.
We define $n_e$ as the last step before the lowest charged level first deviates by more than $10\%$ from its critical value at the input $K$, and use this threshold as our \emph{breakdown} criterion.
\subsection{Seed}
\label{sec:window}
\begin{figure}[tb]
  \centering
  \includegraphics[width=\columnwidth]{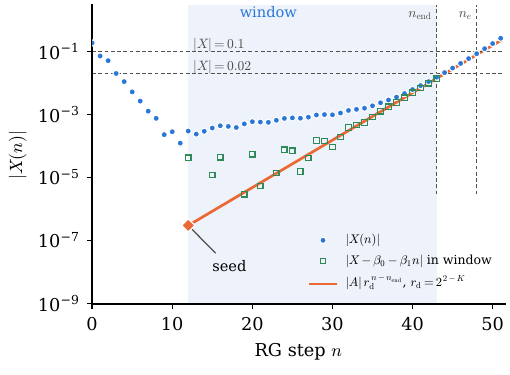}
  \caption{The protocol on the unperturbed run at $K=1.5$ and $\chi=40$ ($O(2)$). Dots: deviation $|X(n)|$ of the lowest charged level. The window (shaded) starts at $n=12$ and ends at $n_{\rm end}=43$, the last step with $|X|\le0.02$; squares: $X$ minus the fitted background $\beta_0+\beta_1n$ in the window; line: the growth term of Eq.~\eqref{eq:fit} with $r_{\rm d}=2^{2-K}$ fixed, extrapolated back to $n=12$, which gives the squared seed ($\gd^2(12)=3.0\times10^{-9}$). The breakdown $|X|=0.1$ comes after $n_e=48$. Compare Ref.~\cite[Fig.~2(c)]{Paper1a}, where the cut is $|X|\le0.1$ and the rate is free.}
  \label{fig:window}
\end{figure}

To read the dimerization coupling we model the signed $X(n)$ as
\begin{equation}
  X(n)\simeq\beta_0+\beta_1n+A\,r_{\rm d}^{\,n-n_{\rm end}}
  \label{eq:fit}
\end{equation}
in a window that starts at $n=12$ and ends at $n_{\rm end}$, the last step before $|X|$ first exceeds $X_{\max}$.
The linear background absorbs the plateau of $X$ and the slow drift of the effective Luttinger parameter $K_n$~\cite[Sec.~IV F]{Paper1a}, and with the rate fixed to its second-order value $r_{\rm d}=2^{2-K}$ the fit has three free parameters, $\beta_0$, $\beta_1$, and $A$.
We obtain
\begin{equation}
  \gd^2(n)=-\frac{A\,r_{\rm d}^{\,n-n_{\rm end}}}{4K\alpha_{0,1}(K)} .
  \label{eq:g2}
\end{equation}
We cut the window at $X_{\max}=0.02$ (Fig.~\ref{fig:window}).

With this cut, a fit in which the rate is left free gives, for each $K$, a median rate within about $1\%$ of $2^{2-K}$.
The cut $X_{\max}=0.1$, which Ref.~\cite{Paper1a} used to measure the growth rate, introduces a bias at large $K$ through the running of $K$ induced by $\gd^2$~\cite[App.~D]{Paper1a}.
With $r_{\rm d}$ fixed, $\gd(n)$ at any step follows from $\gd(12)$, which we call the \emph{seed}; the fit determines only $\gd^2(12)$, and the sign of $\gd(12)$, the sign of the seed, is read in Secs.~\ref{sec:compensation} and \ref{sec:signed}.
It is the amplitude of the growth in the window referred back to $n=12$ at the fixed rate, not the coupling present at step 12, which does not yet contain the part injected later (Sec.~\ref{sec:branches}); at $K=1.75$ the extrapolation spans several tens of steps.
The seeds quoted in Ref.~\cite{Paper1a} were read with a free rate and $X_{\max}=0.1$ and differ from the present ones by a factor that depends on $K$ (Appendix~\ref{app:data}).

In the linearized BKT flow~\cite{Kosterlitz1974} the dimerization coupling has the RG eigenvalue $2-K$~\cite{Giamarchi2004}.
We describe the signed change of $\gd$ generated by the truncation over an RG interval $d\ell$ as $\sigma_{\mathrm d}(\ell)\,d\ell$, so that
\begin{equation}
  \frac{d\gd}{d\ell}=(2-K)\gd+\sigma_{\mathrm d}(\ell),
  \label{eq:sourceflow}
\end{equation}
with $\ell=\ln L$.
An amplitude observed in the window can then come from injection in the first steps or from sources acting later; since an injection at step $n$ is amplified by $(\sqrt2)^{(2-K)(n_{\rm end}-n)}$, the early steps dominate unless $K$ is close to $2$.
Section~\ref{sec:branches} separates the two contributions.

\begin{figure}[tb]
  \centering
  \includegraphics[width=\columnwidth]{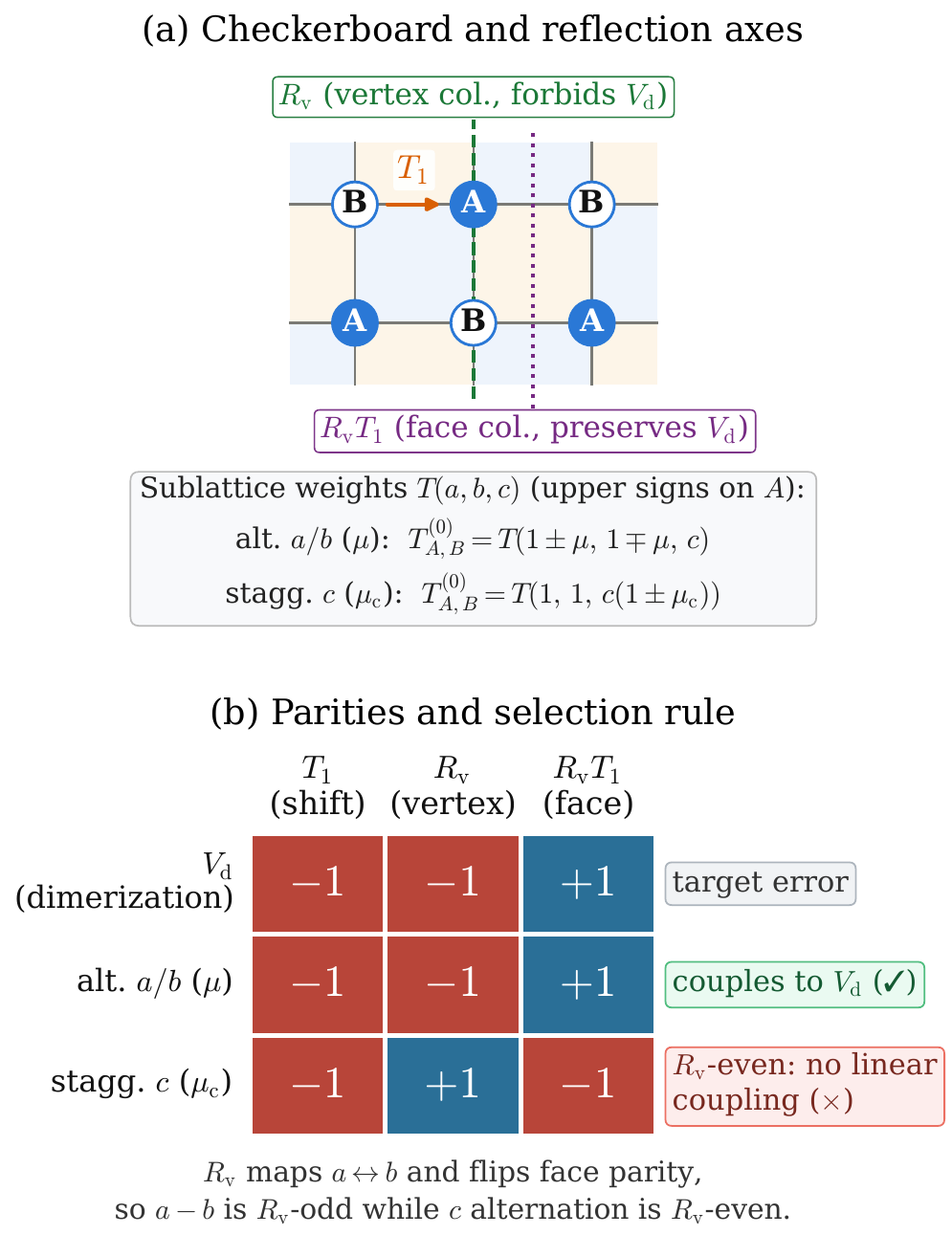}
  \caption{Initial deformations and lattice symmetries (Secs.~\ref{sec:window}, \ref{sec:compensation}, and~\ref{sec:discussion}).
    (a)~Checkerboard placement of the sublattice tensors $T_A^{(0)}$ and $T_B^{(0)}$ with the two face sublattices shaded.
    The one-site translation $\Tr$ exchanges $A$ and $B$; the reflection $R_{\rm v}$ about a column of vertices exchanges $a$ and $b$ and the face sublattices, whereas the reflection $R_{\rm v}\Tr$ about a column of faces preserves the face sublattices and leaves $\Vd$ invariant.
    (b)~Parities under $\Tr$, $R_{\rm v}$, and $R_{\rm v}\Tr$: the alternating $a/b$ deformation ($\mu$) has those of $\Vd$, whereas the staggered $\cv$ ($\mu_{\rm c}$) is even under $R_{\rm v}$ and does not couple linearly in the lattice model; in LoopTNR its coupling is below detection (Sec.~\ref{sec:compensation}).}
  \label{fig:initial_ab_symmetry}
\end{figure}

To vary the seed at fixed $\chi$, some runs in Secs.~\ref{sec:xi}--\ref{sec:signed} and \ref{sec:branches} start from a deformed initial network.
Before the first coarse-graining step, two six-vertex tensors with $(a,b)=(1+\mu,1-\mu)$ and $(1-\mu,1+\mu)$, and $\cv$ unchanged, are placed on the two checkerboard sublattices of the square lattice [Fig.~\ref{fig:initial_ab_symmetry}(a)], so that $a-b$ changes sign between neighboring vertices; at $\mu=0$ both carry the same tensor.
This alternation has the same lattice quantum numbers as $\Vd$ [Fig.~\ref{fig:initial_ab_symmetry}(b)] and shifts the seed linearly in $\mu$, which Sec.~\ref{sec:compensation} uses to cancel it.

\section{The seed sets the correlation length}
\label{sec:xi}

In this section we define the correlation length of the gapped phase that a run eventually reaches and derive its prediction from the seed read in the weak-coupling window (Sec.~\ref{sec:window}).
As discussed in Sec.~\ref{sec:circularity}, the comparison tests the amplitude of the prediction (Sec.~\ref{sec:prediction}), whether the coupling read in the window predicts the flow after it (Sec.~\ref{sec:collapse}), and whether the gapped phase is that of the sine-Gordon model (Sec.~\ref{sec:sgspectrum}).

Once the dimerization has grown, a run flows into a gapped phase, and the energy $E=-\ln|\lambda/\lambda_0|/(\text{tube height})$ of the lowest level with $|m|=1$, read with the cylinder of Sec.~\ref{sec:setup} (circumference $L$, height $L/2$), no longer decreases with $n$.
This level is the lightest charged particle of the gapped phase, which we identify with a soliton of the sine-Gordon model in Sec.~\ref{sec:formula}.
Its inverse mass $\xi$ is the longest correlation length for $K>1/2$ (Sec.~\ref{sec:formula}).
We take $\xi=1/E$, averaged over the last two steps before the levels reach the numerical floor of the readout, as the measured correlation length.
This length does not depend on the circumference of the readout (Appendix~\ref{app:data}).
We use the $\chi$ scan ($\chi=16$--$64$) at $K=0.75$--$1.75$, the integer scan (every $\chi$ from 24 to 48 at $K=1$ and $1.5$), the runs at $\chi=96$ at $K=1$ and $1.5$, and the runs with the alternating deformation of Sec.~\ref{sec:window}, described in Sec.~\ref{sec:control}.
Altogether, 233 runs reach the gapped phase (Table~\ref{tab:xi}); at the last step used, the circumference is $L=8.7$--$16\,\xi$, so that finite-size corrections to the soliton energy are negligible.

\begin{figure}[tb]
  \centering
  \includegraphics[width=\columnwidth]{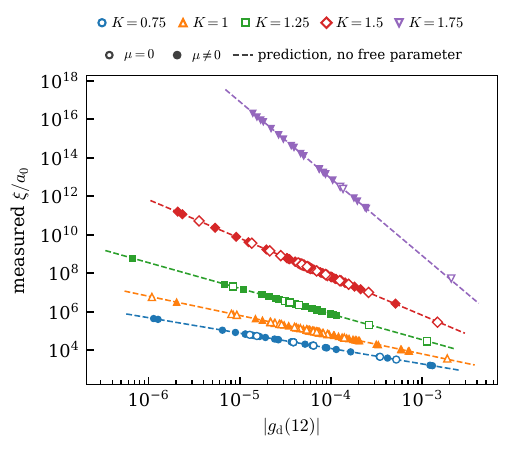}
  \caption{Measured correlation length against the coupling $|\gd(12)|$ read in the weak-coupling window ($r_{\rm d}=2^{2-K}$, $|X|\le0.02$), for all 233 runs that reach the gapped phase: open symbols, the 80 unperturbed runs ($\mu=0$); filled, the 153 runs with the alternating deformation (Sec.~\ref{sec:control}). Dashed lines: Eq.~\eqref{eq:xi} with no free parameter. As explained in Sec.~\ref{sec:circularity}, the slope does not test Eq.~\eqref{eq:xi}, and the test is the height of each line, shown in Fig.~\ref{fig:ratio}.}
  \label{fig:xi_vs_g}
\end{figure}

\begin{table*}[tb]
  \caption{Ratio of the predicted to the measured correlation length for the runs that reach the gapped phase. Number of runs and median over all runs, and range excluding $\chi=16$, with the protocol of Sec.~\ref{sec:window} ($r_{\rm d}=2^{2-K}$, $|X|\le0.02$); median extrapolated to a vanishing window cut and slope of the bias of $\ln\gd^2$ with the cut (see text), both excluding $\chi=16$. Last two columns: $L/\xi$ at $n_e$ with the measured $\xi$, the last step with $|X|\le0.1$, and the value at which the second-order curve reaches $|X|=0.1$ (Sec.~\ref{sec:collapse}).}
  \label{tab:xi}
  \begin{ruledtabular}
    \begin{tabular}{rrcccccc}
      $K$  & runs & median  & range             & cut $\to0$    & slope       & $L/\xi$ at $n_e$ & 2nd order \\
      \hline
      0.75 & 37   & $1.004$ & $0.999$--$1.010$  & $1.001$       & $-0.40$     & $0.98$--$1.33$   & $1.35$    \\
      1.00 & 66   & $1.005$ & $0.999$--$1.029$  & $1.003$       & $-0.17$     & $0.60$--$0.84$   & $0.84$    \\
      1.25 & 30   & $1.008$ & $0.998$--$1.023$  & $1.009$       & $+0.13$     & $0.36$--$0.49$   & $0.50$    \\
      1.50 & 76   & $1.006$ & $0.948$--$1.054$  & $1.018$       & $+0.60$     & $0.18$--$0.25$   & $0.26$    \\
      1.75 & 24   & $1.032$ & $0.969$--$1.109$  & $\approx1.1$  & $\approx+2$ & $0.05$--$0.08$   & $0.13$    \\
    \end{tabular}
  \end{ruledtabular}
\end{table*}

\begin{figure*}[tb]
  \centering
  \includegraphics[width=\textwidth]{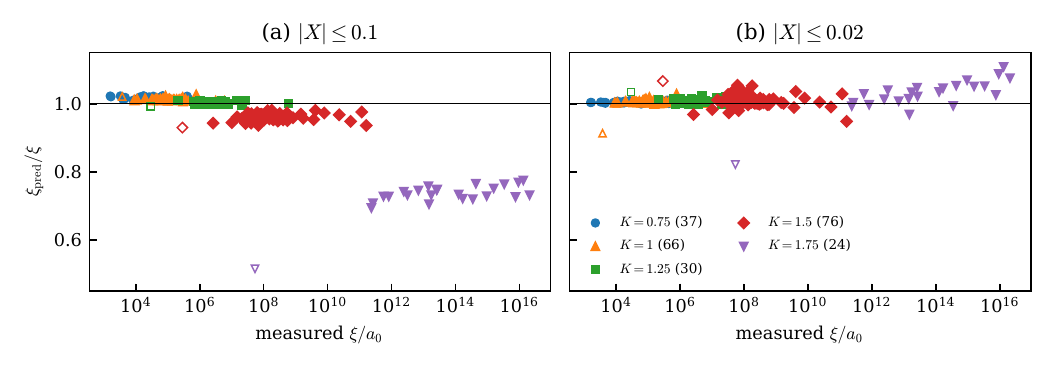}
  \caption{Ratio of the predicted to the measured correlation length against the measured $\xi$ for $K=0.75$--$1.75$, with the coupling read at the end $n_{\rm end}$ of the fit window and $r_{\rm d}=2^{2-K}$ (number of runs in parentheses). (a) Window cut $|X|\le0.1$; (b) $|X|\le0.02$, the protocol of Sec.~\ref{sec:window}. Open symbols: $\chi=16$.}
  \label{fig:ratio}
\end{figure*}

\subsection{Prediction from the sine-Gordon model}
\label{sec:formula}

When the dimerization dominates, the long-distance theory is the sine-Gordon model with the interaction $\tilde g_{\mathrm d}\int d^2x\,\Vd$.
Its lightest charged particle is the soliton, with $|m|=1$, and for $K<1$ the spectrum also contains breathers~\cite{Dashen1975}.
In the normalization of Ref.~\cite{Lukyanov1997}, with the action $\int d^2x\,[(\partial\varphi)^2/16\pi-2\lambda_\SG\cos\beta\varphi]$ and $\langle e^{i\beta\varphi}(x)e^{-i\beta\varphi}(0)\rangle\simeq|x|^{-4\beta^2}$ as $|x|\to0$, the coupling and the soliton mass $M$ are related by~\cite{Zamolodchikov1995}
\begin{equation}
  \lambda_\SG=\kappa_{\rm SG}(p)\,M^{2/(p+1)},\qquad p=\frac{\beta^2}{1-\beta^2},
  \label{eq:zam}
\end{equation}
with
\begin{equation}
  \kappa_{\rm SG}(p)=\frac{\Gamma\!\left(\frac{p}{p+1}\right)}{\pi\,\Gamma\!\left(\frac{1}{p+1}\right)}
  \left[\frac{\sqrt\pi\,\Gamma\!\left(\frac{p+1}{2}\right)}{2\,\Gamma\!\left(\frac p2\right)}\right]^{2/(p+1)} .
  \label{eq:kappa}
\end{equation}
Reference~\cite{Zamolodchikov1995} uses the field $\varphi/\sqrt{8\pi}$ and the parameter $\sqrt{8\pi}\,\beta$, with the same $\lambda_\SG$, $p$, and $\kappa_{\rm SG}(p)$.
The fields $e^{\pm i\beta\varphi}$ are then unit-normalized with dimension $2\beta^2$, so they are $V_{\pm1,0}$ for $\beta^2=K/2$, i.e., $p=K/(2-K)$, and $2\lambda_\SG\cos\beta\varphi=\sqrt2\,\lambda_\SG\Vd$.
Hence $|\tilde g_{\mathrm d}|=\sqrt2\,\lambda_\SG$, and since $2/(p+1)=2-K$,
\begin{equation}
  \xi=\frac1M=\frac{L}{2\pi}\left[\frac{\sqrt2\,\kappa_{\rm SG}(p)}{|\gd(n)|}\right]^{1/(2-K)}
  \label{eq:xi}
\end{equation}
in the lattice units of the readout, with unit velocity and the modular parameter $\tau=i/2$ of the standard tube.
We write $\xi_{\rm pred}$ for Eq.~\eqref{eq:xi} evaluated with the coupling read in the window.
A residual anisotropy of the coarse-grained tensors would enter as a common relative error of all levels~\cite[Sec.~III A]{Paper1a}; it is $1.5\times10^{-4}$ at step 12 in the run at $K=1$ and $\chi=24$ of Appendix~\ref{app:twist}, well below the accuracy of the comparison.
Reading $\xi$ with wider cylinders changes it by less than $0.06\%$ (Appendix~\ref{app:data}); this tests the dependence on the width of the readout, but not a common anisotropy, which would shift all widths alike.
At $K=0.75$, for example, $p=0.6$ and $\kappa_{\rm SG}=0.1390$.
We have checked the whole chain from Eq.~\eqref{eq:g2} to Eq.~\eqref{eq:xi} at the free-fermion point $K=1$, where $\kappa_{\rm SG}=1/2\pi$ and $M=\sqrt2\pi|\tilde g_{\mathrm d}|$.
For the dimerized XX chain, the coupling read from the second-order shift of $x_{0,1}$ predicts the exact gap to within $2\times10^{-5}$ (Appendix~\ref{app:conventions}).

The sine-Gordon model also fixes the neutral levels, which we compare in Sec.~\ref{sec:sgspectrum}.
For $p<1$, i.e., $K<1$, the breathers $B_n$ with $n<1/p$ are neutral bound states of mass $2M\sin(n\pi p/2)$, even or odd under $C$ ($\varphi\to-\varphi$) as $(-1)^n$~\cite{Dashen1975,Zamolodchikov1979}, and in both $C$ sectors the neutral soliton--antisoliton states start at $2M$.
For $K>1/2$ the soliton is therefore the lightest particle, since $2\sin(\pi p/2)>1$ for $p>1/3$.
On a cylinder with $L\gg\xi$, the lowest level for each $m\ne0$, and the lowest excited level for each $C$ at $m=0$, give the lowest excitation energies in the corresponding sectors, while the soliton--antisoliton levels lie above $2M$ by the kinetic energy of a relative momentum of order $2\pi/L$.

\begin{figure*}[tb]
  \centering
  \includegraphics[width=\textwidth]{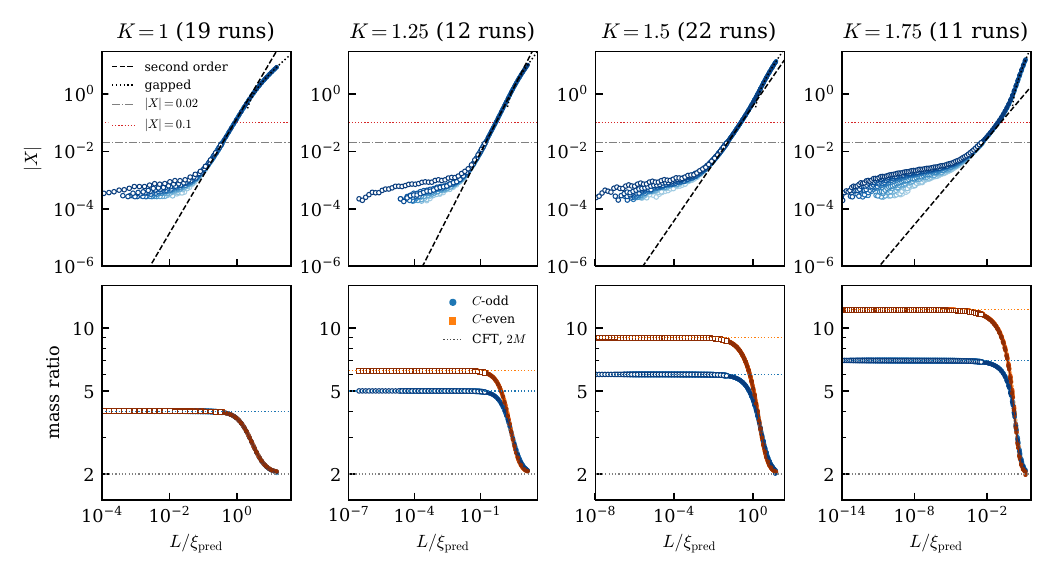}
  \caption{Runs with the alternating deformation at $\chi=40$ for $K=1$--$1.75$ (for $K<1$ see Fig.~\ref{fig:mass}), plotted against $L/\xi_{\rm pred}$ with $\xi_{\rm pred}$ from the window; darker: longer $\xi$. Top: deviation $|X|$ of the lowest charged level; dashed, second order; dotted, the gapped behavior $4KL/(2\pi\xi_{\rm pred})-1$; dash-dotted, the window cut $|X|=0.02$; red dotted, the breakdown threshold $|X|=0.1$. Bottom: mass ratios to the soliton of the lowest $C$-odd neutral level (blue circles; critical value $4K$, the current level, which is the lowest $C$-odd level for $K\ge1$) and of the lowest excited $C$-even neutral level (orange squares; critical value $4K^2$, the $C$-even $e=\pm1$ state), with these critical values and the two-soliton threshold $2$ (dotted); at $K=1$ the two coincide. Open symbols: steps in the fit window; dots: steps not used.}
  \label{fig:collapse}
\end{figure*}

\subsection{Scope of the comparison}
\label{sec:circularity}

The measured $\xi$ and the coupling $\gd$ come from different steps of the same run, and this limits what the comparison can test.
Since $\tilde g_{\mathrm d}$ has dimension $2-K$, any relation in which $\xi$ is set by the dimerization alone gives $\xi\propto|\gd(12)|^{-1/(2-K)}$, a line of slope $-1/(2-K)$ on logarithmic scales.
The slope therefore does not test Eq.~\eqref{eq:xi}.
The comparison tests, first, the height of the line, one amplitude per value of $K$, which is fixed by $\alpha_{0,1}$, $\kappa_{\rm SG}(p)$, and the factors $\sqrt2$ and $2\pi$ in the definition of the coupling (Sec.~\ref{sec:prediction}).
Second, it tests whether the coupling read in the window, the only input of the prediction, predicts the flow after it: $\xi$ is measured in the gapped phase, about 9 ($K=0.75$) to 24 ($K=1.75$) steps after the end of the window (Sec.~\ref{sec:collapse}).
The twisted readout at $n=16$--$24$ shows independently that the coupling is present well before the breakdown (Sec.~\ref{sec:signed}).
Third, the neutral levels, which do not enter the prediction, test the sine-Gordon spectrum (Sec.~\ref{sec:sgspectrum}).

\begin{figure*}[tb]
  \centering
  \includegraphics[width=\textwidth]{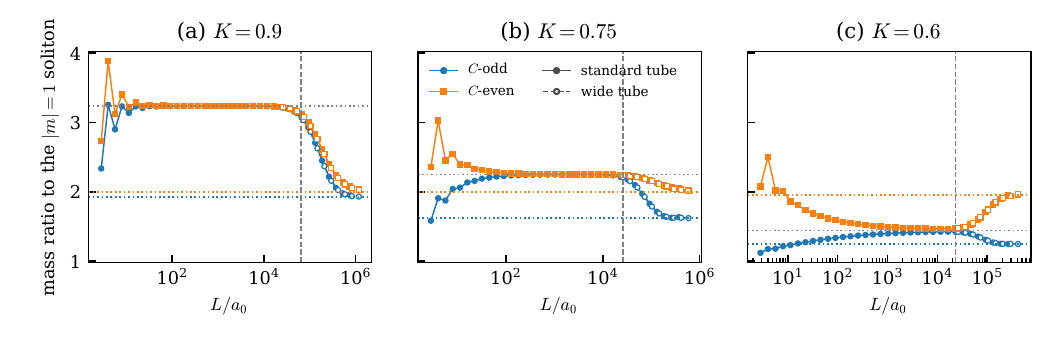}
  \caption{Mass ratios of the lowest $C$-odd (blue circles) and the lowest excited $C$-even (orange squares) neutral levels to the $|m|=1$ soliton for the runs with saved tensors at $K=0.9$, $0.75$, and $0.6$ ($\chi=40$), read with the standard tube (filled, solid) and the wide tube of Appendix~\ref{app:data} (circumference $\sqrt5\,L$; open, dashed). Dotted lines: critical value $4K^2$ (gray) and sine-Gordon values (colored). Vertical dashed line: measured $\xi$. Steps at which a level reaches the numerical floor (the criterion of Appendix~\ref{app:data}) are omitted.}
  \label{fig:mass}
\end{figure*}

\subsection{Predicted and measured correlation lengths}
\label{sec:prediction}

The prediction agrees with the measured $\xi$ to within a few percent (Figs.~\ref{fig:xi_vs_g} and \ref{fig:ratio}(b), Table~\ref{tab:xi}).
With the protocol of Sec.~\ref{sec:window}, the median of the ratio of predicted to measured $\xi$ deviates from unity by less than $1\%$ for $K\le1.5$ and by $3\%$ at $K=1.75$, where all runs but one come from the deformation scans at $\chi=32$ and $40$; the remaining run, at $\chi=16$, deviates by $18\%$, and the unperturbed runs with $\chi\ge24$ do not reach the gapped phase within 80 steps (Appendix~\ref{app:data}).
For $K\le1.5$ the ratio shows no trend over the two to five decades of $\xi$ at each $K$ [Fig.~\ref{fig:ratio}(b); slopes of at most $0.003$ per decade, consistent with zero, excluding $\chi=16$].
The runs at $K=0.6$ and $0.9$ whose tensors were saved for readouts with wider cylinders (Appendix~\ref{app:data}) give $1.004$ and $1.001$, and the runs at $K=0.6$ with $\chi=32$--$64$ give $1.004$--$1.014$ in the second analysis described next; for the BKT point $K=1/2$ see Appendix~\ref{app:bkt}.

The systematic uncertainty of this comparison is a few percent.
Two of us analyzed the runs independently, with different choices of the window ($|X|\le0.02$ and $|X|\le0.05$) and of the estimator of $\xi$, and, for $K<0.75$, of the background (Appendix~\ref{app:conventions}).
For the 20 runs of the $\chi$ scan at $K=0.75$--$1.5$ that both analyses contain, the measured lengths agree to $5\times10^{-4}$, the couplings to $2.1\%$, and the predicted lengths to $2.7\%$ (Table~\ref{tab:crosscheck}).

The ratio also depends on the window cut, most strongly at large $K$.
With $|X|\le0.1$ [Fig.~\ref{fig:ratio}(a)] it stays within $3\%$ of unity for $K\le1.25$ but drops to $0.936$--$0.980$ at $K=1.5$ and to $0.694$--$0.775$ at $K=1.75$ (excluding $\chi=16$, as in Table~\ref{tab:xi}).
We attribute this deviation to a bias of the readout, which is linear in the cut for $K\le1.5$ (Table~\ref{tab:xi}) and produces a larger error in $\xi$ as $K$ approaches $2$, since $\xi\propto|\gd|^{-1/(2-K)}$ (Appendix~\ref{app:conventions}).
When the cut is extrapolated to zero, a small unexplained excess remains, which grows with $K$ to about $2\%$ at $K=1.5$ and $10\%$ at $K=1.75$.
At $K=1.75$ the ratio also depends on the length of the window and on the step at which the slowly drifting $K_n$ is taken, and together with the dependence on the cut this sets a systematic uncertainty of about $10\%$ (Appendix~\ref{app:conventions}); for $K\le1.5$ the drift changes the medians by at most $2\%$, the size of the remaining excess at $K=1.5$.

\subsection{Collapse in \texorpdfstring{$L/\xi_{\rm pred}$}{L/xi\_pred} and the breakdown}
\label{sec:collapse}

At the same $K$, the values of $\xi$ for runs with different $\chi$ and $\mu$ differ by up to five orders of magnitude for $\chi\ge24$.
When their spectra are plotted against $L/\xi_{\rm pred}$, they collapse onto one curve beyond the window (Fig.~\ref{fig:collapse}, which shows the deformed runs at $\chi=40$ for $K=1$--$1.75$), as in the finite-$\chi$ scaling plots of Ref.~\cite{Ueda2023}.
In the top row of Fig.~\ref{fig:collapse}, beyond the window, the deviation of the charged level follows the second-order curve up to $|X|\approx0.1$ for $K\le1.5$ and lies above it near $|X|=0.1$ at $K=1.75$, and for $L/\xi_{\rm pred}\gtrsim10$ reaches $|X|\simeq4KL/(2\pi\xi_{\rm pred})-1$, the value for a particle of mass $1/\xi$, since $x_{0,1}=EL/2\pi$ on the standard tube.
At $K=1.75$ the window corresponds to $L/\xi\lesssim0.005$, so the prediction extends more than three decades in $L$ beyond the data it uses.
In Fig.~\ref{fig:collapse} $\xi$ is predicted from the steps in the window (open symbols), which do not collapse since the background terms $\beta_0+\beta_1n$ in Eq.~\eqref{eq:fit} are not set by $\xi$.
The prediction is tested by the steps after the window (dots), which it does not use.
The breakdown ($|X|=0.1$, Sec.~\ref{sec:setup}; red dotted line in the top row of Fig.~\ref{fig:collapse}) also comes at a fixed $L/\xi_{\rm pred}$, so that at fixed $K$ the number of usable RG steps is set by the seed read in the window.
For $K\le1.5$ the threshold is crossed where the second-order curve reaches it: $L/\xi$ (measured $\xi$) at the last step before $|X|$ exceeds $0.1$ lies within one step ($\sqrt2$ in $L$) below the second-order value (Table~\ref{tab:xi}).
At $K=1.75$ the crossing comes about one to two steps earlier ($L/\xi=0.05$--$0.08$ at $n_e$ against $0.13$), where the running of $K$ induced by $\gd^2$ is largest~\cite[App.~D]{Paper1a}.

\subsection{Sine-Gordon spectrum of the gapped phase}
\label{sec:sgspectrum}

The neutral levels, which are not used for the prediction, test the sine-Gordon spectrum (bottom row of Fig.~\ref{fig:collapse} and Fig.~\ref{fig:mass}).
We divide their energies $E$ by that of the soliton at the same step; this ratio is the ratio of scaling dimensions at criticality and the mass ratio for $L\gg\xi$ (Sec.~\ref{sec:formula}).
For $K\ge1$ ($p\ge1$) the sine-Gordon spectrum has no breathers, and both neutral levels move from their critical values, in units of the charged level, to the two-soliton threshold $2$ (bottom row of Fig.~\ref{fig:collapse}).
The critical values are $4K$ for the current level, which is the lowest $C$-odd level, and $4K^2$ for the $C$-even $e=\pm1$ state.
For $K<1$ the lightest neutral particle is the first breather $B_1$, which is $C$-odd (Sec.~\ref{sec:formula}).
Figure~\ref{fig:mass} shows this for the three runs with saved tensors at $K=0.9$, $0.75$, and $0.6$ (Appendix~\ref{app:data}).
In all three the $C$-odd ratio approaches $B_1$ ($1.930$ against $1.919$ at $K=0.9$, $1.618$ at $L\approx22\,\xi$ at $K=0.75$, and $1.244$--$1.246$ against $1.247$ at $K=0.6$).
At $K=0.9$ and $0.75$ the $C$-even level approaches the two-soliton value $2$ from above; at $K=0.75$ the deviation halves between $L\approx16$ and $22\,\xi$, as expected for two solitons with the smallest relative momentum.
At $K=0.6$ ($p<1/2$) this level is the second breather $B_2$.
Its ratio is $1.937$--$1.961$, compared with the predicted $1.950$.
At the largest $L$ ($\approx18\,\xi$) the ratio read with a wider cylinder ($\sqrt5$ times the circumference, Appendix~\ref{app:data}) rises to $1.961$; relaxing the truncation of that readout from $\chi$ to $2\chi$ lowers it to $1.949$, while at $L\approx13\,\xi$ it changes by only $0.001$, and at larger $L$ the level falls below the floor of this readout.

\section{Correlation length versus bond dimension, and control at fixed \texorpdfstring{$\chi$}{chi}}
\label{sec:control}

\subsection{Bond-dimension dependence of the unperturbed correlation length}
\label{sec:xichi}

\begin{figure}[tb]
  \centering
  \includegraphics[width=\columnwidth]{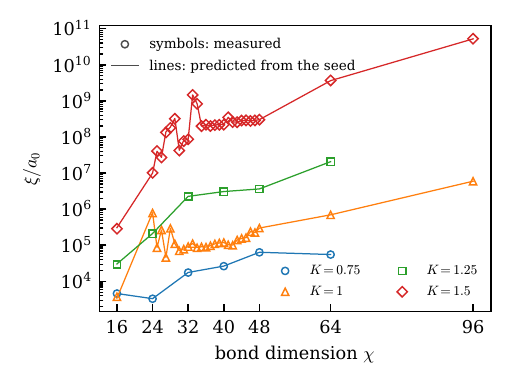}
  \caption{Measured (symbols) and predicted (lines) correlation length of the unperturbed runs ($\mu=0$) against $\chi$, for $K=0.75$--$1.5$. At $K=1$ and $1.5$ the scan includes every integer $\chi$ from $24$ to $48$ and $\chi=64$ and $96$. The correlation length is not a simple power of $\chi$: it changes by up to a factor of $17$ between neighboring $\chi$, not always in the same direction, and the prediction from the seed of each run follows these changes.}
  \label{fig:xi_vs_chi}
\end{figure}

\begin{table}[tb]
  \caption{Effective exponent of $\xi\propto\chi^{\kappa_{\rm FE}}$ for the unperturbed runs. Fit: runs with $\chi\ge24$, repeated runs at the same $\chi$ entered separately (Appendix~\ref{app:seed}). From $\nu$: $\kappa_{\rm FE}=\nu/[2(2-K)]$ with the common seed exponent $\nu=5.1\pm0.8$ of Ref.~\cite[Sec.~VI]{Paper1a}.}
  \label{tab:kappa}
  \begin{ruledtabular}
    \begin{tabular}{ccc}
      $K$ & $\kappa_{\rm FE}$ (fit) & $\kappa_{\rm FE}$ (from $\nu$) \\
      \hline
      0.75 & $2.8\pm0.5$ & $2.0\pm0.3$ \\
      1.00 & $2.0\pm0.5$ & $2.6\pm0.4$ \\
      1.25 & $4.0\pm0.7$ & $3.4\pm0.5$ \\
      1.50 & $4.5\pm0.6$ & $5.1\pm0.8$ \\
    \end{tabular}
  \end{ruledtabular}
\end{table}

Section~\ref{sec:xi} predicted the correlation length of each run from its seed.
Figure~\ref{fig:xi_vs_chi} tests how this prediction follows the measured length when the bond dimension of the unperturbed runs ($\mu=0$) is varied.
At $K=1.5$, $\xi$ grows by more than five orders of magnitude between $\chi=16$ and $96$, but not monotonically: the integer scan contains decreases with increasing $\chi$ as well as a 17-fold rise from $\chi=32$ to $33$.
The predictions follow both the rises and the decreases, because they come from the seed of each run, which scatters by up to two orders of magnitude between neighboring $\chi$ at $K=1$ (Fig.~\ref{fig:xi_vs_chi} and Appendix~\ref{app:seed}; see also Ref.~\cite{Paper1a}).

A finite-entanglement scaling $\xi\propto\chi^{\kappa_{\rm FE}}$ can therefore describe at most the overall trend (Table~\ref{tab:kappa}).
Since $\xi\propto|\gd|^{-1/(2-K)}$ (Sec.~\ref{sec:circularity}), a seed that decreases as $\gd^2(12)\propto\chi^{-\nu_K}$ gives $\kappa_{\rm FE}=\nu_K/[2(2-K)]$; with the common exponent $\nu=5.1\pm0.8$ of Ref.~\cite{Paper1a} in place of $\nu_K$, this predicts a steeper dependence at larger $K$, broadly consistent with the direct fits.
The value depends on the selection of runs.
At $K=1.5$, where $\kappa_{\rm FE}=\nu_K$, the six bond dimensions $\chi=24$, $32$, $40$, $48$, $64$, and $96$ of Ref.~\cite{Paper1a} give $6.0\pm0.4$ with the fixed rate of the present protocol, and gave $6.2\pm0.4$ with a free rate~\cite[Fig.~13]{Paper1a}, whereas the integer scan lowers the estimate to $4.5\pm0.6$, so that the exponent is determined only to $20$--$30\%$ (Appendix~\ref{app:seed}).
The irregularity remains when the loop optimization is run with four times as many sweeps, although it then appears at different $\chi$ (Appendix~\ref{app:seed}).
Since changing $\chi$ changes the seed in an uncontrolled way, we next ask whether reducing the seed at fixed $\chi$ lengthens $\xi$.

\subsection{Locating the compensation point}
\label{sec:compensation}

\begin{table*}[tb]
  \caption{Compensation of the dimerization seed by the alternating $a$/$b$ deformation ($O(2)$). $\rho$ and $g_{\rm seed}$ from the fit $\gd(12)=g_{\rm seed}+\rho\mu$ (sign relative to $\mu$), $\mu^*=-g_{\rm seed}/\rho$, and the run closest to $\mu^*$ compared with $\mu=0$. In parentheses, the jackknife uncertainty of $\rho$ over the runs of each scan in units of the last digit; the relative jackknife uncertainty of $g_{\rm seed}$ is at most $3\%$ except $9\%$ at $(K,\chi)=(1.5,24)$ and $13\%$ at $(1.75,32)$, and a window cut $|X|\le0.05$ instead of $0.02$ changes $\rho$ and $g_{\rm seed}$ by at most $5\%$ (Appendix~\ref{app:data}). Reduction factors quoted in the text are computed from the unrounded values. Last column: measured delay of the breakdown and, in parentheses, the delay $\ln F/\ln2^{2-K}$ expected from the reduction factor $F$ of $\gd^2(12)$ alone. Seeds read with the protocol of Sec.~\ref{sec:window} ($r_{\rm d}=2^{2-K}$, $|X|\le0.02$).}
  \label{tab:b9}
  \begin{ruledtabular}
    \begin{tabular}{rrccccccc}
      $K$  & $\chi$ & $\rho$ & $g_{\rm seed}$      & $\mu^*$             & $\gd^2(12)$, $\mu=0$ & $\gd^2(12)$, closest & $n_e$      & delay (pred.) \\
      \hline
      0.75 & 32     & 25.4(1)      & $-6.5\times10^{-5}$ & $2.6\times10^{-6}$  & $4.1\times10^{-9}$   & $1.6\times10^{-12}$  & $26\to35$  & 9 (9.1)       \\
      0.75 & 40     & 25.20(3)     & $-3.9\times10^{-5}$ & $1.5\times10^{-6}$  & $1.5\times10^{-9}$   & $1.3\times10^{-12}$  & $27\to35$  & 8 (8.1)       \\
      1.00 & 32     & 13.03(2)     & $-7.0\times10^{-5}$ & $5.4\times10^{-6}$  & $4.9\times10^{-9}$   & $7.2\times10^{-11}$  & $29\to35$  & 6 (6.1)       \\
      1.00 & 40     & 13.00(1)     & $-5.4\times10^{-5}$ & $4.2\times10^{-6}$  & $2.9\times10^{-9}$   & $4.1\times10^{-12}$  & $30\to39$  & 9 (9.5)       \\
      1.25 & 32     & 5.83(7)       & $-4.2\times10^{-5}$ & $7.2\times10^{-6}$  & $2.0\times10^{-9}$   & $4.7\times10^{-11}$  & $37\to44$  & 7 (7.2)       \\
      1.25 & 40     & 5.84(4)       & $-3.6\times10^{-5}$ & $6.1\times10^{-6}$  & $1.2\times10^{-9}$   & $4.5\times10^{-13}$  & $38\to53$  & 15 (15.2)     \\
      1.50 & 24     & 2.5(2)        & $+2.5\times10^{-4}$ & $-1.0\times10^{-4}$ & $6.7\times10^{-8}$   & $4.4\times10^{-12}$  & $39\to67$  & 28 (27.8)     \\
      1.50 & 32     & 2.4(2)        & $-8.8\times10^{-5}$ & $3.6\times10^{-5}$  & $7.7\times10^{-9}$   & $5.5\times10^{-12}$  & $45\to66$  & 21 (20.9)     \\
      1.50 & 40     & 2.53(2)       & $-5.6\times10^{-5}$ & $2.2\times10^{-5}$  & $3.0\times10^{-9}$   & $2.9\times10^{-11}$  & $48\to61$  & 13 (13.4)     \\
      1.75 & 32     & 1.2(2)        & $-1.4\times10^{-4}$ & $1.2\times10^{-4}$  & $1.6\times10^{-8}$   & $2.9\times10^{-10}$  & $72\to95$  & 23 (23.2)     \\
      1.75 & 40     & 1.11(3)       & $-1.3\times10^{-4}$ & $1.2\times10^{-4}$  & $1.8\times10^{-8}$   & $2.6\times10^{-11}$  & $71\to109$ & 38 (37.9)     \\
    \end{tabular}
  \end{ruledtabular}
\end{table*}

\begin{figure}[tb]
  \centering
  \includegraphics[width=\columnwidth]{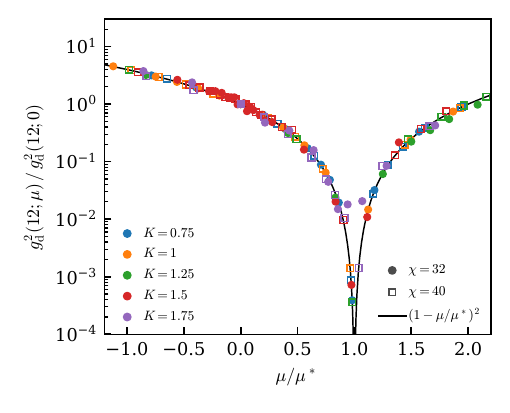}
  \caption{Calibration with the alternating $a$/$b$ deformation ($O(2)$, $K=0.75$--$1.75$; filled circles: $\chi=32$, open squares: $\chi=40$): $\gd^2(12)$ relative to $\mu=0$ against $\mu/\mu^*$, with $\mu^*=-g_{\rm seed}/\rho$ from the fit of each scan (Table~\ref{tab:b9}). Line: $(1-\mu/\mu^*)^2$. Points at $\mu/\mu^*<-1.2$ or $>2.2$ (26 of 154) are not shown, nor is the case $\chi=24$ of Table~\ref{tab:b9}.}
  \label{fig:seed_collapse}
\end{figure}

We use the alternating deformation of Sec.~\ref{sec:window}, which changes $\Delta_{6\mathrm v}$ only at $O(\mu^2)$.
Breaking $\Tr$ alone is not enough.
A staggered vertex weight $\cv(1\pm\mu_{\rm c})$, which also breaks only $\Tr$ but is even under $R_{\rm v}$, produces no parabola for $|\mu_{\rm c}|\le3\times10^{-2}$; its response is at least four orders of magnitude below the naive estimate $2^{6(2-K)}$ for a lattice-scale coupling amplified over 12 steps (Appendix~\ref{app:seed}).

If the response to the deformation is linear, $\gd(12)=g_{\rm seed}+\rho\mu$, where $g_{\rm seed}$ is the seed at $\mu=0$ with its sign relative to $\mu$.
For each $\mu$ we read $\gd^2(12)$ with the protocol of Sec.~\ref{sec:window} and fit the parabola $(g_{\rm seed}+\rho\mu)^2$, whose minimum at $\mu^*=-g_{\rm seed}/\rho$ cancels the seed; the measured $\xi$ does not enter this fit.
With this deformation the parabola is resolved for $K=0.75$--$1.75$ at $\chi=32$ and $40$ (Fig.~\ref{fig:seed_collapse} and Table~\ref{tab:b9}).
In all but one case, $K=1.75$ and $\chi=32$, $\gd^2(12)$ follows $(g_{\rm seed}+\rho\mu)^2$ over two to five decades.

The response $\rho$ varies with $\chi$ by at most $4\%$ for $K\le1.5$ and $9\%$ at $K=1.75$, but falls from $25$ at $K=0.75$ to $1.1$ at $K=1.75$, $2.3$ to $7.2$ times (LoopTNR) below the amplification $2^{6(2-K)}$.
It agrees with the response of the lattice model itself.
Exact diagonalization of the alternating six-vertex model on cylinders of up to 22 edges, read with the twisted splitting (see Sec.~\ref{sec:twist} and Appendix~\ref{app:seed}) and extrapolated with $\gd\propto L^{2-K}$ to the circumference at $n=12$, gives $\rho=25.3$, $12.97$, $5.79$, $2.48$, and $1.05$ at $K=0.75$, $1$, $1.25$, $1.5$, and $1.75$, within $2\%$ of the LoopTNR values for $K\le1.5$ and consistent with them at $K=1.75$, where the fit at $\chi=32$ is poorly determined, with no input from LoopTNR (Appendix~\ref{app:seed}).
At $K=1$, where the lattice value $128/\pi^2=12.969$ is exact, LoopTNR gives $13.03$ and $13.00$; the offset of $0.2$--$0.5\%$ exceeds the jackknife errors and is comparable to the dependence on the window cut ($0.2\%$), which they do not include.
LoopTNR thus carries the lattice dimerization to step 12 without measurable loss (within $2\%$), and the seed, as an asymptotic amplitude, acts like an alternation of the initial vertex weights by $-\mu^*$, of order $10^{-6}$--$10^{-4}$ (Table~\ref{tab:b9}).
The seed includes a part injected after step 12 (Sec.~\ref{sec:branches}); since the response to $\mu$ is linear with the lattice value of $\rho$, this part does not depend on $\mu$ and is cancelled at $\mu^*$ together with the early part.
The sign of the seed relative to $\mu$, which a single untwisted run cannot determine but the position of the minimum $\mu^*$ does, is the same at $\chi=32$--$64$ but opposite at $\chi=24$ for $K=1.5$, where runs at larger $|\mu|$ give $\mu^*\simeq-1\times10^{-4}$ (Appendix~\ref{app:seed}); the campaign of Sec.~\ref{sec:signed} gives the same at $\chi=24$ for $K=0.75$ and $1.25$ (Table~\ref{tab:campaign}).
Like its magnitude~\cite{Paper1a}, the sign of the seed thus varies irregularly with $\chi$.

\subsection{Correlation length and critical window after compensation}
\label{sec:compeffect}

\begin{figure*}[tb]
  \centering
  \includegraphics[width=\textwidth]{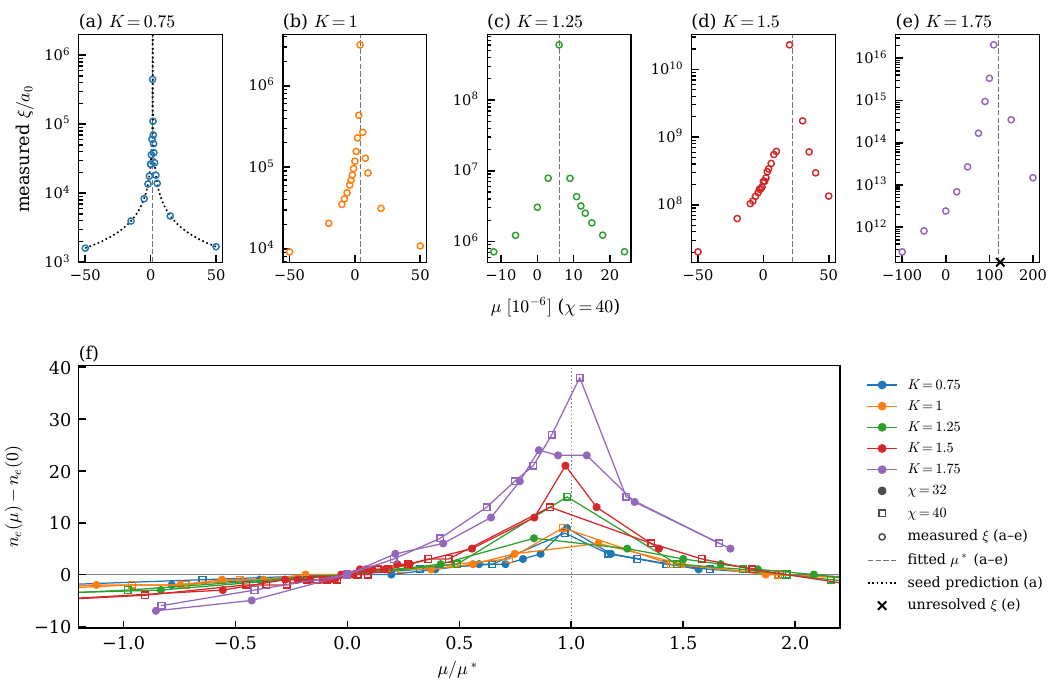}
  \caption{Effect of the compensation ($O(2)$). (a)--(e) Measured $\xi$ against $\mu$ at $\chi=40$ for $K=0.75$--$1.75$; dashed vertical lines, $\mu^*$ of Fig.~\ref{fig:seed_collapse}. In (a), dotted curve: Eq.~\eqref{eq:xi} with $|g_{\rm seed}+\rho\mu|$ fitted to the squared seeds of the 19 runs at $\chi=40$ (the 12 runs of the scan in Table~\ref{tab:b9} and seven additional runs, Appendix~\ref{app:data}); this fit differs slightly from that of Table~\ref{tab:b9}. The cross in (e) marks the run at $\mu=1.25\times10^{-4}$, whose $\xi$ is not resolved within the computed steps. (f) Change of the breakdown step $n_e$ relative to $\mu=0$ against $\mu/\mu^*$ at $\chi=32$ (filled circles) and $40$ (open squares). As in Fig.~\ref{fig:seed_collapse}, points outside the plotted range are not shown.}
  \label{fig:comp_effects}
\end{figure*}

If the seed sets the gap, the compensation should lengthen $\xi$ at fixed $\chi$.
Figure~\ref{fig:comp_effects}(a)--(e) shows $\xi$ against $\mu$ at $\chi=40$ for all five $K$.
At each $K$ the longest resolved length occurs near $\mu^*$.
At $K=0.75$, $\xi$ rises from $2.6\times10^4$ at $\mu=0$ to $4.5\times10^5$ near $\mu^*$ and falls to $(1.6$--$1.7)\times10^3$ at $|\mu|=5\times10^{-5}$, all at the same bond dimension, and follows Eq.~\eqref{eq:xi} evaluated with $|g_{\rm seed}+\rho\mu|$ from the squared seeds alone.
At $K=1.75$ the longest resolved length is $2.1\times10^{16}$ at $\mu=1.10\times10^{-4}$; the run at $\mu=1.25\times10^{-4}$, closest to $\mu^*$, has no settled soliton energy within the computed steps, but its breakdown is delayed from $n_e=71$ to $109$.
The deformed runs at all $K$ are included in Figs.~\ref{fig:xi_vs_g}--\ref{fig:collapse} and follow the prediction as closely as the unperturbed ones.

For a calculation at the critical point, what matters is that the spectrum stays close to its critical values for more steps.
Reducing $\gd^2(12)$ by a factor $F$ delays the breakdown step $n_e$ (Sec.~\ref{sec:setup}) by $\ln F/\ln2^{2-K}$ steps, since the breakdown lies at a fixed $L/\xi$ (Sec.~\ref{sec:collapse}) and $\xi\propto|\gd(12)|^{-1/(2-K)}$.
Near $\mu^*$, $\gd^2(12)$ is reduced by factors of $42$--$2800$ at $\chi=32$ and $40$, and the breakdown is delayed by $6$--$38$ steps [Fig.~\ref{fig:comp_effects}(f) and Table~\ref{tab:b9}], while the growth rate stays the same for all $\mu$.
The estimate reproduces the measured delays to within one step in all eleven cases of Table~\ref{tab:b9}, including the scan at $\chi=24$.
The delays quoted are those of the sampled run closest to the fitted $\mu^*$.
Runs closer to $\mu^*$ may delay the breakdown further, but the parabola does not guarantee this because the response contains an irregular component (Sec.~\ref{sec:signed}).

Tuning one parameter of the initial tensor extends the usable range of RG steps as far as, or further than, doubling $\chi$ does.
The compensated runs at $\chi=32$ reach $n_e=35$, $35$, $44$, and $66$ at $K=0.75$, $1$, $1.25$, and $1.5$, compared with $29$, $35$, $43$, and $56$ for the uncompensated runs at $\chi=64$ (the main runs of Ref.~\cite{Paper1a}).
At $K=1.5$ and $\chi=32$, the run closest to $\mu^*$ has $\xi=1.2\times10^{11}$, compared with $8.6\times10^{7}$ without the deformation and $3.7\times10^{9}$ for the uncompensated run at $\chi=64$.
The compensation lengthens the critical window for the levels not used in the calibration as well, without changing their accuracy, and it does not shift $K$ at the level of its slow drift (Appendix~\ref{app:seed}).
Each parabola consists of 12 to 22 full-length runs at $\chi=32$ and $40$.
Three of them, at $\mu=0$ and $\pm\mu_0$, would locate $\mu^*$ to within a few percent for most groups at a cost comparable to the signed calibration, but only if $0.5\lesssim\mu_0/|\mu^*|\lesssim10$ (Appendix~\ref{app:twist}); $|\mu^*|$ is not known in advance, varies over two decades in Table~\ref{tab:b9}, and changes sign with $\chi$.
In Sec.~\ref{sec:signed} we determine the counterterm from short runs that do not require this knowledge.

\section{Signed readout and one-shot compensation}
\label{sec:signed}

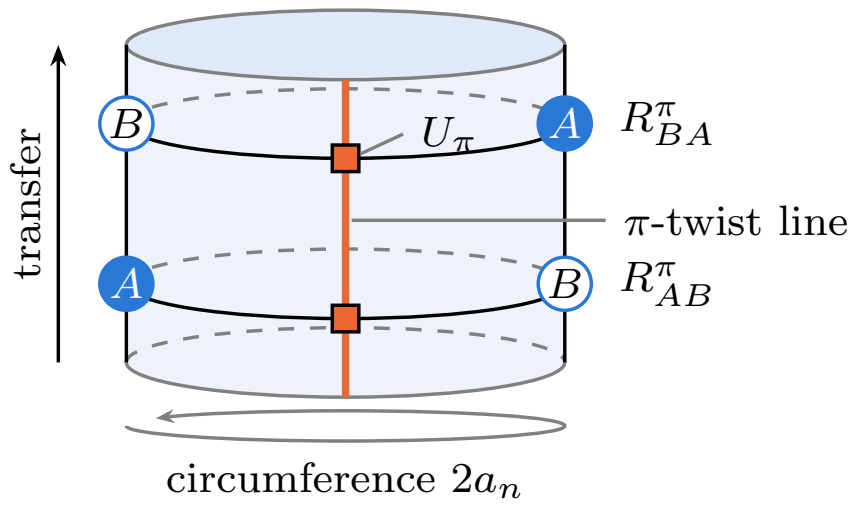
\begin{figure}[!htbp]
  \centering
  \definecolor{tensorblue}{HTML}{2A78D6}
  \definecolor{twistorange}{HTML}{EB6834}
  \definecolor{cylinderwall}{HTML}{EEF3FB}
  \definecolor{cylinderrim}{HTML}{DFE9F7}
  \begin{tikzpicture}[
    font=\footnotesize, line width=.65pt,
    tensor/.style={circle, draw=tensorblue, minimum size=10pt, inner sep=0pt},
    A/.style={tensor, fill=tensorblue, text=white},
    B/.style={tensor, fill=white},
    insertion/.style={rectangle, draw=black, fill=twistorange,
                      minimum size=5pt, inner sep=0pt}]
    \fill[cylinderwall] (-1.45,0) arc (180:360:1.45 and .23)
      -- (1.45,2.10) arc (0:180:1.45 and .23) -- cycle;
    \draw[gray,dashed] (-1.45,0) arc (180:0:1.45 and .23);
    \draw[gray] (-1.45,0) arc (180:360:1.45 and .23);
    \draw[fill=cylinderrim, draw=gray] (0,2.10) ellipse (1.45 and .23);
    \foreach \x in {-1.45,1.45} \draw (\x,0) -- (\x,2.10);
    \foreach \y in {.52,1.58} {
      \begin{scope}[shift={(0,\y)}]
        \draw[gray,dashed] (-1.45,0) arc (180:0:1.45 and .23);
        \draw (-1.45,0) arc (180:360:1.45 and .23);
      \end{scope}
    }
    \draw[twistorange,line width=1.4pt] (0,-.23) -- (0,1.87);
    \foreach \y/\lefttensor/\righttensor in {.52/A/B,1.58/B/A} {
      \node[\lefttensor] at (-1.45,\y) {$\lefttensor$};
      \node[\righttensor] at (1.45,\y) {$\righttensor$};
      \node[insertion] at (0,{\y-.23}) {};
      \node[right] at (1.69,\y) {$R^\pi_{\lefttensor\righttensor}$};
    }
    \draw[gray] (.10,1.38) -- (.38,1.51) node[right,text=black] {$U_\pi$};
    \draw[gray] (.04,.94) -- (1.72,.94) node[right,text=black] {$\pi$-twist line};
    \draw[-{Stealth[length=4pt]}] (-1.90,0) -- (-1.90,2.10)
      node[midway,above,sloped] {transfer};
    \draw[gray,-{Stealth[length=3.5pt]}] (-1.45,-.42) arc (180:510:1.45 and .10);
    \node at (0,-.78) {circumference $2a_n$};
  \end{tikzpicture}
  \caption{Two-row transfer matrix $Q_\pi=R^\pi_{BA}R^\pi_{AB}$ at RG step $n$ (Appendix~\ref{app:twist}).
    The squares mark the two bonds crossed by the same twist line, one in the lower $AB$ row and one in the upper $BA$ row. Each crossing carries $U_\pi$.}
  \label{fig:twisted_cylinder}
\end{figure}

\begin{figure*}[!tp]
  \centering
  \includegraphics[width=\textwidth]{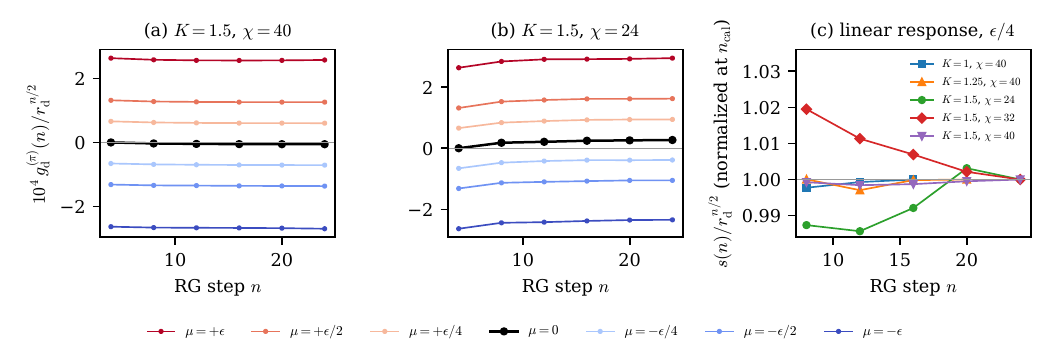}
  \caption{Signed readout of the dimerization coupling with a $\pi$ twist ($O(2)$). (a), (b) Estimator $\gpi$ divided by its first-order growth $(\sqrt2)^{(2-K)n}$ at $K=1.5$ for $\mu=0,\pm\epsilon,\pm\epsilon/2,\pm\epsilon/4$ with $\epsilon=10^{-3}$, at $\chi=40$ and $\chi=24$; the uncompensated run ($\mu=0$, black) is negative at $\chi=40$ and positive at $\chi=24$. (c) Linear response $s(n;\epsilon/4)$, divided by $(\sqrt2)^{(2-K)n}$ and normalized at the calibration step, for the five groups of Table~\ref{tab:signed}.}
  \label{fig:signed}
\end{figure*}

\begin{table*}[tb]
  \caption{One-shot compensation from the signed readout ($O(2)$): calibration step $n_{\rm cal}$, sign of the seed from $\gpi(n_{\rm cal})$ at $\mu=0$, predicted counterterm $\mupred$, counterterm $\mu^*$ from the parabola (Table~\ref{tab:b9}), and $n_e$ without and with $\mupred$ (with $\mu$ closest to $\mu^*$ in parentheses), all on the workstation of the main runs (Appendix~\ref{app:twist}; near $\mu^*$, $n_e$ differs by up to four steps between machines). Last columns: twisted splitting at $n_{\rm cal}$ relative to its uncompensated value with $\mupred$ and after refinement, and $n_e$ with the refined counterterm (see text). The full campaign, fixed in advance, is in Table~\ref{tab:campaign}.}
  \label{tab:signed}
  \begin{ruledtabular}
    \begin{tabular}{rrrcccccc}
      $K$ & $\chi$ & $n_{\rm cal}$ & seed & $\mupred$ & $\mu^*$ & $n_e$ & residual & refined \\
      \hline
      1.00 & 40 & 16 & $-$ & $4.12\times10^{-6}$   & $4.2\times10^{-6}$  & $30\to44$ $(39)$ & $0.11\%$ & --- \\
      1.25 & 40 & 20 & $-$ & $6.00\times10^{-6}$   & $6.1\times10^{-6}$  & $38\to53$ $(53)$ & $1.0\%\to0.05\%$ & 55 \\
      1.50 & 24 & 24 & $+$ & $-1.015\times10^{-4}$ & $-1.0\times10^{-4}$ & $39\to58$ $(67)$ & $5.2\%\to2.7\%$ & 58 \\
      1.50 & 32 & 24 & $-$ & $3.46\times10^{-5}$   & $3.6\times10^{-5}$  & $45\to59$ $(66)$ & $8.2\%\to0.6\%$ & 64 \\
      1.50 & 40 & 24 & $-$ & $2.13\times10^{-5}$   & $2.2\times10^{-5}$  & $48\to61$ $(61)$ & $8.8\%\to0.2\%$ & 71 \\
    \end{tabular}
  \end{ruledtabular}
\end{table*}

A $\pi$ twist in the readout gives the dimerization coupling at first order, with its sign, so that the counterterm follows from short runs at a fixed amplitude of the alternation that need not be adapted to the unknown $\mu^*$; their total length is comparable to that of three unsigned runs (Appendix~\ref{app:twist}).

\subsection{Twisted readout}
\label{sec:twist}

On the standard cylinder, the leading shift of the lowest charged level is proportional to $\gd^2$.
To measure the sign of $\gd$, we compare two neutral levels on a cylinder with a $\pi$ twist.
At RG step $n$, we build the transfer matrix from two rows of tensors on a cylinder of circumference $2a_n$ ($\tau=i$), as shown in Fig.~\ref{fig:twisted_cylinder}.
A single twist line runs along the cylinder and crosses the closing bond of each row, where we insert the $U(1)$ rotation $U_\pi|_m=(-1)^m$.

The twist shifts the dual charge by one half~\cite{Alcaraz1988}.
Twisted boundary conditions have been used to locate Gaussian and BKT transitions in spin chains~\cite{Kitazawa1997,NomuraKitazawa1998} and, in tensor networks, to compute defect spectra~\cite{Hauru2016} and the helicity modulus of the $O(2)$ model~\cite{Akiyama2026}.
Here the shift gives a neutral doublet with $e=\pm\frac12$, $m=0$, and scaling dimension $K/4$ at the Gaussian fixed point.
The dimerization operator $\Vd$ changes $e$ by one and mixes these states; its eigenvalues within the doublet are $+1/\sqrt2$ for the $C$-even combination and $-1/\sqrt2$ for the $C$-odd combination.
To first order, the splitting is
\begin{equation}
  \delta_\pi\coloneqq x^{(\pi)}_{C=-1}-x^{(\pi)}_{C=+1}
    =-2\sqrt2\,\pi\,\gd .
  \label{eq:split}
\end{equation}
In the convention fixed by the alternating $a/b$ deformation, the $C$-odd level therefore lies lower for $\gd>0$.
Appendix~\ref{app:twist} derives Eq.~\eqref{eq:split} and checks both the charge shift and the splitting by exact diagonalization of the dimerized XXZ chain.

The signed estimator $\gpi\coloneqq-\delta_\pi/(2\sqrt2\,\pi)$ gives the coupling at circumference $2a_n$, with leading growth factor $(\sqrt2)^{2-K}$ per RG step.
A spin-two perturbation shifts both twisted levels equally to first order, so its contribution cancels in $\delta_\pi$ (Appendix~\ref{app:twist}).
Its coupling rotates with the lattice orientation, which repeats up to a $180^\circ$ rotation every four RG steps~\cite{Paper1a}; since the doublet center and the untwisted vacuum reference depend on it, we compare readings four steps apart.

At $K=1.5$, the uncompensated coupling is negative at $\chi=40$ [Fig.~\ref{fig:signed}(a)] and positive at $\chi=24$ [Fig.~\ref{fig:signed}(b)].
For all five groups in Table~\ref{tab:signed}, the sign at the calibration step agrees with the sign inferred from the parabolic fits.
An earlier reading can give the opposite sign if little dimerization has accumulated by that step (Sec.~\ref{sec:branches}).

To compare the magnitude with the charged-level estimate, we convert $\gpi$ to the standard circumference $2\sqrt2\,a_n$ by multiplying it by $(\sqrt2)^{2-K}$.
At $n=24$, the two estimates agree to within about $3\%$ in all five groups.
Inserting this single early twisted readout into Eq.~\eqref{eq:xi} of Sec.~\ref{sec:xi} also predicts the correlation length measured later in the gapped regime to within $8\%$ (Appendix~\ref{app:twist}).
We next use the signed splitting and its response to a small alternation of the vertex weights $a$ and $b$ to determine the compensating input.

\subsection{One-shot compensation}
\label{sec:oneshot}

At the calibration step $n_{\rm cal}$, we measure the twisted splitting in seven short runs with input alternations $\mu=0,\pm\epsilon/4,\pm\epsilon/2,\pm\epsilon$.
The innermost pair determines the slope $s=[\delta_\pi(n_{\rm cal},\epsilon/4)-\delta_\pi(n_{\rm cal},-\epsilon/4)]/(\epsilon/2)$, and the larger amplitudes test linearity.
Setting the linearized splitting to zero gives the counterterm
\begin{equation}
  \mupred=-\frac{\delta_\pi(n_{\rm cal},0)}{s} .
  \label{eq:mupred}
\end{equation}
The conversion to the standard circumference multiplies both the intercept and the slope by $(\sqrt{2})^{2-K}$, so the factor cancels in their ratio.

We specified the calibration step, probe amplitudes, and acceptance criteria in advance for 19 groups $(K,\chi)$ spanning $K=0.75$--$1.75$ and $\chi=24$--$64$, and fixed each counterterm before testing it in an independent long run.
One group ($K=1.75$, $\chi=24$) failed the acceptance criteria and one ($K=1.5$, $\chi=64$) was calibrated only; compensation delayed the breakdown by 5--36 RG steps in 16 of the remaining 17 groups (Table~\ref{tab:campaign}).
The exception was $K=1$, $\chi=24$, where $n_e$ fell from 35 to 34; at $n_{\rm cal}=16$ the coupling already had its eventual sign, but much of the seed accumulated later (Sec.~\ref{sec:branches}).

For the five groups in Table~\ref{tab:signed}, the predicted counterterms agree to within $4\%$ with $\mu^*$ from the parabolic fits in Table~\ref{tab:b9}.
At $K=1.5$, a residual splitting of $5$--$9\%$ of its uncompensated magnitude remains at $n_{\rm cal}$.
Near the predicted zero, small changes in the input tensor produce an irregular response that limits the accuracy of a single calibration (Appendix~\ref{app:twist}).

\subsection{Refinement}
\label{sec:refine}

Starting from the one-shot prediction $\mupred$, we adjust the alternation $\mu$ in the initial tensor to reduce the residual twisted splitting at $n_{\rm cal}$.
We use the calibrated slope for the first Newton step and secant updates thereafter; each trial runs to the same $n_{\rm cal}$.
We then select the input with the smallest $|\delta_\pi(n_{\rm cal})|$ and measure its breakdown step $n_e$ in a long run.

The one-shot residual exceeds $0.1\%$ in all five groups. Refinement was not applied at $K=1$, where the residual, $0.11\%$, is at the stopping threshold of Appendix~\ref{app:twist}; in the other four groups it reduced the splitting and delayed breakdown beyond the one-shot result in three (Table~\ref{tab:signed}).
At $K=1.5$, $\chi=24$, the secant updates oscillated, and even with a smaller residual splitting, breakdown occurred at the same step.
In the three runs with a longer critical window, the residual splitting alone overestimates the delay relative to the uncompensated runs; the observed delays are consistent with additional dimerization generated after $n_{\rm cal}$ (Sec.~\ref{sec:branches} and Appendix~\ref{app:twist}).
Because the twist gives $\gd$ at every measured step, it also allows compensation during the RG (Sec.~\ref{sec:iterative}).

\subsection{Computational cost}
\label{sec:cost}

We compare the wall-clock time needed to reach a given breakdown step.
In our runs, the time per RG step increases steeply with $\chi$.
The counterterm changes only the initial tensor, so the cost of an RG step at fixed $\chi$ is the same.
The compensated runs in Table~\ref{tab:cost} come from a different batch and take about $25\%$ longer per step than the uncompensated ones; within the same batch, per-step timings are similar with and without compensation.
The extra cost comes from the short runs used to calibrate the counterterm and, if needed, to refine it.
At $K=1.5$, the refined $\chi=40$ run reaches $n_e=71$ in about one hour including calibration, while the uncompensated $\chi=96$ run reaches $n_e=64$ in about 30 hours.
With refinement, $\chi=32$ reaches the same $n_e=64$ in about half an hour.
At $K=1$, one-shot compensation at $\chi=40$ also takes the flow beyond the breakdown step of the uncompensated $\chi=96$ run (Tables~\ref{tab:signed} and~\ref{tab:cost}).

These comparisons concern the number of usable RG steps.
At $K=1.5$, before dimerization grows, the charge-two and charge-three scaling dimensions follow the expected quadratic dependence on charge more closely at larger $\chi$; at fixed $\chi$, compensation has little effect on these deviations (Appendix~\ref{app:seed}; see also Sec.~\ref{sec:iterative}).
The additional steps are useful for following logarithmic corrections, reaching the scale of a weak imposed perturbation, or measuring a long correlation length.
More precise scaling dimensions still require a larger $\chi$.

\begin{table}[tb]
  \caption{Cost of compensation compared with a larger bond dimension ($O(2)$): breakdown step $n_e$ (last step with $|X|\le0.1$), wall-clock RG time up to $n_e$ (5--11 cores per run; the calibration of Sec.~\ref{sec:signed}, about $0.3$\,h at $\chi=32$ and $0.6$\,h at $\chi=40$ at $K=1.5$, is not included), and the measured correlation length. The compensated entries use inputs from the deformation scans; Table~\ref{tab:signed} gives the breakdown steps after refinement.}
  \label{tab:cost}
  \begin{ruledtabular}
    \begin{tabular}{rrcrrc}
      $K$ & $\chi$ & $\mu$ & $n_e$ & time (h) & $\xi/a_0$ \\
      \hline
      1.0 & 40 & 0                   & 30 & 0.11 & $1.2\times10^{5}$ \\
      1.0 & 40 & $4\times10^{-6}$    & 39 & 0.18 & $3.2\times10^{6}$ \\
      1.0 & 64 & 0                   & 35 & 0.87 & $7.0\times10^{5}$ \\
      1.0 & 96 & 0                   & 41 & 14.4 & $6.0\times10^{6}$ \\
      1.5 & 32 & 0                   & 45 & 0.07 & $8.6\times10^{7}$ \\
      1.5 & 32 & $3.5\times10^{-5}$  & 66 & 0.13 & $1.2\times10^{11}$ \\
      1.5 & 40 & 0                   & 48 & 0.16 & $2.2\times10^{8}$ \\
      1.5 & 40 & $2\times10^{-5}$    & 61 & 0.26 & $2.3\times10^{10}$ \\
      1.5 & 64 & 0                   & 56 & 1.37 & $3.7\times10^{9}$ \\
      1.5 & 96 & 0                   & 64 & 29.6 & $5.3\times10^{10}$ \\
    \end{tabular}
  \end{ruledtabular}
\end{table}

\section{Compensation during the RG}
\label{sec:iterative}

In Sec.~\ref{sec:signed} we used the splitting of the twisted levels to tune a counterterm in the initial tensor, but the dimerization can grow again after the calibration step.
Here we start without an initial counterterm; at intervals during the RG, we measure the same splitting and adjust the coarse-grained tensors to cancel it.
Unlike the Newton method~\cite{Ebel2025} and algebraic or analytic constructions of fixed-point tensors~\cite{Gu2026b,Cheng2026}, the correction does not solve for a fixed point.
It only removes one relevant direction, measured by the twisted splitting, from an otherwise unchanged LoopTNR flow.
We use it as a diagnostic rather than as a general method to find fixed points: with the dominant relevant error removed, it exposes the next limit of the truncation, the drift of the effective Luttinger parameter $K_n$ along the critical line for $K>1/2$ and a marginal source that slows the logarithmic flow at the BKT point.

\subsection{Correction of the coarse-grained tensors}

A counterterm in the initial tensor cancels the coupling at one step.
The residual, together with any later injection (Sec.~\ref{sec:branches}), grows again at the rate $2^{(2-K)/2}$ of $\gd$.
No simple local deformation of the step-$n$ tensors carries the quantum numbers of $\Vd$, since the original $\Tr$ has no local action on the coarse-grained network~\cite[App.~A]{Paper1a}.
The direction of the correction can, however, be taken from the readout itself.
Let $\mathcal T_n=(T_A,T_B)$ be the tensor pair after RG step $n$.
At each correction we choose the amplitude $\gamma_n$ to cancel the measured twisted splitting $\delta_\pi$ to first order along a direction $\Delta\mathcal T_n$ defined below:
\begin{equation}
  \mathcal T_n\to\mathcal T_n+\gamma_n\,\Delta\mathcal T_n,\qquad
  \gamma_n=-\frac{\delta_\pi(\mathcal T_n)}{\delta_\pi'},
  \label{eq:midrg-update}
\end{equation}
where $\delta_\pi'=d\delta_\pi(\mathcal T_n+\gamma\Delta\mathcal T_n)/d\gamma|_{\gamma=0}$.
To obtain $\Delta\mathcal T_n$ we take the gradient of $\delta_\pi$ with respect to the tensor entries, obtained by differentiating through the tensor contraction~\cite{Liao2019}, and project out the components that change, to first order, four quantities:
\begin{itemize}
  \item the normalizations of $T_A$ and of $T_B$ (two constraints);
  \item the center $\bar x_\pi$ of the twisted pair, i.e., its mean dimension measured from the vacuum eigenvalue $\Lambda_0$ of the untwisted two-row transfer matrix;
  \item the anisotropy $\ln(\Lambda_0^{(90^\circ)}/\Lambda_0)$, where $\Lambda_0^{(90^\circ)}$ is the vacuum eigenvalue of the same transfer matrix built from the tensors rotated by $90^\circ$.
\end{itemize}
The direction is recomputed at every correction.
We corrected every two steps from step 24 at $K=1.5$, from step 16 at $K=1$, and from step 12 at the BKT point, for $\chi=24$, $32$, and $40$, up to step 160.
All other settings are as in the rest of this paper.

The interval between corrections is not constrained by the orientation of the lattice, which repeats (up to a rotation by $180^\circ$) only every four steps, because the spin-two couplings of $T,\bar T$ shift the two twisted levels equally at first order and do not enter $\delta_\pi$ (Sec.~\ref{sec:twist} and Appendix~\ref{app:twist}).
It is limited instead by the growth of the dimerization between corrections, by $2^{2-K}$ per step in $\gd^2$, which is largest at the BKT point.
In separately scripted runs at the BKT point and $\chi=24$ with the same constraints (Appendix~\ref{app:numerics}), corrections every four steps did not keep the twisted splitting small: its value before each correction grew from $5\times10^{-3}$ at step 16 to $7\times10^{-2}$ at step 64, and the triplet split.
With corrections every two steps it stayed below $5\times10^{-4}$ from step 16 on, as it does up to step 160 in our run at $\chi=24$.
For this reason we use the shorter interval.

\subsection{Results}

\begin{figure*}[!tp]
  \centering
  \includegraphics[width=\textwidth]{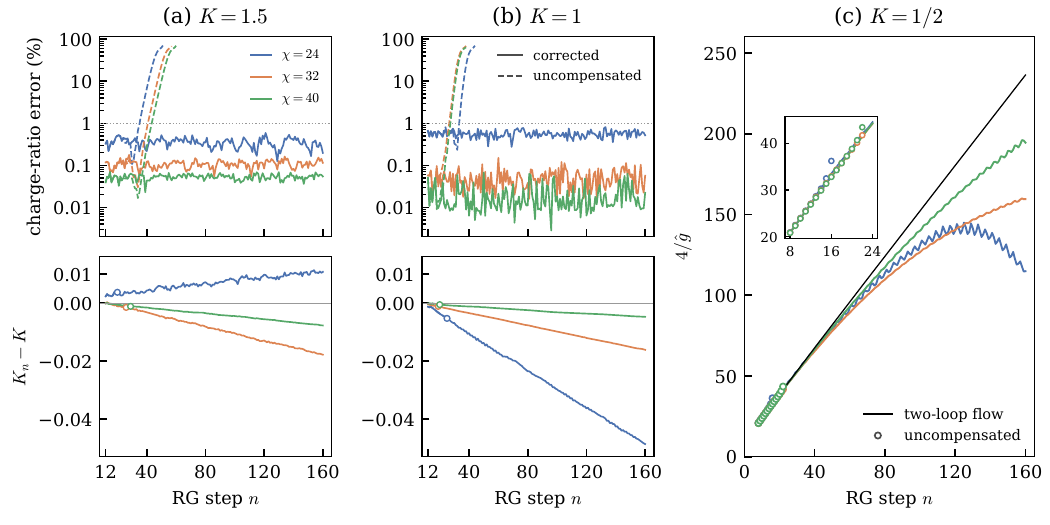}
  \caption{Corrections during the RG with four constraints, every two steps ($O(2)$; $\chi=24$, $32$, $40$). (a), (b) $K=1.5$ and $1$, with corrections (solid, from step 24 and 16) and without (dashed, the main runs of Ref.~\cite{Paper1a}). Top: larger of the relative deviations of $x_{0,2}/x_{0,1}$ and $x_{0,3}/x_{0,1}$ from $4$ and $9$, levels not used for the correction; dotted, $1\%$. Bottom: drift $K_n-K$ of the effective Luttinger parameter; without corrections only up to the end of the floor window of Ref.~\cite[Sec.~IV F]{Paper1a} (open circles). (c) BKT point: $4/\hat g$ from the triplet of levels of dimension $1/2$, with corrections from step 12 (solid) and without, up to the growth of the dimerization (open circles; inset, steps 8--24), compared with the two-loop flow started from the $\chi=40$ value at step 12.}
  \label{fig:midrg}
\end{figure*}

At $K_0=1$ and $1.5$ (we write $K_0$ for the input $K$ in this section), correcting every two RG steps keeps the lowest charged level within the $10\%$ tolerance of Sec.~\ref{sec:setup} through the last computed step, $n=160$: in all six runs, $|X(n)|\le0.054$ over steps 12--160 ($\le0.017$ for $\chi\ge32$).
With the same criterion, the uncompensated runs reach only $n_e=29$--$48$, and at $K_0=1.5$ a counterterm applied only to the initial tensor extends this range to $n_e=58$--$71$ (Table~\ref{tab:signed}).

The upper panels of Figs.~\ref{fig:midrg}(a) and \ref{fig:midrg}(b) test the Gaussian relations $x_{0,2}=4x_{0,1}$ and $x_{0,3}=9x_{0,1}$ as the flow proceeds.
The charge-two and charge-three levels enter neither the correction nor the estimate of $K_n$, and the common scale cancels in the ratios.
Both ratios stay within $1\%$ of their Gaussian values over steps 12--160 in all six corrected runs; at $K_0=1.5$ and $\chi=40$ the rms relative deviations over steps 24--160 are $0.018\%$ and $0.054\%$, compared with $0.017\%$ and $0.049\%$ over steps 12--32 without compensation.
The lower panels show the drift of $K_n=\tfrac12\sqrt{x_e/x_{0,1}}$, where $x_e$ is the mean dimension of the neutral $e=\pm1$ pair~\cite[Sec.~IV F]{Paper1a}.
At $K_0=1$ and $\chi=24$, for example, $K_n$ falls to $0.951$ by step 160 while the larger of the two ratio errors is $0.52\%$.
Over this range the charge spectrum shows that the flow stays on the Gaussian critical line, while $K_n$, which specifies its position along the line, gradually drifts from the input value $K_0$.

At $K_0=1.5$, $K_n$ drifts upward for $\chi=24$ and downward for $\chi=32$ and $40$.
A straight-line fit of $K_n$ against $n$ from the first correction through step 160 gives, in five of the six runs, a slope within $25\%$ of the uncompensated slope fitted before the dimerization grows~\cite[Sec.~IV F]{Paper1a}, e.g., $dK_n/dn=-1.2\times10^{-4}$ ($-1.1\times10^{-4}$ without) at $K_0=1.5$ and $\chi=32$ and $-3.2\times10^{-4}$ ($-2.7\times10^{-4}$) at $K_0=1$ and $\chi=24$.
In the remaining run ($K_0=1.5$, $\chi=24$) the uncompensated slope depends on the fit window~\cite[App.~D]{Paper1a}; both slopes are positive but differ by a factor of three ($+5.5\times10^{-5}$ with and $+1.5\times10^{-4}$ without the correction).
When the corrections are stopped, the twisted splitting regrows at the first-order rate of $\gd$ (Appendix~\ref{app:twist}), so keeping it small requires repeated corrections.
At $\chi=40$ one correction takes about as long as one RG step ($11$--$14$\,s against $13$--$15$\,s), so correcting every two steps raises the RG time by $36$--$45\%$.

All six corrected runs satisfy $|\ln(\Lambda_0^{(90^\circ)}/\Lambda_0)|<1.5\times10^{-3}$ from step 12 through step 160 ($<10^{-3}$ except at $K_0=1$ and $\chi=24$, and $<10^{-4}$ at $\chi=40$).
Earlier runs without the anisotropy constraint, corrected every four steps and, at $K_0=1.5$, started from the initial counterterm of Sec.~\ref{sec:signed}, also stay on the critical line up to step 512 at $\chi=24$, with deviations of the unused levels of about $1\%$ and a similar drift of $K_n$, to $0.879$ at $K_0=1$ and $1.529$ at $K_0=1.5$; at $K_0=1$ the deviation $X$ defined with the input $K$ therefore exceeds $0.1$ well before step 512.
Without the anisotropy constraint, however, the metric anisotropy associated with the spin-two operators $T$ and $\bar T$ reaches $1.4\%$ at step 512 ($K_0=1$, $\chi=24$; Appendix~\ref{app:twist}).

At the BKT point the corrected runs follow the logarithmic approach of Ref.~\cite[Sec.~IV E]{Paper1a} [Fig.~\ref{fig:midrg}(c)].
Over steps 12--22 the inverse $4/\hat g$ of the marginal coupling $\hat g$ of Ref.~\cite[Sec.~IV E]{Paper1a}, read from the triplet of levels of dimension $1/2$ as $\hat g=4(\tfrac12-\bar x_t)$ with $\bar x_t$ their mean, grows by $1.44$--$1.48$ per step for $\chi=24$--$40$, in agreement with the two-loop value $1.47$.
The triplet stays degenerate to $6\times10^{-3}$, $8\times10^{-4}$, and $6\times10^{-4}$ at step 64 for $\chi=24$, $32$, and $40$; since it decreases with $\chi$, it is set by the truncation rather than by the correction.
Beyond step $40$, however, the approach slows down: between steps 40 and 80, $4/\hat g$ grows by $1.1$--$1.3$ per step instead of $1.42$, and later more slowly still.
A marginal source in the flow, $d\hat g/d\ell=-\hat g^2+\sigma_g$, has this effect, since it stops $\hat g$ near $\sqrt{\sigma_g}$.
The values of $\sigma_g$ estimated from steps 100--160, $5\times10^{-4}$ and $1.9\times10^{-4}$ for $\chi=32$ and $40$, decrease with $\chi$, as does the direct drift of $K$ in Ref.~\cite[Sec.~IV F]{Paper1a}.
At $\chi=24$, where $4/\hat g$ even decreases after step 125 and the triplet splitting is no longer small, this description fails.

Once the dimerization is removed, a run is thus limited by the marginal error of the truncation, the drift of $K_n$ for $K>1/2$ and the source $\sigma_g$ at the BKT point, which the repeated correction exposes.

\section{Where is the seed injected?}
\label{sec:branches}

This section uses the calibration campaign of Sec.~\ref{sec:signed} and branched runs to ask when the seed is injected (Sec.~\ref{sec:accumulate}) and whether the contributions of individual steps add (Sec.~\ref{sec:additive}).

\subsection{When does the seed accumulate?}
\label{sec:accumulate}

The seed of Sec.~\ref{sec:window} is the amplitude of the growing dimerization referred back to step 12.
How much of it is already present at that step?
The signed readout of Sec.~\ref{sec:signed} compares the uncompensated splitting $\delta_\pi(n;0)$ with the response $s(n)$ to the alternation, which grows at the first-order rate,
\begin{equation}
  s(n)\propto r^n,\qquad r=(\sqrt2)^{2-K}=r_{\rm d}^{1/2},
  \label{eq:seed-response-growth}
\end{equation}
from $n=8$ on, to within $3\%$ in all groups except the held group $K=1.75$, $\chi=24$ ($7\%$) [Fig.~\ref{fig:signed}(c) for $K\le1.5$ and Appendix~\ref{app:twist}].
Applying the calibration of Sec.~\ref{sec:oneshot} at every step therefore gives the alternation that would cancel the splitting at that step,
\begin{equation}
  \mupred(n)=-\frac{\delta_\pi(n;0)}{s(n)} ,
  \label{eq:seed-running-counterterm}
\end{equation}
for all groups of the calibration campaign (Table~\ref{tab:campaign}).
If the uncompensated splitting grew only by the amplification of the dimerization already present, $\mupred(n)$ would stay constant.

Instead, $\mupred(n)$ keeps changing.
For $n\ge12$ it follows
\begin{equation}
  \mupred(n)\simeq\mu_\infty+\mu_1\,r^{-n}
  \label{eq:seed-counterterm-fit}
\end{equation}
to $0.2$--$1.1\%$ at $K=1.25$ and $1.5$ and $1.6$--$2.7\%$ at $K=1.75$ (Appendix~\ref{app:twist}); at $K=1$ only $n=12$ and $16$ are available, so the two parameters are determined without a test of the form.
This is the solution of Eq.~\eqref{eq:sourceflow} with a constant source $\sigma_{\mathrm d}$, which adds a non-growing coupling to the amplified one.
An injection at step $n$ is weighted by $r^{-n}$ in the asymptotic amplitude $\mu_\infty$, so the early steps dominate it, increasingly less as $K\to2$.
The fraction of $\mu_\infty$ injected after $n=12$ is $1.5$--$4\%$ at $K=1$ ($\chi=32$--$48$; $19\%$ at $\chi=64$), between $-7\%$ and $+11\%$ at $K=1.25$, $19$--$36\%$ at $K=1.5$, and about $48\%$ at $K=1.75$, so that for $K\le1.5$ most of the seed accumulates over the first truncated steps, $n=5$--$11$, whereas at $K=1.75$ nearly half is injected later.
Where the early injection is small, the later one dominates and reverses the sign of the coupling during the flow ($K=1$ and $1.75$ at $\chi=24$).

Both the early and the later injection enter the seed.
In the linearized flow a counterterm tuned to the full seed cancels the growing component, since the response to $\mu$ has no transient; a counterterm that only cancels the splitting at one calibration step leaves the later injection (Secs.~\ref{sec:refine} and \ref{sec:iterative}).

\subsection{Which steps contribute, and do their contributions add?}
\label{sec:additive}

\begin{table}[tb]
  \caption{Effect of changing one setting from step $n_0$ onward: $\gd^2(12)$ of the branch divided by that of the unbranched $\chi=40$ run at the same $K$ (Appendix~\ref{app:seed}).}
  \label{tab:branch}
  \begin{ruledtabular}
    \begin{tabular}{@{}lcccc@{}}
      & \multicolumn{2}{c}{$K=1$} & \multicolumn{2}{c}{$K=1.5$} \\
      \cline{2-3}\cline{4-5}
      Setting changed & $n_0=6$ & $n_0=10$ & $n_0=6$ & $n_0=10$ \\
      \hline
      $\chi=40\to32$                      & 2.7  & 1.4  & 3.7  & 1.7  \\
      $\chi=40\to48$                      & 0.39 & 0.83 & 0.63 & 0.85 \\
      Sweeps $20\to80$                    & 0.74 & 0.92 & 0.45 & 0.72 \\
      Filter cutoff $10^{-14}\to10^{-12}$ & 1.00 & 1.00 & 1.01 & 1.00 \\
    \end{tabular}
  \end{ruledtabular}
\end{table}

We first ask from which step a change of the settings affects the seed.
We continued the $\chi=40$ runs at $K=1$ and $1.5$ from their checkpoints at $n_0=6$ and $10$, changing one setting from that step on: the bond dimension, the number of loop-optimization sweeps, or the threshold of the entanglement filtering that removes short-range correlations before each truncation (Table~\ref{tab:branch}).
At $K=1.5$, for example, lowering $\chi$ to $32$ increases $\gd^2(12)$ by a factor of $3.7$ from $n_0=6$ and $1.7$ from $n_0=10$.
More sweeps reduce the seed, again more strongly when begun earlier, the filtering threshold has no effect beyond the variation of a restart with unchanged settings, and none of the branches changes the growth rate (Appendix~\ref{app:seed}).
No single step dominates: replacing one of steps 5--10 by a step with $\chi=64$ reduces the seed by only $28$--$35\%$ ($10\%$ for step 11), whereas skipping the loop optimization in one step changes it by factors of $0.4$ to $23$ (Appendix~\ref{app:seed}).

To test whether the effects of individual steps add, we used the signed readout at $n=24$.
At $K=1.5$ and $\chi=40$, with 80 instead of 20 sweeps in each of steps 5--11 in seven separate runs, the individual changes of $\gpi(24)$ add up to a reduction of $46.5\%$, whereas the same change in all seven steps in one run reduces it by $24.8\%$.
The same comparison for other groups of steps, for $\chi=32$, and for replacements by steps with $\chi=64$ gives ratios of the joint change to the sum of the single ones between $0.01$ and $0.98$ (Table~\ref{tab:replace} in Appendix~\ref{app:seed}); replacing all of steps 5--11 reduces $|\gpi|$ by $25$--$40\%$, whereas the single reductions add up to $47$--$137\%$.
A replacement by a step with $\chi=64$ is followed by a truncation back to the bond dimension of the run, but the replacements with 80 sweeps keep the bond dimension and are also non-additive in most groups (ratios $0.01$--$0.98$).
A change of one step thus changes what the following steps inject, as the irregular component of Sec.~\ref{sec:signed} also indicates, and the seed is in general not a sum of independent contributions of individual steps.

\section{Summary and discussion}
\label{sec:discussion}

For LoopTNR with $O(2)$ symmetry on the critical line of the six-vertex model, the correlation length of a finite-$\chi$ run is set by the seed of the dimerization, and the seed can be cancelled at fixed $\chi$.
The main numbers are collected in Table~\ref{tab:summary}.
The seed predicts $\xi$ through the sine-Gordon relation without adjustable parameters, the response to the $a$/$b$ alternation agrees with that of the lattice model, and the counterterm located from the parabola delays the breakdown by 6 to 38 steps at fixed $\chi$; predicted from one signed calibration, it delays it in 16 of 17 compensated groups of a campaign fixed in advance, and after refinement a run at $\chi=40$ outlasts one at $\chi=96$ at a few percent of its cost.
Repeating the correction keeps the run critical at $K=1$ and $1.5$ up to step 160, the last step computed, while $K$ drifts slowly.
For $K\le1.5$ the seed is injected mostly during the first RG steps after the first truncation and partly by a weaker later source, whose share grows to about half at $K=1.75$, and a change of one early step changes what the following steps inject.

\begin{table}[tb]
  \caption{Main results by $K$ ($O(2)$). Median ratio of predicted to measured $\xi$ with the protocol of Sec.~\ref{sec:window} and extrapolated to a vanishing window cut (Table~\ref{tab:xi}); response $\rho$ of the seed to the $a$/$b$ alternation from LoopTNR at $\chi=40$ (Table~\ref{tab:b9}) and from exact diagonalization of the lattice model (Appendix~\ref{app:seed}); systematic uncertainty of the ratio (spread between the two analyses, Table~\ref{tab:crosscheck}, and dependence on the cut and, at $K=1.75$, on the treatment of the drift of $K$, Appendix~\ref{app:conventions}); measured delay of the breakdown for the sampled run closest to $\mu^*$ at $\chi=40$ (Table~\ref{tab:b9}).}
  \label{tab:summary}
  \begin{ruledtabular}
    \begin{tabular}{ccccccc}
      $K$  & $\xi_{\rm pred}/\xi$ & cut $\to0$   & syst. & $\rho$ (LoopTNR) & $\rho$ (ED) & delay \\
      \hline
      0.75 & $1.004$             & $1.001$      & $1\%$ & $25.20(3)$        & $25.3$      & 8     \\
      1.00 & $1.005$             & $1.003$      & $1\%$ & $13.00(1)$       & $12.97$     & 9     \\
      1.25 & $1.008$             & $1.009$      & $3\%$ & $5.84(4)$        & $5.79$      & 15    \\
      1.50 & $1.006$             & $1.018$      & $3\%$ & $2.53(2)$        & $2.48$      & 13    \\
      1.75 & $1.032$             & $\approx1.1$ & $10\%$ & $1.11(3)$        & $1.05$      & 38    \\
    \end{tabular}
  \end{ruledtabular}
\end{table}

A finite-$\chi$ run is thus described by one relevant coupling together with a slow marginal drift that accumulates at every step.
The relevant coupling grows perturbatively~\cite{Paper1a} and eventually takes the run into the massive sine-Gordon phase.
In practice, $\xi$, and with it the number of RG steps for which a run stays critical, should be estimated from one weak-coupling readout of the run and not from $\chi$; over the integer $\chi$ scans the twisted readout at $n=24$ predicts it with a median ratio of measured to predicted $\xi$ of $0.99$ ($K=1$) and $0.92$ ($K=1.5$) and a residual scatter of $2$--$4\%$ (Appendix~\ref{app:twist}).
A finite-entanglement scaling $\xi\propto\chi^{\kappa_{\rm FE}}$ describes at most the trend over many runs, with an exponent $\kappa_{\rm FE}=\nu_K/[2(2-K)]$ set by the decrease $\gd^2(12)\propto\chi^{-\nu_K}$ of the seed and by the dimension of $\Vd$ rather than by the central charge alone as for matrix product states~\cite{Tagliacozzo2008,Pollmann2009}; at fixed $\nu_K$ the factor $1/(2-K)$ makes it grow as $K$ approaches $2$, although the fitted values are not monotonic.
The difference reflects what is being approximated. A matrix product state represents the whole critical state at once, so its entanglement entropy, bounded by $\ln\chi$, limits $\xi$, and the perturbation generated by finite $\chi$ adjusts itself to the scale $\xi$~\cite{Schneider2024}. In tensor network renormalization each step only has to represent a block near an approximately scale-invariant tensor; the truncation then acts as a source along a single relevant direction, and a bare coupling can largely cancel it (Secs.~\ref{sec:compensation}--\ref{sec:iterative}).
For given settings the seed is reproducible, but it changes irregularly with $\chi$ and the optimization, and differences in the floating-point environment (Julia and library versions) change $\xi$ by up to $0.7\%$ between the repeated runs of our data set and by $11\%$ in a check at $K=1$, $\chi=35$ (Appendix~\ref{app:data}), where the predicted $\xi$ changes by the same factor, and $n_e$ near the compensation point by up to four steps. The seed, and with it $\xi(\chi)$, is thus reproducible but irregular as a function of the algorithmic details.

A counterterm must match all the lattice quantum numbers of the operator, including those that are not imposed on the tensors.
The staggered vertex weight, for example, breaks $\Tr$ as the dimerization does but does not measurably couple to it, because it is even under $R_{\rm v}$.
The precision of a one-shot counterterm is limited by an irregular component of the response and by the injection after the calibration step; refinement reduces the residual splitting to $0.05$--$0.6\%$ of its uncompensated value in three of four cases.
Repeating the correction during the RG removes this limit, and what then remains is the marginal error of the truncation, which appears as the drift of $K$ and, at the BKT point, as a source that slows the logarithmic flow.
Other routes are a LoopTNR scheme that preserves the original $\Tr$ and hence forbids $\Vd$~\cite[App.~A]{Paper1a}, an improved loop optimization, for instance with nuclear-norm regularization~\cite{Homma2024}, or other truncation schemes~\cite{Evenbly2015,Hauru2018}.
Lattice reflection and rotation symmetries can also be imposed on the tensors~\cite{Lyu2026a,Lyu2026b}; such a scheme forbids $\Vd$ only if a preserved reflection axis passes through a column of vertices, since the reflection about a line between two columns, $R_{\rm v}\Tr$, leaves $\Vd$ invariant.
More generally, our results bear on the construction and stabilization of critical fixed-point tensors~\cite{Ebel2025,Gu2026b,Cheng2026}, for which the seed identifies the direction that has to be controlled.

Several questions remain open.
We do not understand what sets the size of the seed.
Its response to the $a$/$b$ alternation is that of the lattice model (Sec.~\ref{sec:compensation}), but the seed itself, $|g_{\rm seed}|=(4$--$14)\times10^{-5}$ at $\chi=32$ and $40$, equivalent to an alternation of order $10^{-6}$--$10^{-4}$, has no such explanation, and its sign, like its magnitude, varies irregularly with $\chi$.
In the comparison of $\xi$, a small excess remains when the window cut is extrapolated to zero, about $2\%$ at $K=1.5$ and $10\%$ at $K=1.75$ (Sec.~\ref{sec:prediction}); at $K=1.5$ the drift of $K$ is of the size of the excess, and at $K=1.75$ it explains much of the dependence on the window length, but the size of the correction is uncertain.
At the BKT point, the matching to the sine-Gordon model has an uncertainty of a few percent (Appendix~\ref{app:bkt}).

We have studied one algorithm, LoopTNR as implemented in TNRKit, and one model.
The two-tensor unit cell and the $135^\circ$ rotation per step are properties of this coarse-graining map, and other schemes such as HOTRG~\cite{Xie2012}, TNR~\cite{Evenbly2015}, or GILT~\cite{Hauru2018} preserve different lattice symmetries and may generate different operators.
The picture of a relevant coupling that is injected early and then amplified should carry over, but its quantitative form has to be checked for each scheme.
With reduced symmetry, the dominant errors are the $U(1)$ current and the charge-two vertex operator $V_{0,\pm2}$~\cite{Paper1a}; for the latter, with only $C$ imposed, the seed of $V_{0,\pm2}$ predicts the lowest $C$-odd gap to within $1.2\%$ in the three runs at $K\ge1.5$ in which it is the only growing coupling, but not when $V_{0,\pm1}$ also grows (Appendix~\ref{app:conly}).
On the antiferroelectric side $\cv>2$, the competition between the order of the uniform model and the dimerization may lead to an Ising-type transition, which we leave to future work.

\begin{acknowledgments}
  This research was supported in part by the World Premier International Research Center Initiative (WPI), MEXT, Japan; JSPS KAKENHI Grant No.~23K25865; JST, Japan (CREST Grant No.~JPMJCR26XA, Moonshot R\&D Grant No.~JPMJMS256E); and IBM-UTokyo-sponsored research.
  This work was carried out as part of the Google UTokyo Quantum Partnership Research Program.
  This research in part used computational resources of Miyabi provided by the Multidisciplinary Cooperative Research Program in the Center for Computational Sciences, University of Tsukuba.
  The authors used AI tools (Anthropic Claude Opus~5.5 via Claude Code; OpenAI Codex/ChatGPT with GPT-6.2 Sol/Astra; Google Gemini~4 Neon/Argon via Antigravity) to draft and revise the text of this manuscript from the authors' notes, to organize the literature, to write and debug the run, analysis, and plotting scripts, and to check the manuscript against the data and scripts.
  All text, claims, and references were checked and revised by the authors.
\end{acknowledgments}

\paragraph*{Data availability.}
The transfer-matrix spectra, run and analysis scripts, and figure data underlying this paper are available in the repository of Ref.~\cite{Paper1a}, \url{https://github.com/masahitoyamazaki/looptnr-c1-spectroscopy}, archived at Zenodo (\href{https://doi.org/10.5281/zenodo.24361530}{doi:10.5281/zenodo.24361530}); paths quoted in this paper are relative to the repository root, and the repository also documents each run type of Appendix~\ref{app:numerics}. The coarse-grained tensors are available from the authors upon reasonable request.

\appendix

\section{Runs, readouts, and data}
\label{app:numerics}
\label{app:data}

All runs use the environment, code, and readout of the main runs of Ref.~\cite[App.~C]{Paper1a}: Julia~1.13.3, TNRKit~0.7.0 (released version), TensorKit~0.16.5, default \texttt{LoopParameters()}, and the readout with fixed $\tau=i/2$.
Loop optimization stops when the normalized squared error between the loop of four tensors before truncation and its truncated approximation falls below $10^{-9}$ or after 20 sweeps, and the entanglement filtering uses a relative singular-value cutoff of $10^{-14}$.
The uncompensated reference runs ($O(2)$, $K=0.75$--$1.75$, $\chi=16$--$96$) are the main and additional runs of Ref.~\cite{Paper1a}.
The runs specific to this paper use the following options of the run script, which are inactive by default.
(a) Every integer $\chi$ from 25 to 47, so that every $\chi$ from 24 to 48 is covered ($O(2)$, $K=1$, $1.5$).
(b) Deformed initial tensors, namely staggered vertex weights $\cv(1\pm\mu_{\rm c})$ on the two sublattices ($O(2)$, $K=1$, $1.5$, $\chi=32$, $40$, $10^{-7}\le|\mu_{\rm c}|\le3\times10^{-2}$) and the alternating $a$/$b$ deformation ($O(2)$; Table~\ref{tab:abruns}).

\begin{table}[tb]
  \caption{Runs with the alternating $a$/$b$ deformation ($O(2)$), all up to 80 steps (120 at $K=1.75$). (a) Used only for the sign of the seed.}
  \label{tab:abruns}
  \begin{ruledtabular}
    \begin{tabular}{rrl}
      $K$ & $\chi$ & $\mu$ \\
      \hline
      0.75 & 32, 40 & $-2\times10^{-6}\le\mu\le4\times10^{-6}$ \\
      1.00 & 24     & $\pm10^{-4}$ \\
      1.00 & 32, 40 & $|\mu|\le5\times10^{-5}$ \\
      1.00 & 48, 64 & $\pm\mu_0$, $\pm2\mu_0$ (a) \\
      1.25 & 32, 40 & $-1.2\times10^{-5}\le\mu\le2.4\times10^{-5}$ \\
      1.50 & 24     & $-1.5\times10^{-4}\le\mu\le10^{-4}$ \\
      1.50 & 32, 40 & $|\mu|\le5\times10^{-5}$ \\
      1.50 & 48, 64 & $\pm\mu_0$, $\pm2\mu_0$ (a) \\
      1.75 & 32, 40 & $-10^{-4}\le\mu\le2\times10^{-4}$ \\
    \end{tabular}
  \end{ruledtabular}
\end{table}
(c) Modified RG steps: branches from the $\chi=40$ checkpoints at $n_0=6$ and $10$, loop optimization with 80 or 320 sweeps and a cost tolerance of $10^{-12}$, and a single step $n=5$--$11$ of the reference trajectory ($O(2)$, $K=1.5$, $\chi=40$) replaced by a step with $\chi=64$, with 80 sweeps, or without loop optimization; for the signed readout of Sec.~\ref{sec:branches}, single steps and the groups of steps 5--7, 8--11, and 5--11 replaced by steps with $\chi=64$ or 80 sweeps, at $\chi=32$ and $40$.
(d) The twisted readout of Sec.~\ref{sec:signed}, with the calibration and compensated runs using the same LoopTNR parameters and readout code as above (Appendix~\ref{app:twist}; calibration data in \nolinkurl{external/fushio_twist_campaign/}); the readout of the integer $\chi$ scans at $n\le24$ was rerun with the same code (\nolinkurl{runs/early_xi/}); the runs at $\chi=48$ and $64$ for the sign of the seed are in \nolinkurl{runs/sign_and_wide_readout/}.
(e) Corrections during the RG with four constraints (Sec.~\ref{sec:iterative}), using a separate projected-gradient code in the same environment, for $K=0.5$, $1$, $1.5$ and $\chi=24$, $32$, $40$, together with uncompensated references at $K=0.5$ (\nolinkurl{runs/midrg_compensation/}); the earlier runs with three constraints (Sec.~\ref{sec:iterative}; $K=1$ and $1.5$ at $\chi=24$ up to step 512, and $K=1.5$ at $\chi=40$ up to step 160) use the same twisted readout and LoopTNR parameters.  The runs at the BKT point and $\chi=24$ with corrections every four and every two steps (Sec.~\ref{sec:iterative}) are separately scripted runs (\nolinkurl{external/cho_midrg_interval/}).
(f) The integer scan at $K=1.5$ ($\chi=24$--$48$) repeated with 80 loop-optimization sweeps and a cost tolerance of $10^{-12}$ (Appendix~\ref{app:seed}).
Runs with the alternating $a$/$b$ deformation at $K=1.5$ and $\chi=24$ with $|\mu|=10^{-3}$ and $2\times10^{-3}$, and the runs at $K=1.5$ with up to 320 loop-optimization sweeps in steps 5--16 ($\chi=16$--$48$), come from separately scripted runs, of which we use only the stored spectra.
The Trotter calibration of Appendix~\ref{app:seed} uses the same TNRKit version with a separate preprocessing of the initial tensor.
The seeds of the runs used only for the sign of the seed are read as in Ref.~\cite{Paper1a}, with a free rate and $|X|\le0.1$.
All other seeds, including those of the branched and modified runs (Sec.~\ref{sec:branches} and Appendix~\ref{app:seed}), use the protocol of Sec.~\ref{sec:window} (\nolinkurl{runs/seed_refit/refit_origin.py}).

\emph{Data sets.}
The $\chi$ scan ($\chi=16$--$64$, $\mu=0$, $K=0.75$--$1.75$), the integer scan ($\chi=25$--$47$, $K=1$ and $1.5$), the runs at $\chi=96$, and the alternating-deformation runs at $\chi=32$ and $40$ (and $24$ at $K=1.5$) are the runs listed above and in Ref.~\cite{Paper1a}.
At $K=0.75$, seven runs at $\chi=40$ ($\mu=0$, $\pm5\times10^{-6}$, $\pm1.5\times10^{-5}$, $\pm5\times10^{-5}$) were generated separately with the same weights and read with the same tube, in a different environment (Julia~1.13.2, TensorKit~0.17.1, a local copy of TNRKit~0.7.0); the run at $\mu=0$ reproduces the main runs to $3\times10^{-5}$ in $\xi$.
The runs at $K=0.5$ and $0.6$ with $\chi=32$--$64$ are the runs at and near the BKT point in Ref.~\cite{Paper1a}.
We use only runs that have reached the gapped phase, as indicated by a ratio $q\coloneqq E(n)/E(n-1)$ above $0.98$ at the last step included.
This excludes six runs at $K=1.75$: the five runs of the $\chi$ scan with $\chi\ge24$, which stop at 80 steps before reaching the gapped phase ($q=0.71$--$0.82$ at the last step; at $\chi=64$ the charged level has not even reached $|X|=0.1$), and one deformed run at $\chi=40$.

The unperturbed condition at $\chi=32$ and $40$ was run more than once, in the $\chi$ scan, in the calibration of the deformation, and at $K=0.75$ in the separate set.
We count these repetitions as separate runs.
Their measured $\xi$ differ by at most $0.7\%$ (at $K=1.5$, $\chi=40$) and typically by $10^{-4}$, which shows how strongly differences in the floating-point environment (Julia and library versions) are amplified together with the seed.
The difference can be larger: in a check outside the data set, the same script with TNRKit~0.7.0 and TensorKit~0.16.5 on two other machines (Julia~1.10 and 1.11) gave, at $K=1$ and $\chi=35$, twisted estimators $\gpi(n)$ that differ by $10\%$ from $n=8$ on and correlation lengths that differ by $11\%$, read from the steps at which $|X|$ crosses $0.1$--$2$; the predicted $\xi$ differs by the same factor, $1.111$, and $n_e=29$ on both machines (\nolinkurl{runs/machine_check/}).
One pair, at $K=1.5$ and $\chi=24$, is the same stored run and is counted once.
After merging repetitions of the same condition, the 233 runs correspond to 224 distinct conditions.

\emph{Measurement of $\xi$.}
As the run enters the gapped phase, the low-lying gaps approach finite values, so $q$ rises from its critical value $1/\sqrt2$ toward $1$.
A level that has reached the numerical floor stops decreasing, and $q$ drops; we use such a drop as an empirical sign of the floor.
We follow the soliton and the lowest $C$-odd and lowest excited $C$-even neutral levels, stop at the first step with $n\ge12$ at which $q$ of any of them drops by more than $0.02$ or the eigenvalue ratio of any of them falls below $10^{-7}$, and average $\xi$ over the last two steps before this point.
For the mass ratios of Fig.~\ref{fig:mass} we also use the wide tube of TNRKit, whose floor is $10^{-4}$.
In their overlap the two tubes give the same $\xi$ to four digits at $K=0.75$.

\emph{Width of the readout.}
To test the dependence of $\xi$ on the width of the readout (Sec.~\ref{sec:xi}), we read the saved checkpoints of the $\chi=40$ runs in the gapped phase with the standard cylinder and with two wider ones.
The first is the wide cylinder of TNRKit ($\tau=0.1+0.2i$, circumference $2\sqrt{10}\,a_n$), with its extra coarse-graining step truncated at $\chi$ and at $2\chi$, and the second is the $[1,8,1]$ cylinder (circumference $8a_n$, extra step truncated at $\chi$).
For the latter, only the vacuum sector and the charge-one sector were diagonalized, with the construction of TNRKit's \texttt{CFTData}.
On a $\chi=16$ checkpoint ($K=1$, step $24$) this restricted readout reproduces the full one to $4\times10^{-15}$ (\nolinkurl{runs/width_check/}).
With the wide cylinder, $\xi$ changes by less than $0.06\%$ at $K=0.6$, $0.75$, $1$, $1.25$, and $1.5$ ($\chi=40$), and the $[1,8,1]$ cylinder of TNRKit, with $2\sqrt2$ times the circumference, gives the same $\xi$ to within $0.04\%$ at $K=0.75$--$1.5$.

\emph{Seed readouts.}
For the unperturbed runs with $\chi\ge24$, the median ratio of the present $\gd^2(12)$ to that read as in Ref.~\cite{Paper1a} is $0.76$, $0.90$, $1.06$, $1.6$, and $4.0$ at $K=0.75$, $1$, $1.25$, $1.5$, and $1.75$ (all unperturbed runs of Sec.~\ref{sec:xi} and, at $K=1.75$, also those that do not reach the gapped phase; \nolinkurl{runs/xi_sine_gordon/} in the data repository).
The two readouts of the seed, the protocol of Sec.~\ref{sec:window} and the free-rate fits with $|X|\le0.1$ of Ref.~\cite{Paper1a}, give the same compensation points.
Across the eleven cases of Table~\ref{tab:b9}, $\mu^*$ changes by at most $0.5\%$.
The response $\rho$, however, changes by a factor that grows with $K$: the free-rate readout gives $\rho$ larger by $14$--$17\%$ at $K=0.75$, smaller by $15$--$20\%$ at $K=1.5$, and smaller by a factor of $2.2$--$2.3$ at $K=1.75$.
The last reflects the bias of the cut $|X|\le0.1$ near $K=2$.
The refits are reproduced by \nolinkurl{runs/seed_refit/refit_b9.py} in the data repository, and the jackknife uncertainties and the dependence on the window cut quoted in Table~\ref{tab:b9} by \nolinkurl{runs/seed_refit/refit_b9_errors.py}.

\begin{table}[tb]
  \caption{The two analyses on the common $\chi$ scan ($\mu=0$). $|\gd(12)|$: this paper's protocol ($|X|\le0.02$); $h(12)$: second analysis ($|X|\le0.05$, Appendix~\ref{app:conventions}); $\xi$: measured length (the two estimators agree to $5\times10^{-4}$); last two columns: ratio of predicted to measured $\xi$ in the two analyses.}
  \label{tab:crosscheck}
  \begin{ruledtabular}
    \begin{tabular}{rrcccrr}
      $K$ & $\chi$ & $|\gd(12)|$ & $h(12)$ & $\xi/a_0$ & \multicolumn{2}{c}{$\xi_{\rm pred}/\xi$}\\
      \hline
            0.75 & 24 & $5.20\times10^{-4}$ & $5.14\times10^{-4}$ & $3.309\times10^{3}$ & 1.005 & 1.015 \\
      0.75 & 32 & $6.43\times10^{-5}$ & $6.38\times10^{-5}$ & $1.758\times10^{4}$ & 1.007 & 1.012 \\
      0.75 & 40 & $3.87\times10^{-5}$ & $3.86\times10^{-5}$ & $2.641\times10^{4}$ & 1.005 & 1.008 \\
      0.75 & 48 & $1.29\times10^{-5}$ & $1.28\times10^{-5}$ & $6.386\times10^{4}$ & 1.003 & 1.010 \\
      0.75 & 64 & $1.54\times10^{-5}$ & $1.53\times10^{-5}$ & $5.540\times10^{4}$ & 1.003 & 1.007 \\
      1 & 24 & $8.09\times10^{-6}$ & $8.09\times10^{-6}$ & $7.790\times10^{5}$ & 1.029 & 1.029 \\
      1 & 32 & $6.98\times10^{-5}$ & $6.98\times10^{-5}$ & $9.200\times10^{4}$ & 1.010 & 1.010 \\
      1 & 40 & $5.41\times10^{-5}$ & $5.41\times10^{-5}$ & $1.190\times10^{5}$ & 1.007 & 1.008 \\
      1 & 48 & $2.16\times10^{-5}$ & $2.16\times10^{-5}$ & $2.998\times10^{5}$ & 1.002 & 1.003 \\
      1 & 64 & $9.28\times10^{-6}$ & $9.26\times10^{-6}$ & $6.970\times10^{5}$ & 1.002 & 1.004 \\
      1.25 & 24 & $2.63\times10^{-4}$ & $2.58\times10^{-4}$ & $2.094\times10^{5}$ & 1.011 & 1.039 \\
      1.25 & 32 & $4.44\times10^{-5}$ & $4.46\times10^{-5}$ & $2.235\times10^{6}$ & 1.014 & 1.009 \\
      1.25 & 40 & $3.53\times10^{-5}$ & $3.55\times10^{-5}$ & $3.054\times10^{6}$ & 1.009 & 1.001 \\
      1.25 & 48 & $3.12\times10^{-5}$ & $3.11\times10^{-5}$ & $3.646\times10^{6}$ & 0.998 & 1.000 \\
      1.25 & 64 & $8.45\times10^{-6}$ & $8.45\times10^{-6}$ & $2.077\times10^{7}$ & 0.998 & 0.998 \\
      1.5 & 24 & $2.58\times10^{-4}$ & $2.61\times10^{-4}$ & $1.016\times10^{7}$ & 0.983 & 0.967 \\
      1.5 & 32 & $8.77\times10^{-5}$ & $8.80\times10^{-5}$ & $8.631\times10^{7}$ & 1.004 & 0.998 \\
      1.5 & 40 & $5.48\times10^{-5}$ & $5.55\times10^{-5}$ & $2.195\times10^{8}$ & 1.013 & 0.987 \\
      1.5 & 48 & $4.70\times10^{-5}$ & $4.75\times10^{-5}$ & $3.012\times10^{8}$ & 1.002 & 0.980 \\
      1.5 & 64 & $1.35\times10^{-5}$ & $1.36\times10^{-5}$ & $3.674\times10^{9}$ & 0.989 & 0.977 \\

    \end{tabular}
  \end{ruledtabular}
\end{table}

\section{Additional controls of the truncation seed}
\label{app:seed}

\emph{Bond-dimension scan.}
The fits of Table~\ref{tab:kappa} use the unperturbed runs with a measured $\xi$ at $\chi\ge24$, entering repeated runs at the same $\chi$ separately.
At $K=0.75$ and $1.25$ the bond dimensions are $24$, $32$, $40$, $48$, and $64$; at $K=1$ and $1.5$ they include every integer from $24$ to $48$ and $64$ and $96$, so that the integer scan dominates the fit.
The scatter of $\log_{10}\xi$ about the fitted lines is $0.15$--$0.36$ decades, and $\nu_K=2(2-K)\kappa_{\rm FE}$ gives $\nu_K=7.0\pm1.3$, $4.0\pm1.0$, $6.0\pm1.0$, and $4.5\pm0.6$ at $K=0.75$, $1$, $1.25$, and $1.5$, consistent with the common exponent $5.1\pm0.8$ of Ref.~\cite[Sec.~VI]{Paper1a}.
At $K=1$, for example, $\chi=24$ gives $\xi=7.8\times10^5$ and $\chi=27$ gives $4.6\times10^4$, and at $K=0.75$ the run at $\chi=64$ is shorter than that at $\chi=48$.
A scan over every integer $\chi$ from 24 to 48 shows that the seed changes in steps.
At $K=1.5$, $\gd^2(n=12)$ is nearly constant for $35\le\chi\le40$ [$(3.0$--$3.3)\times10^{-9}$] and for $41\le\chi\le48$ [$(1.9$--$2.6)\times10^{-9}$].
For $\chi\le34$ it changes between neighboring $\chi$ by factors of up to $17$ (e.g., $7.7\times10^{-9}$ at $\chi=32$ and $4.6\times10^{-10}$ at $\chi=33$).
The same pattern is found at $K=1$.
The single-step replacements below show that the seed is sensitive to the variational step, in which the truncated loop is optimized, and the scan with 80 sweeps below connects this sensitivity to the $\chi$-to-$\chi$ irregularity.

\emph{Branches.} Up to $n=4$ the bond dimension grows without truncation (to $16$), and the first step that truncates and optimizes the loop is $n=5$ for all runs with $\chi\le64$ and for $K=1$ at $\chi=96$, but $n=6$ at $K=1.5$ for $\chi=64$ and $96$ (\nolinkurl{runs/seed_refit/first_truncation.py}).
A branch from a common checkpoint with identical settings reproduces the reference to $10^{-15}$ in the first steps.
This round-off difference is amplified at the same rate as the dimerization, but its effect on the fitted seed stays below $3\%$.
Lowering the bond dimension to $\chi=32$ at $n_0=6$ increases $\gd^2(12)$ by a factor of $2.7$ ($K=1$) and $3.7$ ($K=1.5$), and raising it to $\chi=48$ reduces it to $0.39$ and $0.63$ of the reference.
At $n_0=10$ the same changes give factors of $1.4$--$1.7$ and $0.83$--$0.85$.
Increasing the number of loop-optimization sweeps from 20 to 80 reduces $\gd^2(12)$ to $0.45$--$0.74$ ($n_0=6$) and $0.72$--$0.92$ ($n_0=10$) of the reference, and a filtering threshold of $10^{-12}$ instead of $10^{-14}$ has no effect.
A branch that switches from $\chi=40$ to $\chi=32$ ($48$) at $n_0=6$ ends up with a seed within a factor of $1.6$ ($2.5$) of that of the run with $\chi=32$ ($48$) from the start, at both $K=1$ and $1.5$.
None of the branches changes the growth rate: a free-rate fit with $|X|\le0.1$, as in Ref.~\cite{Paper1a}, gives $r/2^{2-K}=0.992$--$0.997$ at $K=1$ and $1.012$--$1.015$ at $K=1.5$, against $0.993$ and $1.014$ for the unbranched runs, whose offsets from unity are the known bias of that cut (\nolinkurl{runs/additional/analysis/b3_b4.csv}).

\emph{Loop-optimization convergence.}
Runs from scratch with 80 sweeps and a cost tolerance of $10^{-12}$ reduce $\gd^2(12)$ to $0.38$--$0.49$ of the default (20 sweeps, $10^{-9}$) and delay the breakdown by $1$--$3$ steps at $\chi=24$ and $40$, again without changing the rate.
Beyond about 80 sweeps there is no systematic gain.
To test whether the irregular $\chi$ dependence is an artifact of the incomplete optimization, we repeated the integer scan at $K=1.5$ ($\chi=24$--$48$) with 80 sweeps and a cost tolerance of $10^{-12}$.
The seed decreases by a median factor of $0.46$, but in four of the 25 runs it increases, by up to a factor of four.
The $\chi$ dependence remains irregular.
The scatter of $\gd^2(12)$ about a power law in $\chi$ decreases only from $0.35$ to $0.24$ decades, the largest jump between neighboring $\chi$ from $1.2$ to $0.7$ decades, and $\xi$ still decreases with increasing $\chi$ for 8 of the 24 pairs of neighboring $\chi$, as with 20 sweeps.
The outliers, however, move to other $\chi$ (for example $\chi=28$, $29$, and $35$ instead of $30$ and $33$), so that the irregularity depends on the loop optimization as well as on $\chi$ (\nolinkurl{runs/sweep80/}).
With 320 sweeps $\gd^2(12)$ is $1.11$ ($\chi=24$, $K=1.5$), $0.41$ ($\chi=40$, $K=1.5$), and $0.36$ ($\chi=40$, $K=1$) of the default, and $n_e$ is $39$, $50$, and $31$ (default: $39$, $48$, $30$), at four times the cost per step of the 80-sweep runs.
Correspondingly, the final loop cost averaged over $n=6$--$20$ at $\chi=40$ decreases only from $3.3\times10^{-6}$ (default) to $2.4\times10^{-6}$ (80 sweeps) and $2.2\times10^{-6}$ (320 sweeps) at $K=1.5$, and from $3.7\times10^{-6}$ to $2.9\times10^{-6}$ and $2.7\times10^{-6}$ at $K=1$.
Separately scripted runs at $K=1.5$ show the same saturation.
With up to 320 sweeps in steps 5--16, none of which reaches the cost $10^{-9}$, the step at which the charged level reaches a $1\%$ error is delayed by only $0$--$5$ steps for $\chi=16$, $32$, $40$, and $48$.
The remaining seed depends on the loop optimization as well as on $\chi$.

\emph{Where in the step is the seed injected?}
To resolve the injection within the RG, we replaced a single step $n=5$--$11$ of the reference trajectory ($O(2)$, $K=1.5$, $\chi=40$) by a modified step and continued with the default settings.
Truncating one step to $\chi=64$ instead of $40$ reduces $\gd^2(12)$ to $0.65$--$0.72$ of the reference for every $n=5$--$10$ ($0.90$ for $n=11$).
No step dominates: after amplification, the part of the seed injected over steps 5--11 accumulates with comparable weights, as the branches of Sec.~\ref{sec:branches} also indicate.
Skipping the loop optimization in one step (keeping only the initial decomposition and the projectors) changes $\gd^2(12)$ by factors between $0.37$ and $23$, mostly upward ($\times18$ for $n=6$ and $\times23$ for $n=9$).
By contrast, 80 instead of 20 sweeps in one step changes it by $0.70$--$1.07$.
The loop optimization removes most of the $\Tr$-odd error of the initial truncation, but the remainder varies strongly in size and sign from step to step.
With the signed readout at $n=24$ (Table~\ref{tab:replace}), $|\gpi|$ decreases in 26 of the 28 single replacements, by up to $49\%$, so that the injections of these steps mostly have the sign of the seed; the exceptions, both with 80 sweeps, are step 6 at $\chi=40$ and step 9 at $\chi=32$, where it increases by $4\%$ and $2\%$ (\nolinkurl{runs/step_injection/}).

\begin{table}[tb]
  \caption{Single and joint replacements at $K=1.5$, read with the signed readout at $n=24$ (Sec.~\ref{sec:additive}). Each entry gives the sum of the reductions $100(1-g_{\rm var}/g_{\rm ref})$ of $\gpi(24)$ (in $\%$, with sign) of the single replacements of the steps in the group, and the reduction measured when the whole group is replaced in one run. A replacement by a step with $\chi=64$ is followed by a truncation back to the bond dimension of the run; the replacements with 80 sweeps keep it.}
  \label{tab:replace}
  \begin{ruledtabular}
    \begin{tabular}{@{}rlccc@{}}
      $\chi$ & replacement & steps 5--7 & steps 8--11 & steps 5--11 \\
      \hline
      32 & $\chi=64$         & 66.0 / 31.4 & 70.6 / 60.4 & 136.6 / 33.7 \\
      32 & 80 sweeps         & 72.0 / 0.5  & 37.0 / 35.8 & 109.0 / 37.2 \\
      40 & $\chi=64$         & 54.1 / 4.8  & 51.0 / 12.9 & 105.1 / 40.3 \\
      40 & 80 sweeps         & 15.9 / 15.6 & 30.6 / 21.6 & 46.5 / 24.8  \\
    \end{tabular}
  \end{ruledtabular}
\end{table}

\emph{Known deformation: calibration.}
A dimerization of known strength can be introduced through the initial tensor.
In a Trotter construction, the initial tensor is obtained from the imaginary-time evolution (time step $\delta\tau=1/8$) of the dimerized XXZ chain $H_{\rm XXZ}=\sum_l[1+(-1)^l\mu_{\rm T}]B_l$, with $B_l\coloneqq S^x_lS^x_{l+1}+S^y_lS^y_{l+1}+\Delta_{\rm XXZ}S^z_lS^z_{l+1}$ and $\Delta_{\rm XXZ}=-\cos[\pi/(2K)]$.
Read at $K=1$ ($\chi=16$ and $24$) from the tensors after two and four LoopTNR steps, the coupling read back from the splitting of the charged levels reproduces the input within a relative error $<10^{-8}$, and the shift of the lowest charged level that it predicts agrees with the exact finite-size free-fermion value to the same accuracy.
Fed directly into LoopTNR, the strongly anisotropic Trotter tensor causes an early failure of the compression for small $\chi$ (at $K=1$ and $n=6$, the error of $x_{0,1}$ relative to its exact finite-size value is $+816\%$, $+48\%$, and $+1.6\%$ for $\chi=16$, $24$, and $40$).
Compressing 32 Trotter rows along the imaginary-time direction into one nearly isotropic cell before the LoopTNR steps (bond dimension $64$ in this compression, $\chi=32$ in LoopTNR) removes this failure.
Read out with a tube of twice the circumference, the injected dimerization can then be followed up to $n\simeq14$ at the nominal $K=1.5$ of the chain; the growth factors below correspond to an effective $K$ of $1.46$--$1.49$ of the Trotter tensor.
After subtraction of the unperturbed run, the couplings read from the charged levels $m=1$, $2$, $3$ agree with each other to within $7\%$ at $n=8$.
The squared couplings read from these three levels grow by $1.423$--$1.433$ per step for $\mu_{\rm T}=0.01$ ($n=6$--$14$) and $1.443$--$1.452$ for $\mu_{\rm T}=0.02$ ($n=6$--$10$), compared with $2^{2-K}=1.414$.
Doubling $\mu_{\rm T}$ multiplies $\gd^2$ by $3.7$--$4.2$ for $n=6$--$10$, consistent with $\gd\propto\mu_{\rm T}$ (\nolinkurl{external/fushio_trotter_injection/}).
This calibrates the readout against an injected dimerization, but a comparison with the truncation seed requires a deformation that acts on the isotropic vertex weights used throughout this paper.

\emph{Response of the lattice model.}
The response to the alternating $a$/$b$ deformation can be computed without LoopTNR.
We diagonalize the diagonal-to-diagonal transfer matrix of the six-vertex model, a brickwork of vertices on a cylinder of $N=6$--$22$ edges, i.e., of circumference $L=Na_0/\sqrt2$.
The vertices of one layer form one sublattice, so the two layers carry $(a,b)=(1+\mu,1-\mu)$ and $(1-\mu,1+\mu)$.
With $\mu=0$ the lowest charged level and the $\pi$-twisted pair approach $1/(4K)$ and $K/4$.
A $\pi$ twist on the seam and Eq.~\eqref{eq:split} give $\gd(L)$, which is linear in $\mu$ to $10^{-5}$ for $|\mu|\le10^{-3}$.
Fits of $\gd(L)/\mu=A(L/2\pi)^{2-K}[1+B(L/2\pi)^{-\omega}]$ with $\omega=1$ or $2$, separately for $N\equiv0$ and $2\pmod4$ and with or without the smallest sizes, give values of $\gd/\mu$ at the circumference $L=2\sqrt2\cdot2^6a_0$ of step 12 that differ by at most $1.4\%$ for $K\le1.5$ and $2.8\%$ at $K=1.75$.
At $K=1$ the result is exactly $\gd(L)/\mu=(\sqrt2/\pi)(L/2\pi)$, so that $\rho=128/\pi^2=12.97$.
The amplitude $A$ lies between $0.38$ and $0.47$ for all $K$; together with the factor $(2\sqrt2/2\pi)^{2-K}$ from the normalization $\gd\propto(L/2\pi)^{2-K}$ of Sec.~\ref{sec:setup}, it accounts for the ratio $1/7.2$--$1/2.7$ (ED) of $\rho$ to the naive estimate $2^{6(2-K)}$.
Only $|\rho|$ is compared, since the sign of $\gd$ is fixed relative to $\mu$ by convention (Appendix~\ref{app:twist}) (\nolinkurl{runs/rho_ed/}).

\emph{Staggered vertex weight.}
With vertex weights $\cv(1\pm\mu_{\rm c})$ on the two sublattices (Sec.~\ref{sec:compensation}) and $10^{-7}\le|\mu_{\rm c}|\le3\times10^{-2}$ ($K=1$, $1.5$ and $\chi=32$, $40$), $\gd^2(12)$ varies only between $0.3$ and $1.8$ times its value at $\mu_{\rm c}=0$, without a systematic dependence on the sign or the magnitude of $\mu_{\rm c}$.
Neither $n_e$ ($28$--$30$ at $K=1$, $45$--$50$ at $K=1.5$) nor the growth rate changes.
A fit of $(g_{\rm seed}+\rho_{\rm c}\mu_{\rm c})^2$ gives only upper bounds, $|\rho_{\rm c}|\lesssim3\times10^{-3}$ ($K=1$) and $(1$--$3)\times10^{-4}$ ($K=1.5$), at least four orders of magnitude below the naive estimate $|\rho_{\rm c}|\sim2^{6(2-K)}$ for a lattice-scale coupling amplified over 12 steps.
As expected from the selection rule of $R_{\rm v}$~\cite[App.~A]{Paper1a}, the staggered $\cv$ is not a practical handle on the truncation seed.
To first order in $\mu_{\rm c}$, the staggered $\cv$ couples only to operators that are odd under $\Tr$ and even under $R_{\rm v}$.
Among the primaries, the $R_{\rm v}$-even combination $V_{1,0}-V_{-1,0}$ is $C$ odd and excluded by $C$.
Its descendants of dimension $K+1$ have spin $\pm1$ and are excluded by the lattice rotations, and the scalars of dimension $K+2$, $\partial\bar\partial V_{\pm1,0}$, are total derivatives, since the level-one space of a vertex operator with $e\ne0$ is one-dimensional ($J_{-1}\propto L_{-1}$).
Non-derivative scalars first appear at dimension $K+4$, so the coupling can arise only from these or from the breaking of $R_{\rm v}$ by the truncation itself.

\emph{Alternating $a$/$b$ deformation.}
At $K=1$ and $\chi=24$, $\mu=\pm10^{-4}$ gives $\gd^2(12)=1.7\times10^{-6}$ for both signs and moves $n_e$ from $35$ to $21$.

In the fits of Table~\ref{tab:b9}, the rms residual in $\ln\gd^2$ is $0.006$--$0.21$, dominated by the points closest to $\mu^*$, where $\gd^2(12)$ is smallest.
The exception is $K=1.75$, $\chi=32$ ($0.61$), where the points near $\mu^*$ scatter: the measured $\gd^2(12)$ is smallest at $\mu=1.0\times10^{-4}$, while the fit gives $\mu^*=1.2\times10^{-4}$.

\emph{Sign of the seed.}
At $K=1.5$ and $\chi=24$ the seed has the sign opposite to that at $\chi=32$ and $40$.
This was first found from separately scripted runs with $\mu=\pm10^{-3}$ and $\pm2\times10^{-3}$ (24 steps), in which the negative $\mu$ gives the smaller $\gd^2(12)$.
These runs, like the others used only for the sign of the seed, are read as in Ref.~\cite{Paper1a} (free rate, $|X|\le0.1$); in this readout the runs at $\chi=32$ and $40$ give $\rho=13.6$--$13.7$ at $K=1$ and $2.0$ at $K=1.5$.
For these runs, $\gd^2(12)$ read from the same fit is $4.2\times10^{-6}$ and $1.9\times10^{-5}$ for negative $\mu$ and $6.6\times10^{-6}$ and $2.1\times10^{-5}$ for positive $\mu$.
Interpreted as $(g_{\rm seed}+\rho\mu)^2$, the pairs give $\rho=2.2$--$2.3$, consistent with $\rho=2.0$ at $\chi=32$ and $40$ given the shorter windows.
They also give $g_{\rm seed}=+(1$--$2.6)\times10^{-4}$, whose magnitude agrees with $|\gd(12)|=2.2\times10^{-4}$ of the $\mu=0$ run at $\chi=24$ (in the free-rate readout; $2.6\times10^{-4}$ with the protocol of Sec.~\ref{sec:window}).
Runs near the predicted $\mu^*\simeq-1\times10^{-4}$, read with the protocol of Sec.~\ref{sec:window}, then give $\mu^*=-1.01\times10^{-4}$ (Table~\ref{tab:b9}).

At $\chi=48$ and $64$, again with the free-rate readout, we determined the sign with $\mu_0=3\times10^{-6}$ and $1.5\times10^{-6}$ ($K=1$) and $4\times10^{-5}$ and $1\times10^{-5}$ ($K=1.5$).
Fits of $(g_{\rm seed}+\rho\mu)^2$ to the five points have residuals of $1$--$5\%$ in $\ln\gd^2$, and the fitted $g_{\rm seed}^2$ agrees with $\gd^2(12)$ of the uncompensated run within $6\%$.
The fits give $\mu^*=+1.7\times10^{-6}$, $+7.2\times10^{-7}$, $+1.8\times10^{-5}$, and $+5.4\times10^{-6}$ for ($K$, $\chi$) $=(1,48)$, $(1,64)$, $(1.5,48)$, and $(1.5,64)$.
The corresponding seeds, $g_{\rm seed}=-2.3\times10^{-5}$ and $-1.0\times10^{-5}$ at $K=1$ and $-3.8\times10^{-5}$ and $-1.1\times10^{-5}$ at $K=1.5$ ($\chi=48$ and $64$), have the same sign as at $\chi=32$ and $40$, with $\rho=13.7$, $14.2$ and $2.11$, $2.02$.
At equal $|\mu|$, $\gd^2(12)$ is smaller by factors of $2.4$--$12$ on the side of $\mu^*$ than on the opposite side, where it exceeds its value at $\mu=0$ by factors of $8$--$32$ and $n_e$ is reduced by $3$--$10$ steps.
Because $\mu_0$ turned out to be about $2\mu^*$, no point is close to $\mu^*$, and these runs are not included in Table~\ref{tab:b9}.
The runs at $\mu=\mu_0$ change $\gd^2(12)$ by factors of $0.65$--$1.5$ and $n_e$ by at most one step.
The growth rate is the same as at $\mu=0$ within $0.5\%$ in all runs.
At $K=1.5$ and $\chi=24$, where a sampled run lies at the fitted $\mu^*$, the delay is the predicted $28$ steps.
The delay is limited by how precisely $\mu^*$ is hit and by the irregular component of the response (Sec.~\ref{sec:signed}), and the gain is largest where the calibration grid comes closest to $\mu^*$.

\emph{Other levels and the drift of $K_n$.}
The compensation also improves levels other than the one used for the calibration.
We count the consecutive steps in which all other assigned levels (24--28 levels in five sectors) stay within a root-mean-square deviation of $10^{-2}$ from their CFT values, with $|\mathsf{c}-1|<5\times10^{-3}$.
Of the ten cases at $\chi=32$ and $40$, such a window exists in nine (all except $K=0.75$, $\chi=32$); in these its length increases from $13$--$48$ to $21$--$84$ steps, while the median deviation within it changes by less than $20\%$ (\nolinkurl{runs/compensation_drift/}).
The longer critical window also improves quantities on which the dimerization has no leading-order effect.
Reference~\cite{Paper1a} measured the slow drift of $K_n$ along the critical line in the floor window, defined as the steps $n\ge12$ before the dimerization term exceeds $10^{-4}$.
Without compensation this window lasts $7$--$21$ steps.
With the same definition, the runs closest to $\mu^*$ have floor windows $1.6$--$2.9$ times longer, and the standard error of the fitted $dK_n/dn$ is smaller by a median factor of $3.6$ and by up to $7.6$.
At $K=1.5$ and $\chi=32$, for example, $dK_n/dn=(-1.01\pm0.03)\times10^{-4}$ from steps $12$--$46$, compared with $(-1.11\pm0.11)\times10^{-4}$ from steps $12$--$25$.
The values with and without compensation agree to within $23\%$ in all cases except $K=0.75$, $\chi=40$, where both are below $5\times10^{-6}$ per step.
The counterterm therefore does not shift $K$ at the level of this drift.
The remaining differences exceed the statistical errors, which suggests that $K_n$ does not drift linearly over the longer windows.

\section{Twisted transfer matrix}
\label{app:twist}

\emph{First-order splitting.}
A twist by $\phi$ of the $U(1)$ symmetry generated by $m$ shifts the dual charge to $e+\phi/2\pi$ in Eq.~\eqref{eq:spectrum}~\cite{Alcaraz1988}.
At $\phi=\pi$ the lowest neutral states are $|{\pm\tfrac12}\rangle\coloneqq|e=\pm\tfrac12,m=0\rangle$, with dimension $K/4$, and $C$ exchanges them.
Since $V_{\pm1,0}$ changes $e$ by $\pm1$, the only nonzero matrix elements of $\Vd$ in this doublet are $\langle{\pm\tfrac12}|V_{\pm1,0}|{\mp\tfrac12}\rangle$, which equal the OPE coefficient of three vertex operators, $1$ in the normalization of Sec.~\ref{sec:setup}.
In the basis $|C\!=\!\pm1\rangle=(|{+\tfrac12}\rangle\pm|{-\tfrac12}\rangle)/\sqrt2$, $\Vd=(V_{1,0}+V_{-1,0})/\sqrt2$ is therefore diagonal with eigenvalues $\pm1/\sqrt2$.
With the first-order shift $2\pi\gd$ times the matrix element on a cylinder of circumference $2\pi$~\cite[Sec.~IV~D]{Paper1a}, the $C$-even level moves by $+\sqrt2\pi\gd$ and the $C$-odd level by $-\sqrt2\pi\gd$, which gives Eq.~\eqref{eq:split}.
A coupling to $T$ or $\bar T$ shifts each state at first order in proportion to its conformal weight $h$ or $\bar h$, which is the same ($K/8$) for the two states of the doublet, so it cancels in $\delta_\pi$.
The sign of $\gd$ in Eq.~\eqref{eq:split} is tied to the phase convention of the vertex operators.
As for $\rho$ in Sec.~\ref{sec:compensation}, we fix it relative to the $a$/$b$ deformation, and both readouts use the same convention.

\emph{Check by exact diagonalization.}
For the periodic XXZ chain with bond alternation $1\pm\delta$, $\Delta_{\rm XXZ}=-\cos[\pi/(2K)]$, and a $\pi$ twist, exact diagonalization ($L_{\rm ch}=14$--$20$) places the twisted pair at $x=K/4$ for $\delta=0$.
The ratio of the squared first-order splitting of this pair to the second-order shift of the lowest charged level must equal $8\pi^2/\alpha_{0,1}$, independently of the normalization of $\gd$.
The second-order shift also contains a term of order $\delta^2$ that does not grow with $L$, from the shift of $K$, so the ratio approaches its limit as $L^{-2(2-K)}$.
Extrapolated with this correction and its square, it is $2.8854$ at $K=1$ and $4.624$ at $K=1.5$ (with the correction alone, $2.8850$ and $4.594$), compared with $8\pi^2/\alpha_{0,1}=2.8854$ and $4.620$; the six-vertex transfer matrix of Appendix~\ref{app:seed} ($N\le22$), extrapolated in the same way from $N\ge10$ or $14$, gives $2.8856$--$2.8857$ and $4.64$--$4.65$ (\nolinkurl{runs/signed_readout/ed/}, \nolinkurl{runs/rho_ed/}).
On the lattice the $C$ labels of the twisted pair alternate with $L_{\rm ch}$ mod $4$, which is another reason to fix the sign relative to the $a$/$b$ deformation.

\emph{Construction.}
With $A$ and $B$ the step-$n$ tensors and $U_\pi$ inserted on the leg that closes each row, a symmetry defect line~\cite{Hauru2016}, the lower row of the twisted transfer matrix is
\begin{equation}
  (R^\pi_{AB})_{j_1j_2;\,i_1i_2}=\sum_{c,a,k}(U_\pi)_{ca}\,A_{aj_1;\,i_1k}\,B_{kj_2;\,i_2c},
\end{equation}
the upper row $R^\pi_{BA}$ follows by exchanging $A$ and $B$, and $Q_\pi=R^\pi_{BA}R^\pi_{AB}$ (shape $[2,2,0]$ in TNRKit~\cite{TNRKit}, $\tau=i$).
Its circumference $2a_n$ is $1/\sqrt2$ times that of the standard cylinder, so the coupling read from it is smaller by $(\sqrt2)^{-(2-K)}$ (Sec.~\ref{sec:signed}).
The two copies of $U_\pi$ mark the same symmetry line where it crosses the two rows and do not cancel.
With $\operatorname{Im}\tau=1$ in Eq.~\eqref{eq:readout}, a splitting is $-\ln|\lambda_-/\lambda_+|/(2\pi)$, where $\lambda_\pm$ are the leading eigenvalues in the two neutral $C$ blocks.
The vacuum of the untwisted two-row transfer matrix at the same step, obtained by replacing $U_\pi$ by the identity, sets the reference.

The contraction was checked against exact diagonalization of a four-site Trotterized XXZ chain with periodic and twisted boundaries for 14 sets of couplings, where all 16 levels and their $C$ sectors agree to within $1.7\times10^{-14}$.
Moving the seam to another bond gives a transfer matrix equal to within $3.6\times10^{-15}$ up to a similarity transformation for ten stored tensors.
The production code, which diagonalizes only the two neutral $C$ blocks, reproduces the full contraction to within $2.8\times10^{-17}$ in $\delta_\pi$.

\begin{figure*}[tb]
  \centering
  \includegraphics[width=\textwidth]{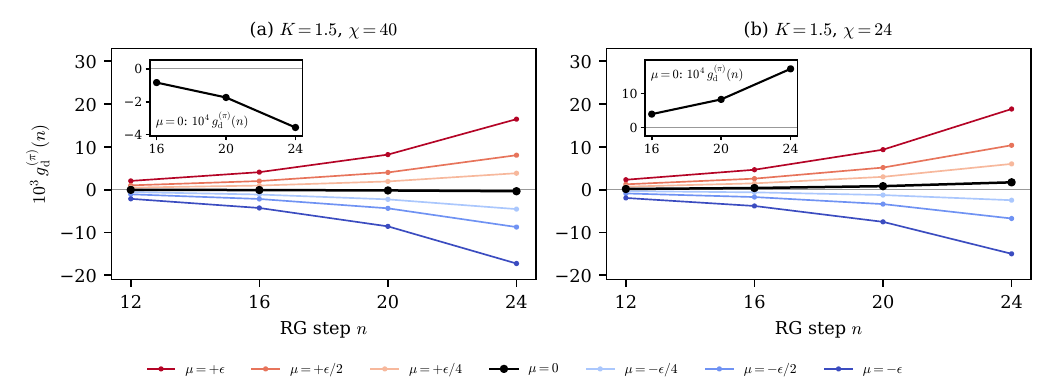}
  \caption{Signed readout of Fig.~\ref{fig:signed}(a), (b) without the rescaling by the first-order growth: the estimator $\gpi(n)$ at $K=1.5$ for $\mu=0,\pm\epsilon,\pm\epsilon/2,\pm\epsilon/4$ ($\epsilon=10^{-3}$) and $n=12$--$24$, at (a) $\chi=40$ and (b) $\chi=24$ ($O(2)$). The uncompensated run ($\mu=0$, black) is small on this scale; the insets show it for $n=16$--$24$ in units of $10^{-4}$. It is negative at $\chi=40$ and positive at $\chi=24$. Every curve roughly doubles over four steps, the first-order factor $(\sqrt2)^{4(2-K)}=2$.}
  \label{fig:signed_raw}
\end{figure*}

\emph{Growth of the estimator.}
Without the rescaling of Fig.~\ref{fig:signed}, the estimator grows by about a factor of $2$ every four steps for every $\mu$ (Fig.~\ref{fig:signed_raw}).
The effective per-step growth factor of the uncompensated $|\gpi|$ exceeds its first-order value $r=(\sqrt2)^{2-K}$ and approaches it as $n$ grows: at $K=1.5$ it exceeds $r$ by $3$--$4\%$ over $n=12$--$16$ and by about $1\%$ over $n=16$--$24$, and at $K=1$ by $0.8\%$ over $n=12$--$16$.
This is not a transient of the estimator.
The response $s(n)$ to the alternation, divided by $(\sqrt2)^{(2-K)n}$, is constant to within $3\%$ from $n=8$ on in all calibration groups except $K=1.75$, $\chi=24$ ($7\%$).
The equivalent alternation $\mupred(n)$ at $n=12$, relative to its value at the last calibration step, is instead $0.97$--$0.99$ at $K=1$ ($\chi=32$--$48$; $0.85$ at $\chi=64$), $0.90$--$1.06$ at $K=1.25$, $0.67$--$0.83$ at $K=1.5$, and $0.59$ at $K=1.75$ ($\chi=32$, $40$); at $K=1$, $\chi=24$ and $K=1.75$, $\chi=24$ it changes sign.
The fits $\mu_\infty+\mu_1\,r^{-n}$ of Sec.~\ref{sec:branches} reproduce $\mupred(n)$, $n\ge12$, to $0.2$--$1.1\%$ at $K=1.25$ and $1.5$ and $1.6$--$2.7\%$ at $K=1.75$; they and the fractions $f_{\rm cal}=-\mu_1\,r^{-n_{\rm cal}}/\mu_\infty$ of the asymptotic amplitude still missing at $n_{\rm cal}$ are given in Table~\ref{tab:campaign} (\nolinkurl{runs/seed_growth/}, \nolinkurl{runs/twist_campaign_summary/}).
A non-growing contribution to $\delta_\pi$ of another origin would have the same form; for the counterterm the two are equivalent.

\emph{Early prediction of $\xi$.}
In the five groups of Table~\ref{tab:signed}, $(\sqrt2)^{2-K}|\gpi|$ at $n=24$ is $0.97$--$1.00$ times $|\gd|$ from the standard readout and $0.96$--$1.00$ times the value inferred from the measured $\xi$ by Eq.~\eqref{eq:xi}.
Inserted into Eq.~\eqref{eq:xi}, the twisted readout at $n=24$, long before the breakdown ($n_e=30$--$48$), thus predicts $\xi$ to within $0.1$--$8\%$, without any fit to the later steps.
In these groups the estimator at $n=16$ is lower by up to $10\%$, because part of the seed is injected after $n=12$ (Sec.~\ref{sec:branches}).
To test the prediction on a larger set, we read the uncompensated runs of the full scans at $K=1$ and $1.5$ (every integer $\chi=24$--$48$ and $\chi=64$, 52 runs; Fig.~\ref{fig:xi_vs_chi}) at fixed $n=16$, $20$, and $24$.
Over these scans the measured $\xi$ varies irregularly by $1.2$ and $2.6$ decades, whereas the ratio of measured to predicted $\xi$ (here the inverse of the ratio in Sec.~\ref{sec:xi}) has a standard deviation in $\ln$ of only $0.05$ and $0.07$ at $n=20$, and $0.02$ and $0.04$ at $n=24$.
The medians of this ratio are $0.98$, $0.99$, and $0.99$ at $n=16$, $20$, and $24$ for $K=1$, and $0.76$, $0.86$, and $0.92$ for $K=1.5$.
At $n=24$ individual runs give $0.97$--$1.00$ at $K=1$ and $0.86$--$1.00$ at $K=1.5$; the exception is $\chi=24$ at $K=1$ ($0.79$ at $n=20$ and $0.90$ at $n=24$), whose twisted splitting is still at the floor at $n=16$.
The bias at $K=1.5$ is the part of the seed injected after the readout (Sec.~\ref{sec:branches}): the twisted coupling divided by its first-order growth is, in the median over $\chi$, $0.91$ and $0.97$ of its value at $n=24$ at $n=16$ and $20$, and $\xi\propto\gd^{-2}$ doubles this deficit.
From $n=20$ on the bias is nearly common to all $\chi$, so a single readout identifies which values of $\chi$ give long runs; at $n=16$ the scatter in $\ln$ is still $0.14$ at $K=1.5$, and a few runs, e.g., $\chi=33$ ($1.10$ at $n=20$), have a later injection of the opposite sign.
These runs were made on a machine other than that of the main runs, with the same software versions, and reproduce the deposited $X(n)$ to $10^{-9}$ up to $n=12$ and to $1.2\times10^{-4}$ up to $n=24$ (\nolinkurl{runs/early_xi/}).

\emph{Number of calibration runs.}
The response to $\mu$ is linear: the slope changes by at most $1.9\%$ between $\epsilon/2$ and $\epsilon/4$ in the groups of Table~\ref{tab:signed} ($2.5\%$ at $K=1.75$, $\chi=32$), and the even part of the response is at most $1.9\%$ of the odd part.
Since $\delta_\pi$ is linear in $\mu$, two short runs, at $\mu=0$ and at one nonzero $\mu$, determine $\mupred$ in principle, and the remaining five runs test the linearity.
With the second run at $\mu=\pm\epsilon/4$, the two-run slope differs from the calibrated one by at most $1.7\%$ in the three groups in which we tested it ($K=1$, $\chi=40$ and $K=1.5$, $\chi=24$, $40$).
The amplitude must, however, be large enough that the irregular component is small compared with the response: at $K=1.5$ and $\chi=40$ the two-run slope is off by up to $10\%$ at $|\mu|=10^{-5}$, by a factor of six at $\mu=10^{-6}$, and has the wrong sign at $\mu=10^{-7}$.
By contrast, the unsigned $\gd^2(12)$ of Sec.~\ref{sec:compensation} needs at least three full-length runs, $\mu=0,\pm\mu_0$, because two points of the parabola leave the sign of $g_{\rm seed}+\rho\mu$ undetermined.
Taking three such runs from the scans of Table~\ref{tab:b9} shows that they would locate $\mu^*$ as well as the signed calibration, to within about $2\%$ ($4.5\%$ for one pair at $K=1.25$, $\chi=32$) for $K\le1.25$ and for $K=1.5$ at $\chi=40$, provided $0.5\lesssim\mu_0/|\mu^*|\lesssim10$, with $0.6$--$0.9$ times the RG steps of the seven signed calibration runs, since each unsigned run must reach the end of the window ($n_{\rm end}=24$--$63$ at $\mu=0$, compared with $n_{\rm cal}=12$--$32$).
They fail when $\mu_0$ is far from $|\mu^*|$: for $\mu_0\lesssim0.2|\mu^*|$ the error at $K=1.5$, $\chi=32$ is $12$--$330\%$, and for $\mu_0\gtrsim10|\mu^*|$ the scatter of $\ln\gd^2$ about the parabola reverses the sign of $g_{\rm seed}$ in up to $39\%$ of noise realizations at $K=1.5$ ($\chi=24$, $32$) and $1.75$; at $K=1.75$ and $\chi=32$ this scatter ($0.6$) keeps the error above about $10\%$ for any $\mu_0$.
Since $|\mu^*|$ ranges from $1.5\times10^{-6}$ to $1.2\times10^{-4}$ in Table~\ref{tab:b9} and $\mu^*$ changes sign with $\chi$, $\mu_0$ cannot be chosen in advance, whereas the signed calibration uses, for each $K$, a fixed amplitude $\epsilon$ chosen in advance that does not depend on $\mu^*$ (\nolinkurl{runs/three_point_check/}).

\emph{Irregular component.}
To check whether the residual splitting of the one-shot runs is a round-off floor, we added relative noise of $10^{-14}$ to the initial tensor.
This changes $\delta_\pi(24)$ by $0.4\%$ and leaves $n_e$ of the compensated run unchanged.
The residual comes instead from a component of the response that is not proportional to $\mu$.
At $K=1.5$ and $\chi=40$ it amounts to $2.4\%$ of the uncompensated splitting already at $\mu=10^{-7}$ and to $1$--$9\%$ for $10^{-5}\le|\mu|\le2.5\times10^{-4}$.
The effective slope then varies between $-133$ and $-152$ (calibrated value $-149$), and between $-148$ and $-163$ at $\chi=24$.
At $K=1$ the component is about $0.3\%$ of the splitting, and the slope varies between $-321$ and $-327$ (calibrated value $-327$).
The longer delays of some parabola scans indicate that their grid points happened to fall where the irregular component was small.
The component presumably has the same origin as the jumps of the seed between neighboring $\chi$, which depend on the loop optimization as well as on $\chi$ (Appendix~\ref{app:seed}).
In the five groups of Table~\ref{tab:signed}, the one-shot counterterm delays the breakdown by $13$--$19$ steps; at $K=1.5$ the delays $\ln F/\ln2^{2-K}$ predicted from the remaining splitting, $17$, $14$, and $14$ steps at $\chi=24$, $32$, and $40$, agree with the measured $19$, $14$, and $13$ steps.

\emph{Delays after refinement.}
The refinement stops when $|\delta_\pi(n_{\rm cal})|$ falls below $0.1\%$ of its uncompensated value or after three additional short runs.
In the three groups in which the refinement succeeds, the delays, 17, 19, and 23 steps, are shorter than the $29$--$35$ steps that $\ln F/\ln2^{2-K}$ predicts from the remaining splitting at $n_{\rm cal}$, whereas at $\chi=24$ the remaining $2.7\%$ predicts 21 steps, close to the measured 19.
The shortfall is the part of the seed injected after $n_{\rm cal}$: a counterterm that cancels the amplitude present at $n_{\rm cal}$ leaves the fraction $f_{\rm cal}$ of Table~\ref{tab:campaign} ($1.3\%$ at $K=1.25$ and $2.4$--$2.8\%$ at $K=1.5$).
Adding the signed residual splitting after refinement gives expected delays of $16$, $17$, $20$, and $22$ steps for $K=1.25$ and $K=1.5$, $\chi=24$, $32$, $40$, against the measured $17$, $19$, $19$, and $23$.
The same extrapolation gives $13$--$16$ steps at $K=1$ ($\chi=32$--$48$) and $28$--$29$ at $K=1.75$, against measured one-shot delays of $11$--$14$ and $29$--$36$; the differences of up to 7 steps reflect the irregular component.

\emph{Acceptance criteria.}
The calibration accepts a group for compensation only if $s$ changes by at most $5\%$ between $\epsilon/2$ and $\epsilon/4$, the even response is at most $5\%$ of the odd one, the numerical checks pass, and $|\mupred|\le2\epsilon$.
All groups pass except $K=1.75$, $\chi=24$, whose slopes at $\epsilon/2$ and $\epsilon/4$ differ by more than $5\%$; it was held without a compensated run.
A criterion added after the fact, that $\mupred(n)$ change by less than $20\%$ between the last two readings before $n_{\rm cal}$, would have flagged the failed group $K=1$, $\chi=24$ and the held group, and no other group with $n_{\rm cal}\ge16$.
The amplitudes are $\epsilon=10^{-5}$ at $K=0.75$, $10^{-4}$ at $K=1$, $5\times10^{-4}$ at $K=1.25$, $10^{-3}$ at $K=1.5$, and $2\times10^{-3}$ at $K=1.75$.
Runs of the campaign with $0.9\,\mupred$ and $1.1\,\mupred$ at $K=1.5$ give $n_e=55$ and $56$ ($\chi=24$), $55$ and $57$ ($\chi=32$), and $58$ and $67$ ($\chi=40$).

\emph{The full campaign.}
The pre-registered campaign, i.e., one fixed in advance of the runs, covered 19 groups (Table~\ref{tab:campaign}).
For each, $n_{\rm cal}$, $\epsilon$, and the acceptance criteria were fixed in advance and $\mupred$ was frozen before the compensated run; 17 groups were compensated, one was held at the acceptance gate, and $K=1.5$, $\chi=64$ was calibrated but not compensated.
The campaign ran on the third machine of the reproducibility check below, so its $n_e$ differ from Table~\ref{tab:signed} by up to four steps.

\begin{table*}[tb]
  \caption{All groups of the twist campaign ($O(2)$): calibration step, frozen counterterm $\mupred$, $\mu^*$ from the parabola where available (Table~\ref{tab:b9}), $n_e$ at $\mu=0$ and with $\mupred$, and the delay. Last two columns: fraction $f_{\rm cal}$ (in percent) of the asymptotic amplitude still missing at $n_{\rm cal}$, from the fit $\mu_\infty+\mu_1\,r^{-n}$ to $\mupred(n)$ for $n\ge12$ (Sec.~\ref{sec:branches}), and the delay $2\ln(1/|f_{\rm cal}|)/\ln2^{2-K}$ expected from it alone. At $K=0.75$ ($n_{\rm cal}=12$) the fit is not possible; ``sign change'': $\mupred(n)$ changes sign after $n=8$.}
  \label{tab:campaign}
  \begin{ruledtabular}
    \begin{tabular}{rrrccccrr}
      $K$ & $\chi$ & $n_{\rm cal}$ & $\mupred$ & $\mu^*$ & $n_e$ & delay & $f_{\rm cal}$ & extrap. \\
      \hline
            0.75 & 24 & 12 & $-1.97\times10^{-5}$ & --- & $21\to28$ & $7$ & --- & --- \\
      0.75 & 32 & 12 & $2.28\times10^{-6}$ & $2.6\times10^{-6}$ & $26\to31$ & $5$ & --- & --- \\
      0.75 & 40 & 12 & $1.52\times10^{-6}$ & $1.5\times10^{-6}$ & $27\to36$ & $9$ & --- & --- \\
      1.00 & 24 & 16 & $9.71\times10^{-8}$ & --- & $35\to34$ & $-1$ & sign change & --- \\
      1.00 & 32 & 16 & $5.27\times10^{-6}$ & $5.4\times10^{-6}$ & $29\to41$ & $12$ & $0.4$ & $16$ \\
      1.00 & 40 & 16 & $4.12\times10^{-6}$ & $4.2\times10^{-6}$ & $30\to44$ & $14$ & $1.0$ & $13$ \\
      1.00 & 48 & 16 & $1.66\times10^{-6}$ & --- & $32\to43$ & $11$ & $0.5$ & $15$ \\
      1.00 & 64 & 16 & $7.69\times10^{-7}$ & --- & $35\to43$ & $8$ & $4.7$ & $9$ \\
      1.25 & 24 & 20 & $-4.23\times10^{-5}$ & --- & $30\to39$ & $9$ & $0.9$ & $18$ \\
      1.25 & 32 & 20 & $7.70\times10^{-6}$ & $7.2\times10^{-6}$ & $37\to47$ & $10$ & $-0.9$ & $18$ \\
      1.25 & 40 & 20 & $6.00\times10^{-6}$ & $6.1\times10^{-6}$ & $38\to53$ & $15$ & $1.3$ & $17$ \\
      1.50 & 24 & 24 & $-1.02\times10^{-4}$ & $-1.0\times10^{-4}$ & $39\to62$ & $23$ & $2.8$ & $21$ \\
      1.50 & 32 & 24 & $3.46\times10^{-5}$ & $3.6\times10^{-5}$ & $45\to58$ & $13$ & $2.4$ & $21$ \\
      1.50 & 40 & 24 & $2.13\times10^{-5}$ & $2.2\times10^{-5}$ & $48\to60$ & $12$ & $2.4$ & $21$ \\
      1.50 & 48 & 24 & $1.81\times10^{-5}$ & --- & $49\to71$ & $22$ & $3.6$ & $19$ \\
      1.50 & 64 & 24 & --- & --- & $56$ (no $\mupred$) & --- & $4.5$ & $18$ \\
      1.75 & 24 & 32 & --- & --- & $75$ (held) & --- & sign change & --- \\
      1.75 & 32 & 32 & $1.06\times10^{-4}$ & $1.2\times10^{-4}$ & $72\to108$ & $36$ & $8.3$ & $29$ \\
      1.75 & 40 & 32 & $1.09\times10^{-4}$ & $1.2\times10^{-4}$ & $72\to101$ & $29$ & $8.5$ & $28$ \\

    \end{tabular}
  \end{ruledtabular}
\end{table*}

\emph{Runs with three constraints.}
In earlier runs the anisotropy was not constrained: only the components of the gradient that change the normalization of each tensor and the center $\bar x_\pi$ were removed, and each correction was followed by at most one further Newton step.
The correction was repeated every four RG steps, which suffices at $K=1$ and $1.5$ but not at the BKT point (Sec.~\ref{sec:iterative}).
At $K=1$ the corrections start at step 12, and at $K=1.5$ they start after the initial counterterm of Sec.~\ref{sec:signed} (Table~\ref{tab:three}).
Every correction brings $|\delta_\pi|$ below $10^{-8}$.
After step 160 the splitting regrows between corrections at $\chi=24$ to at most $2.1\times10^{-3}$ at $K=1$ (step 176) and $1.2\times10^{-3}$ at $K=1.5$ (step 248), and the relative size of each update stays below $10^{-3}$.

\begin{table}[tb]
  \caption{Runs with three constraints, corrected every four steps: range of steps, number of corrections, $K_n$ at the last step, and deviation of the levels not used for the correction from the compact boson at $K_n$, determined as in Ref.~\cite[Sec.~IV F]{Paper1a} from the lowest charged level and the $e=\pm1$ pair. (a) Central charge from the metric of four vacuum eigenvalues in two orientations, at step 512; one excited charge-two level deviates by about $1.5\%$. (b) At step 512; 16 of the 477 states before correction (steps 36--512) exceed $1\%$, with a maximum of $1.22\%$. (c) Charge-one and charge-two excitations over all 152 stored states (121 steps and the 31 corrected states); at step 160, $x_{0,2}$, $x_{0,3}$, and the current level deviate by $0.02\%$, $0.04\%$, and $0.01\%$, and $\mathsf c=0.99999$.}
  \label{tab:three}
  \begin{ruledtabular}
    \begin{tabular}{rrrrcl}
      $K$ & $\chi$ & steps & corrections & $K_n$ & unused levels \\
      \hline
      1.0 & 24 & 12--512 & 126 & 0.879 & $\mathsf c=0.9997$ (a) \\
      1.5 & 24 & 36--512 & 120 & 1.529 & $<0.8\%$ (b) \\
      1.5 & 40 & 40--160 &  31 & 1.492 & $\le0.05\%$, $\le0.2\%$ (c) \\
    \end{tabular}
  \end{ruledtabular}
\end{table}

The levels not used for the correction remain those of the compact boson at the slowly drifting $K_n$ (Table~\ref{tab:three}).
At $K=1.5$ and $\chi=24$ the drift is upward, whereas it is downward at $\chi=32$ and $40$; in the uncompensated runs the sign of the slow drift at this $\chi$ depends on the window~\cite[App.~D]{Paper1a}.
At $K=1$ and $\chi=24$ the central charge read with the fixed modular parameter $\tau=i/2$ oscillates with period four in $n$, as expected for the spin-two operators $T$ and $\bar T$, whose couplings rotate by $270^\circ$ per step~\cite{Paper1a}.
Determining the metric from the four vacuum eigenvalues of the tensor read in two orientations gives $\mathsf c=0.9997$ at step 512, but an excited charge-two level still deviates by about $1.5\%$ there, which is not yet understood.

\begin{figure*}[tb]
  \centering
  \includegraphics[width=0.8\textwidth]{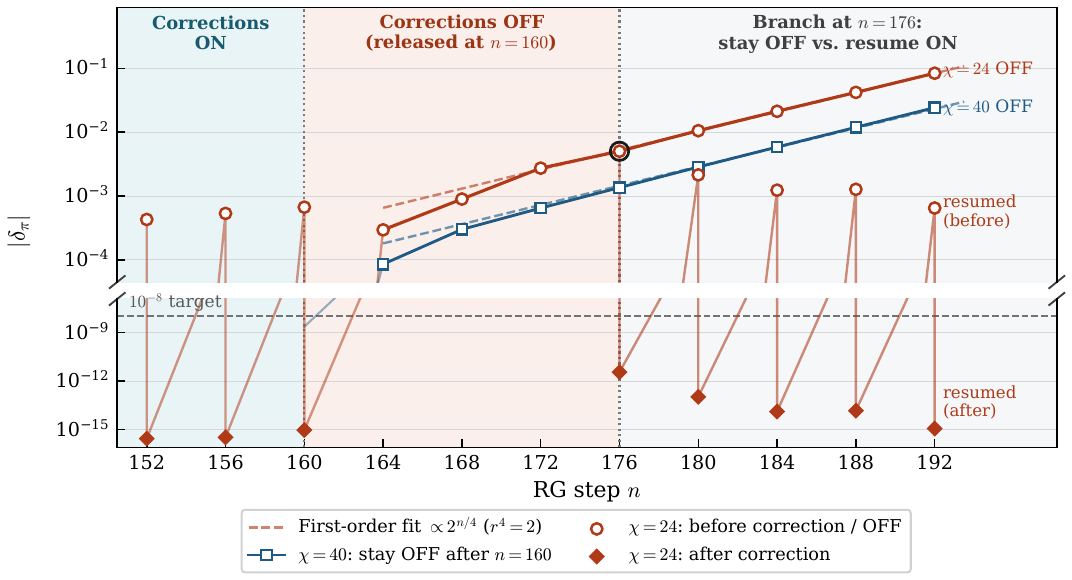}
  \caption{Release and recovery of the corrections with three constraints at $K=1.5$. Up to $n=160$ (blue shading) the corrections every four steps keep $|\delta_\pi|$ below $10^{-8}$ (dashed line) after each correction. Without corrections after $n=160$ (red shading), $|\delta_\pi|$ regrows at the first-order rate $\propto2^{n/4}$ (dashed fits) and reaches $0.084$ ($\chi=24$) and $0.024$ ($\chi=40$) at $n=192$. In a branch at $n=176$ ($\chi=24$, circled), the corrections resume: two Newton updates take $|\delta_\pi|$ from $5.0\times10^{-3}$ to $4.5\times10^{-6}$ and $3.6\times10^{-12}$, and the corrections every four steps then continue.}
  \label{fig:release}
\end{figure*}

Continuing the runs without corrections from the compensated tensors at step 160 confirms that the correction is needed (Fig.~\ref{fig:release}).
The splitting grows to $|\delta_\pi|=0.08$ ($\chi=24$) and $0.02$ ($\chi=40$) at step 192; in the later steps it grows by about a factor of two per four steps, the first-order rate $(2^{(2-K)/2})^4=2$ of $\gd$ at $K=1.5$.
In a branch resumed at step 176 ($\chi=24$), the first correction returns $|\delta_\pi|$ to $4\times10^{-12}$, and the levels not used for the correction then stay within $1\%$ of their predicted values up to step 256.
The procedure does not correct the marginal drift of $K$, which proceeds at about the same rate before and after the corrections start, and the direction of the correction is not purely that of $\Vd$.
The anisotropy of the coarse-grained tensors grows steadily under the corrections.
We measure it by the deviation from its isotropic value of the modular parameter $\tau_0$ obtained from the vacuum eigenvalues.
Like the measure $\ln(\Lambda_0^{(90^\circ)}/\Lambda_0)$ of Sec.~\ref{sec:iterative}, which compares the vacuum eigenvalues of the tensor and of its $90^\circ$-rotated copy, it is of first order in the anisotropy of the coarse-grained tensors and vanishes for an isotropic tensor; the four-constraint runs use the latter because it can be constrained directly.
At $K=1$ and $\chi=24$ it rises from $0.015\%$ at step 12 to about $1.4\%$ at step 512, without taking the run off the critical line.

\emph{Reproducibility.}
Near the compensated point, $n_e$ depends on the floating-point environment.
Rerunning the input $\mu=-1.0\times10^{-4}$ of Table~\ref{tab:b9} ($K=1.5$, $\chi=24$) on the workstation of the main runs gives $n_e=67$, as on the Miyabi-C system at JCAHPC, whereas a third machine gives $65$.
The values of Table~\ref{tab:b9} are thus reproduced on two of the three machines.
With $\mupred$ the workstation and the third machine give $n_e=44$ and $44$ ($K=1$, $\chi=40$), $58$ and $62$ ($K=1.5$, $\chi=24$), $59$ and $58$ ($\chi=32$), and $61$ and $60$ ($\chi=40$), so the two machines differ by up to four steps.

\section{Conventions and checks of the prediction}
\label{app:conventions}

\emph{Equivalent forms.}
Two of us analyzed the runs independently (Sec.~\ref{sec:prediction}).
The second analysis, which uses the window $|X|\le0.05$, writes the coupling as $h(n)=\tilde h\,R_n^{2-K}$ with $R_n=L/2\pi$, which is $\gd(n)$ of Eq.~\eqref{eq:g2}, and uses the mass formula in the form of Ref.~\cite{Takayoshi2010},
\begin{equation}
  \begin{gathered}
    M=C_K\,[B_K|\tilde h|]^{1/(2-K)},\\
    C_K=\frac{2\Gamma(\frac p2)}{\sqrt\pi\,\Gamma(\frac{1+p}2)},\qquad B_K=\frac{\pi}{\sqrt2}\frac{\Gamma(1-\frac K2)}{\Gamma(\frac K2)} .
  \end{gathered}
\end{equation}
Solving Eq.~\eqref{eq:zam} for $M$ with $\lambda_\SG=|\tilde g_{\mathrm d}|/\sqrt2$ and $\beta^2=K/2$ gives exactly this expression, since $\Gamma(\beta^2)/\Gamma(1-\beta^2)=\Gamma(\frac K2)/\Gamma(1-\frac K2)$ and $[\sqrt\pi\,\Gamma(\frac{p+1}2)/2\Gamma(\frac p2)]^{-1}=C_K$.
Numerically the two forms agree to machine precision for all $K$ used here.
The two analyses therefore differ only in how the coupling and the length are read.
They use different windows ($|X|\le0.02$ against $|X|\le0.05$), the second analysis includes an umklapp term in the background for $K<0.75$, and the estimators of $\xi$ differ (average of the last two steps against the median of the last three converged steps).
Table~\ref{tab:crosscheck} compares them on the runs they share.

\emph{Bias of the window cut.}
With the cut $|X|\le0.1$ the ratio of predicted to measured $\xi$ drops at large $K$ (Sec.~\ref{sec:prediction}).
This shortfall cannot be explained by a shift of $K$ in Eq.~\eqref{eq:xi}, since matching at $K=1.5$ would require $\delta K\approx+0.006$--$0.022$ (median $+0.013$), obtained by varying $K$ in Eq.~\eqref{eq:xi} and in $4K\alpha_{0,1}(K)$ of Eq.~\eqref{eq:g2} at fixed amplitude $|A|\approx0.07$--$0.10$ of the growing term at $n_{\rm end}$, opposite in sign to the drift estimated in Ref.~\cite{Paper1a}.
Comparing the cuts $0.02$, $0.05$, and $0.1$ shows instead that it is a bias of the readout.
We define the bias as the difference between $\ln\gd^2$ read in the window and $\ln\gd^2$ inferred from the measured $\xi$ through Eq.~\eqref{eq:xi}, which equals $-2(2-K)$ times the logarithm of the ratio.
For $K\le1.5$ the medians of this difference, excluding $\chi=16$, are linear in the cut to within $0.0015$, with the slopes of Table~\ref{tab:xi}.
The signs match those of the deviations of the free-rate growth rate in Ref.~\cite{Paper1a}, which also disappear as the cut is lowered: there the rate is too low for $K\le1$, where terms beyond second order enter as the coupling grows, and too high for $K\ge1.5$, where $\gd^2$ makes $K$ run.
At $K=1.75$ the ratio also increases with the length of the window, from an average of $1.014$ for the eleven runs whose window ends at $n=56$--$69$ to $1.030$ for the five ending at $n=70$--$79$ and $1.067$ for the seven ending at $n=80$--$89$.
The longest windows belong to runs close to the compensation point $\mu^*$ of Sec.~\ref{sec:compensation}, with $\xi$ up to $10^{16}$.
To test whether the drift of $K$ causes this, we determined $K_n$ at each step of the window from the lowest charged level and one level of the $e=\pm1$ pair ($C$ odd; $C$ even at $K=1$), solving for $K_n$ and $\gd^2$ with the second-order coefficients of both levels~\cite[App.~B]{Paper1a}, for all runs of the deformation scans at $\chi=32$ and $40$.
At $K=1.75$, in all 23 runs of these scans, $K_n$ decreases linearly, by $(6$--$10)\times10^{-5}$ per step, from $1.750$--$1.754$ at $n=12$ to $1.745$--$1.749$ at the end of the window.
With $K_n$ at the end of the window in place of the input $K$ in Eq.~\eqref{eq:xi} and in $4K\alpha_{0,1}(K)$ of Eq.~\eqref{eq:g2}, the three averages become $0.974$, $0.967$, and $0.994$, and the individual ratios $0.91$--$1.04$.
The result depends on the step at which $K_n$ is taken: the median ratio over these 23 runs, $1.034$ with the input $K$ ($1.032$ over the 24 runs of Table~\ref{tab:xi}), becomes $1.017$, $0.978$, and $0.945$ with $K_n$ at the center of the window, at its end, and extrapolated to the last step used for $\xi$.
At $K\le1.5$ the same corrections change the medians by at most $1\%$ with $K_n$ at the end of the window (at $K=1.5$, over the 45 runs of the deformation scans, from $1.007$ to $0.998$) and $2\%$ at the last step.
A second possible cause is the later injection of Sec.~\ref{sec:branches}.
With the constant source found there, $\gd(n)\propto\mu_\infty r^n+\mu_1$, where $r=(\sqrt2)^{2-K}$ and $\mu_\infty+\mu_1 r^{-n}$ is the campaign fit to $\mupred(n)$, so $\gd^2$ contains, besides the term $\propto r^{2n}=r_{\rm d}^{\,n}$ of Eq.~\eqref{eq:fit}, a cross term $\propto2\mu_\infty\mu_1 r^n$.
In $X(n)$ this term is $B\,r_{\rm d}^{(n-n_{\rm end})/2}$, where $B/A=2\mu_1/(\mu_\infty r^{n_{\rm end}})$ is its ratio to the growing term at the end of the window.
With $\mu_1/\mu_\infty$ from the campaign fit at the same $\chi$ (Sec.~\ref{sec:branches}) and $\mu_\infty$ scaled by $(g_{\rm seed}+\rho\mu)/g_{\rm seed}$ (Table~\ref{tab:b9}), $B/A$ lies between $-0.03$ and $+0.03$.
Adding it to Eq.~\eqref{eq:fit} with this fixed $B$ changes the median ratio from $1.034$ to $1.022$ and leaves the window dependence ($1.003$, $1.027$, and $1.051$), so the cross term is a $1\%$ effect.
A fit with $B$ free gives $B/A\approx-0.1$ even at $\mu=0$, ten times the measured source, and is not used.
The systematic uncertainty of about $10\%$ at $K=1.75$ therefore comes from the choice of $K_n$ and from the dependence on the cut (\nolinkurl{runs/kdrift_K175/}).

\emph{Free-fermion check.}
At $K=1$ the six-vertex model is equivalent to free fermions, and the dimerized XX chain $H=\sum_j[1+\delta(-1)^j](S^x_jS^x_{j+1}+S^y_jS^y_{j+1})$ provides an independent test of the chain from Eq.~\eqref{eq:g2} to Eq.~\eqref{eq:xi}.
The velocity is $1$ at $\delta=0$, and the exact single-particle gap is $\delta$, so $M=\delta$.
For chains of $L_{\rm ch}=200$--$3200$ sites with periodic spin boundary conditions, we obtain $x_{0,1}$ from the free-fermion levels with the appropriate fermion boundary conditions and read $\tilde g_{\mathrm d}$ from its second-order shift at $\delta=10^{-5}$ with $\alpha_{0,1}(1)=4\pi^2\ln2$~\cite{Paper1a}.
Equation~\eqref{eq:xi}, which at $K=1$ reads $M=\sqrt2\pi|\tilde g_{\mathrm d}|$, then gives $M/\delta=0.99998$.
This checks the factor $\sqrt2$, the factor $2\pi$, $\alpha_{0,1}$, and $\kappa_{\rm SG}(1)=1/2\pi$ together.
The $K$ dependence of Eq.~\eqref{eq:kappa} is that of Refs.~\cite{Zamolodchikov1995,Lukyanov1997}.

\section{Correlation length at the BKT point}
\label{app:bkt}

\begin{figure*}[tb]
  \centering
  \includegraphics[width=\textwidth]{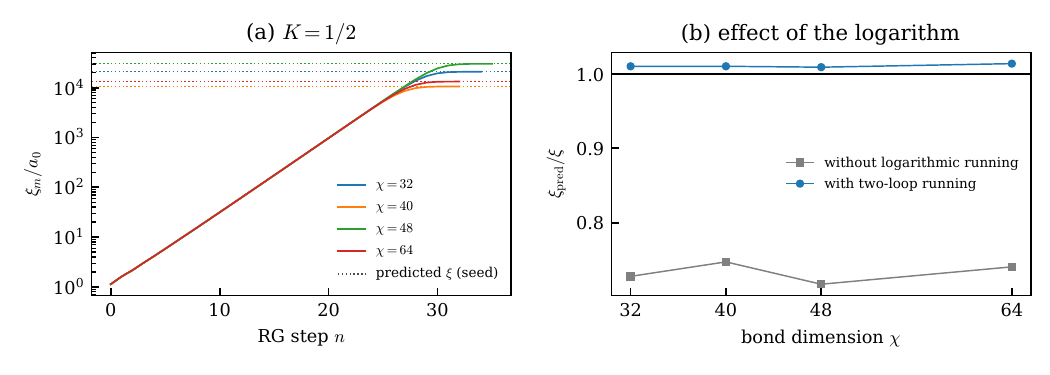}
  \caption{BKT point $K=1/2$, $\chi=32$--$64$. (a) Correlation length $\xi$ of the charged channel against the RG step; dotted lines, prediction from the seed. (b) Ratio $\xi_{\rm pred}/\xi$ of predicted to measured $\xi$, as in Fig.~\ref{fig:ratio}, without (squares) and with (circles) the two-loop running of the marginal coupling.}
  \label{fig:bkt}
\end{figure*}

At $K=1/2$ the umklapp operator is marginal, and together with the current--current coupling it forms the marginally irrelevant coupling $\hat g$ of the $SU(2)_1$ theory~\cite{Affleck1989,Barzykin1999}.
Reference~\cite{Paper1a} found that the levels approach their CFT values logarithmically~\cite{Cardy1986,Affleck1989} before the dimerization grows.
The dimerization is then the $SU(2)$ singlet, whose dimension runs as $x_s=1/2+3\hat g/4$, so its coupling grows as $d\gd/d\ell=(3/2-3\hat g/4)\gd$.
Integrated with the two-loop flow of $\hat g$~\cite[Eq.~(9)]{Barzykin1999}\cite{Affleck1998}, the growth acquires a logarithmic factor, $\gd\propto R^{3/2}/(\ln R)^{3/4}$ at one loop, with $R=L/2\pi$, and the length predicted from a given seed becomes longer.

To read the seed, we first fit the triplet and singlet over steps $8$--$18$ to determine $\hat g(12)$ and the second-order background.
With $\ell=\ln(R/R_{12})$ and $\hat g(\ell)$ from the numerical integration of $d\hat g/d\ell=-\hat g^2-\hat g^3/2$, the growth equation above integrates to
\begin{equation}
  \gd(R)=\gd(12)\left(\frac{R}{R_{12}}\right)^{3/2}\left[\frac{f(\hat g(R))}{f(\hat g(12))}\right]^{3/4}
\end{equation}
with $f(g)=g/(1+g/2)$,
and we fit the growth with this $R$ dependence in the window $n\ge18$ to obtain $\gd(12)$.
For the pure sine-Gordon model, Eq.~\eqref{eq:xi} at $K=1/2$ ($p=1/3$) reads $MR=[|\gd(R)|/\sqrt2\,\kappa_{\rm SG}(1/3)]^{2/3}$~\cite{Zamolodchikov1995}, the relation also used for dimerized XXZ chains~\cite[Eq.~(11)]{Takayoshi2010}.
We set $\xi_{\rm pred}$ to the radius $R_*$ at which the running coupling gives $MR_*=1$; no coefficient is adjusted to the measured length.
The prediction without the logarithmic running inserts the same $\gd(12)$ into Eq.~\eqref{eq:xi}, i.e., omits the factor with $f$.
Only the flow of $\hat g$ is kept to two loops; terms of order $\hat g^2\gd$ in the growth equation and the feedback of the dimerization on $\hat g$ are omitted.

Without the logarithmic running, the measured $\xi$ is $1.34$--$1.39$ times the prediction ($\xi_{\rm pred}/\xi=0.72$--$0.75$).
With it, $\xi_{\rm pred}/\xi=1.009$--$1.014$ (Fig.~\ref{fig:bkt}).
Moving the matching condition from $MR=1$ to $MR=1/2$ or $2$ changes the prediction by up to $2.8\%$, which sets the uncertainty of this comparison.
The marginal coupling at step 12 is $\hat g(12)=0.143$ from the fit over steps 8--18 and about $0.148$ from the fit over steps 8--20 of Ref.~\cite{Paper1a}, and $0.070$--$0.076$ at the matching step, so the matching to the pure sine-Gordon model remains an approximation.
As at larger $K$, $\xi$ is not monotonic in $\chi$: $\chi=32$ and $40$ give $20\,843$ and $10\,549$, and $\chi=48$ and $64$ give $30\,108$ and $13\,256$, in the order predicted by the seeds.

\section{Correlation length with only \texorpdfstring{$C$}{C} imposed}
\label{app:conly}

With only $C$ imposed~\cite[Sec.~V C]{Paper1a}, the charge-$(\pm1)$ primaries appear as a $C$-even and a $C$-odd level.
The $C$-even coupling $g_{0,2}$ of $(V_{0,2}+V_{0,-2})/\sqrt2$, of dimension $x=1/K$, splits them at first order, $|x_{\rm even}-x_{\rm odd}|=2\sqrt2\,\pi|g_{0,2}|$, by the same argument as for the twisted splitting (Appendix~\ref{app:twist}).
We read $g_{0,2}(12)$ from this splitting in the window of Ref.~\cite{Paper1a} ($n\ge10$, $10^{-5}<8K|\Delta x|<10^{-2}$, with $\Delta x=\tfrac12(x_{\rm even}-x_{\rm odd})$) with the rate fixed to $2^{1-1/(2K)}$, and insert it into Eq.~\eqref{eq:xi} with $K$ replaced by $x$ and $p=x/(2-x)=1/(2K-1)$.
The interaction $V_{0,\pm2}\propto\cos2\theta$ has two minima, and its kinks carry a half-integer $e$, which the untwisted cylinder does not contain.
The gapped phase then has a nearly degenerate pair of ground states, and the lightest excitation in the $C$-odd sector is the first breather, of mass $2M\sin(\pi p/2)$ for $p<1$.
Table~\ref{tab:conly} compares this mass with the lowest $C$-odd level above the ground states at the last step of the $C$-only runs of Ref.~\cite[Sec.~V C]{Paper1a} with $K>1$, where the dimerization is less relevant than $V_{0,\pm2}$.
$C$-odd levels below $0.3M$, at most $0.003M$ in these runs, are counted as ground states, and the excitations lie above $0.76M$.
Over the last two steps $m$ changes by less than $0.2\%$, by $1.2\%$ at $K=1.75$ and $\chi=40$, and by $12\%$ at $K=1.25$ and $\chi=40$, where it is still decreasing.
The last column is the factor by which the residual of the joint fit of Ref.~\cite{Paper1a} decreases when the second-order $V_{0,\pm1}$ term is added, and it is large when $V_{0,\pm1}$ grows; we call a gain below $3$ no signal, $3$--$10$ a weak signal, and above $10$ a strong signal.
For $K\ge1.5$, the three runs without a $V_{0,\pm1}$ signal agree with the prediction to within $1.2\%$, and the two with a weak signal to within $12\%$.
With a strong signal, the gap is larger by factors of $1.9$ to $19$.
At $K=1.25$ the deviations are $-6\%$ and $+26\%$.
There the dimerization, of dimension $1.25$, is only moderately less relevant, and the run at $\chi=32$ shows no pair of nearly degenerate ground states.
The identification of the lowest $C$-odd level with the first breather and the restriction to runs without $V_{0,\pm1}$ are the assumptions of this comparison, which covers only a few runs.

\begin{table}[!htb]
\caption{Only $C$ imposed: coupling $g_{0,2}(12)$ read from the splitting of the charge-$(\pm1)$ levels, predicted soliton length $\xi=1/M$, measured lowest $C$-odd mass $m$ in units of $M$, the first-breather value $2\sin(\pi p/2)$, their ratio, and the gain of the joint fit from the $V_{0,\pm1}$ term (\nolinkurl{runs/conly_correlation_length/}).}
\label{tab:conly}
\begin{ruledtabular}
\begin{tabular}{rrccccrr}
$K$ & $\chi$ & $g_{0,2}(12)$ & $\xi/a_0$ & $m/M$ & $2\sin\frac{\pi p}{2}$ & ratio & gain\\
\hline
1.25 & 32 & $4.1\times10^{-6}$ & $2.4\times10^{5}$ & 2.177 & 1.732 & 1.257 & 1.2\\
1.25 & 40 & $2.1\times10^{-6}$ & $4.2\times10^{5}$ & 1.633 & 1.732 & 0.943 & 1.6\\
1.50 & 24 & $3.8\times10^{-6}$ & $9.1\times10^{4}$ & 3.309 & 1.414 & 2.340 & 28\\
1.50 & 32 & $9.6\times10^{-6}$ & $4.6\times10^{4}$ & 1.431 & 1.414 & 1.012 & 1.3\\
1.50 & 40 & $7.9\times10^{-7}$ & $3.0\times10^{5}$ & 17.50 & 1.414 & 12.37 & 66\\
1.75 & 24 & $6.8\times10^{-7}$ & $1.6\times10^{5}$ & 22.83 & 1.176 & 19.42 & 153\\
1.75 & 32 & $1.6\times10^{-5}$ & $1.8\times10^{4}$ & 1.172 & 1.176 & 0.997 & 2.6\\
1.75 & 40 & $2.0\times10^{-6}$ & $7.6\times10^{4}$ & 1.033 & 1.176 & 0.879 & 3.6\\
2.50 & 24 & $1.2\times10^{-5}$ & $8.2\times10^{3}$ & 1.452 & 0.765 & 1.897 & 12\\
2.50 & 32 & $5.7\times10^{-5}$ & $3.1\times10^{3}$ & 0.769 & 0.765 & 1.005 & 1.9\\
2.50 & 40 & $3.3\times10^{-6}$ & $1.9\times10^{4}$ & 0.826 & 0.765 & 1.079 & 7.3\\
\end{tabular}
\end{ruledtabular}
\end{table}

\bibliography{main_en}

\end{document}